\documentclass[prb,reprint,preprintnumbers,amsmath,amssymb,showpacs,superscriptaddress,citeautoscript,aps,floatfix]{revtex4-2}
\usepackage{amssymb}

\usepackage{graphicx}% Needed to include png/jpg/pdf figure files
\usepackage{float}
\usepackage{amsmath}
\usepackage{bm}
\usepackage[normalem]{ulem}
\usepackage{siunitx}
\usepackage{xcolor}
\usepackage{upgreek}
\usepackage{booktabs}

\usepackage{xr}
\makeatletter

\newcommand*{\addFileDependency}[1]{% argument=file name and extension
\typeout{(#1)}% latexmk will find this if $recorder=0
\@addtofilelist{#1}
\IfFileExists{#1}{}{\typeout{No file #1.}}
}\makeatother

\newcommand*{\myexternaldocument}[1]{%
\externaldocument[SI_]{#1}%
\addFileDependency{#1.tex}%
\addFileDependency{#1.aux}%
}
\myexternaldocument{SI}

\usepackage[breaklinks]{hyperref}%hyperef package does internal and external linking (labels, urls)
\usepackage[all]{hypcap}%addition to hyperref, it ensures figure is correctly displayed (and not only the caption) when link is clicked

\begin{document}

\title{Spin injection and emission helicity switching in a 2D \texorpdfstring{perovskite/WSe$_2$}{perovskite/WSe2} heterostructure}
%\title{All-optical control of spin polarization in a (BA)$_2$PbI$_4$/WSe$_2$ heterostructure}

\author{Jakub Jasi{\'n}ski}
\altaffiliation{These authors contributed equally to the work\\Current affiliation: Institute of Applied Physics and W{\"u}rzburg-Dresden Cluster of Excellence ct.qmat, Technische Universit{\"a}t Dresden, 01062 Dresden, Germany}
\affiliation{Department of Experimental Physics, Faculty of Fundamental Problems of Technology, Wroclaw University of Science and Technology, 50-370 Wroclaw, Poland}
\affiliation{Laboratoire National des Champs Magn\'etiques Intenses, EMFL, CNRS UPR 3228, Universit{\'e} Grenoble Alpes, Universit{\'e} Toulouse, Universit{\'e} Toulouse 3, INSA-T, Grenoble and Toulouse, France}

\author{Francesco Gucci}
\altaffiliation{These authors contributed equally to the work}
\affiliation{Department of Physics, Politecnico di Milano, Piazza Leonardo da Vinci 32, 20133 Milan, Italy}

\author{Thomas Brumme}
\affiliation{Chair of Theoretical Chemistry, Technische Universit{\"a}t Dresden, Bergstra{\ss}e 66, 01069 Dresden, Germany}

\author{Swaroop Palai}
\affiliation{Laboratoire National des Champs Magn\'etiques Intenses, EMFL, CNRS UPR 3228, Universit{\'e} Grenoble Alpes, Universit{\'e} Toulouse, Universit{\'e} Toulouse 3, INSA-T, Grenoble and Toulouse, France}

\author{Armando Genco}
\affiliation{Department of Physics, Politecnico di Milano, Piazza Leonardo da Vinci 32, 20133 Milan, Italy}

\author{Alessandro Baserga}
\affiliation{Department of Physics, Politecnico di Milano, Piazza Leonardo da Vinci 32, 20133 Milan, Italy}

\author{Jonas D.\ Ziegler}
\affiliation{Institute of Applied Physics and W{\"u}rzburg-Dresden Cluster of Excellence ct.qmat, Technische Universit{\"a}t Dresden, 01062 Dresden, Germany}

\author{Takashi Taniguchi}
\affiliation{Research Center for Materials Nanoarchitectonics, National Institute for Materials Science,  1-1 Namiki, Tsukuba 305-0044, Japan}

\author{Kenji Watanabe}
\affiliation{Research Center for Electronic and Optical Materials, National Institute for Materials Science, 1-1 Namiki, Tsukuba 305-0044, Japan}

\author{Mateusz Dyksik}
\affiliation{Department of Experimental Physics, Faculty of Fundamental Problems of Technology, Wroclaw University of Science and Technology, 50-370 Wroclaw, Poland}

\author{Christoph Gadermaier}
\affiliation{Department of Physics, Politecnico di Milano, Piazza Leonardo da Vinci 32, 20133 Milan, Italy}

\author{Micha{\l} Baranowski}
\affiliation{Department of Experimental Physics, Faculty of Fundamental Problems of Technology, Wroclaw University of Science and Technology, 50-370 Wroclaw, Poland}

\author{Duncan K.\ Maude}
\affiliation{Laboratoire National des Champs Magn\'etiques Intenses, EMFL, CNRS UPR 3228, Universit{\'e} Grenoble Alpes, Universit{\'e} Toulouse, Universit{\'e} Toulouse 3, INSA-T, Grenoble and Toulouse, France}

\author{Alexey Chernikov}
\affiliation{Institute of Applied Physics and W{\"u}rzburg-Dresden Cluster of Excellence ct.qmat, Technische Universit{\"a}t Dresden, 01062 Dresden, Germany}

\author{Giulio Cerullo}
\affiliation{Department of Physics, Politecnico di Milano, Piazza Leonardo da Vinci 32, 20133 Milan, Italy}

\author{Agnieszka Kuc}
\affiliation{Helmholtz-Zentrum Dresden-Rossendorf, HZDR, Bautzner Landstra{\ss}e 400, 01328 Dresden, Germany}
\affiliation{Center for Advanced Systems Understanding, CASUS, Conrad-Schiedt-Stra{\ss}e 20, 02826 G\"orlitz, Germany}

\author{Stefano Dal Conte}\email{stefano.dalconte@polimi.it}
\affiliation{Department of Physics, Politecnico di Milano, Piazza Leonardo da Vinci 32, 20133 Milan, Italy}

\author{Paulina Plochocka}\email{paulina.plochocka@lncmi.cnrs.fr}
\affiliation{Department of Experimental Physics, Faculty of Fundamental Problems of Technology, Wroclaw University of Science and Technology, 50-370 Wroclaw, Poland}
\affiliation{Laboratoire National des Champs Magn\'etiques Intenses, EMFL, CNRS UPR 3228, Universit{\'e} Grenoble Alpes, Universit{\'e} Toulouse, Universit{\'e} Toulouse 3, INSA-T, Grenoble and Toulouse, France}

\author{Alessandro Surrente}\email{alessandro.surrente@pwr.edu.pl}
\affiliation{Department of Experimental Physics, Faculty of Fundamental Problems of Technology, Wroclaw University of Science and Technology, 50-370 Wroclaw, Poland}

\date{\today}

\begin{abstract}
The initialization and control of a long-lived spin population in lead halide perovskites are prerequisites for their use in spintronic applications. Here, we demonstrate circular polarization of the interlayer exciton emission in a (BA)$_2$PbI$_4$/WSe$_2$ monolayer heterostructure. The helicity of this emission is controlled by tuning the energy of the excitation laser through the manifold of exciton resonances of the WSe$_2$ monolayer, together with an emerging interlayer absorption feature of the heterostructure. Theoretical calculations show that this resonance arises from hybridized (BA)$_2$PbI$_4$/WSe$_2$ states in the valence band. This hybrid character enables its observation in both linear absorption and ultrafast pump-probe spectroscopies, and plays a key role in controlling the sign of the helicity of the interlayer exciton emission. The tunable spin polarization demonstrated here, with the WSe$_2$ monolayer effectively acting as a tunable spin filter, represents an important step toward the use of 2D perovskites in opto-spintronic applications.
\end{abstract}

%Word count abstract: 149

\maketitle

%\section*{Introduction}
The use of a binary degree of freedom, such as carrier spin, to transport, store, and process information is the basic working principle of spintronic devices \cite{vzutic2004spintronics}. A prerequisite for a material to possess spintronic functionalities \cite{hirohata2020review} suitable for optoelectronics \cite{nishizawa2017pure,holub2007electrical} is the ability to initialize a long-lived population of spin-polarized charge carriers. Lead halide perovskites exhibit a spin-polarized band structure \cite{zhai2017giant,niesner2018structural,niesner2016giant,yin2021manipulation}, which is intrinsically related to their strong spin-orbit coupling \cite{even2013importance}, and helicity-dependent optical selection rules \cite{odenthal2017spin,chen2018impact,zhan2022stimulating,long2018spin,singh2023valley,giovanni2015highly,kopteva2024highly}. These features make them excellent candidates for applications in opto-spintronics \cite{lu2024spintronic}. While a large spin-orbit coupling is essential to efficiently initialize the spin polarization, it also leads to very short spin lifetimes, usually of the order of a few picoseconds \cite{giovanni2015highly,chen2018impact,yumoto2022rapidly,bourelle2020exciton,chen2021tuning}. 

Multiple attempts have been made to overcome this limitation by tailoring structural parameters \cite{chen2018impact,kopteva2024highly} or electronic structure \cite{chen2021tuning}, exploiting polaronic effects \cite{bourelle2022optical}, incorporating chiral organic spacers in two-dimensional (2D) lead halide perovskites \cite{long2020chiral,long2018spin,ma2019chiral,lu2020highly,jana2020organic,wang2023chirality,chen2019circularly}, or decoupling the generation of spin-polarized carriers in moieties with chiral character and their radiative recombination in materials with a high optical quality \cite{kim2021chiral,ye2022core,zhang2024efficient,zhan2022stimulating,chen2020manipulation,chen2020robust,huang2021enhancing}. However, these efforts have often resulted in a relatively low degree of circular polarization of the emitted photoluminescence (PL) \cite{kim2021chiral,ye2022core,zhang2024efficient,zhan2022stimulating}, fundamentally limited by the strong spin-orbit coupling present in these materials \cite{hautzinger2024room}, and a low PL quantum yield \cite{long2018spin,ma2021recent,wang2024spin}. This strongly motivates the exploration of alternative approaches in which spin injection can be deliberately tailored for specific applications.

A class of materials that naturally lend themselves to acting as spin filters is TMD monolayers, due to their chiral optical selection rules \cite{xiao2012coupled}. TMD/2D perovskite heterostructures \cite{baranowski2022two} have been extensively investigated for applications in high-efficiency photodetectors \cite{fu2019ultrathin,fang2019control,wang2020optoelectronic,zhou2022van} and for the study of charge and energy transfer \cite{zhang2020excitonic,chen2020robust,karpinska2021nonradiative,karpinska2022interlayer,singh2023valley,soni2024long,ghosh2025exciton,qin2025carrier}. However, their use in the generation and transfer of spin-polarized carriers has remained limited \cite{singh2023valley}. The combination of chiral optical selection rules \cite{xiao2012coupled} and type II band alignment \cite{karpinska2021nonradiative,karpinska2022interlayer} enables efficient photoexcitation of spin-polarized carriers with a prolonged spin lifetime \cite{rivera2016valley}. Additionally, this approach offers improved control over the sign of the spin polarization, which is determined purely optically, via the helicity of the excitation light. These factors help overcome the limitations inherent to lead halide perovskites, which makes TMD/2D perovskite heterostructures highly promising for optospintronics.

Here, we exploit the chiral optical selection rules of WSe$_2$ monolayers \cite{xiao2012coupled} to inject a population of carriers with an externally controllable spin polarization into a (BA)$_2$PbI$_4$/WSe$_2$ monolayer heterostructure. This leads to a circularly polarized PL of the interlayer exciton (IX). We achieve control over the sign of the PL circular polarization of the IX by tuning the excitation laser energy. Strikingly, we observe counter-polarized emission when the excitation laser is tuned in resonance with a heterostructure-related absorption feature, attributed to an interlayer charge transfer exciton (X$^{\text{CT}}$) transition at higher energies. While this feature has been recently ascribed to hot electron transfer \cite{ghosh2025exciton}, we reveal its interlayer excitonic nature using a combination of density functional theory (DFT), helicity-resolved pump-probe spectroscopy, and linear absorption spectroscopy. Our results thus demonstrate a highly tunable approach for injecting spin-filtered carriers into a 2D perovskite and achieving all-optical control over the helicity of the emitted PL.

\begin{figure*}[!ht]
\centering
\includegraphics[width=1.0\linewidth]{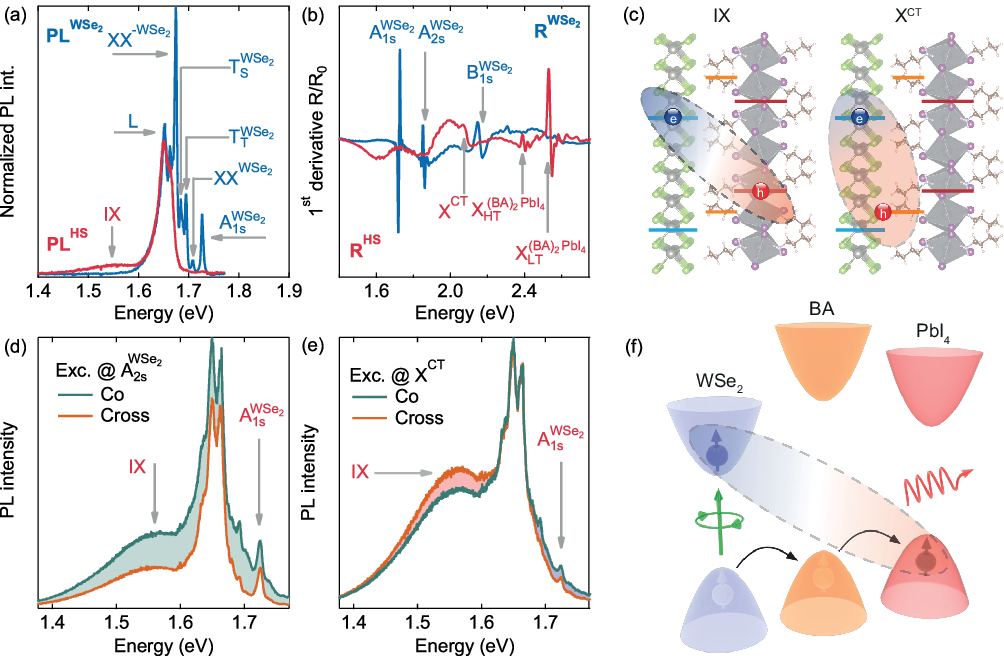}
\caption{(a) PL and (b) reflectivity spectrum of the WSe$_2$ monolayer and the heterostructure. A$_{1\text{s}}^{\text{WSe}_2}$ indicates the 1s state of neutral exciton of WSe$_2$, A$_{2\text{s}}^{\text{WSe}_2}$ the 2s state, B$_{1\text{s}}^{\text{WSe}_2}$ the B exciton, XX$^{\text{WSe}_2}$ the biexciton, T$_{\text{T}}^{\text{WSe}_2}$ and T$_{\text{S}}^{\text{WSe}_2}$ the triplet and singlet charged excitons, respectively, XX$^{\text{-WSe}_2}$ the charged biexciton, and L the localized excitons. X$^{\text{CT}}$ designates the charge transfer interlayer exciton, and X$_{\text{LT}}^{\text{(BA)}_2\text{PbI}_4}$ and X$_{\text{HT}}^{\text{(BA)}_2\text{PbI}_4}$ the exciton of the low and high temperature phases of (BA)$_2$PbI$_4$, respectively. (c) Ball and stick model of the WSe$_2$ monolayer/(BA)$_2$PbI$_4$ heterostructure with a schematic depiction of IX and X$^{\text{CT}}$. The horizontal lines indicate the relative position of the band edges.  Helicity-resolved PL spectrum of the WSe$_2$ monolayer/(BA)$_2$PbI$_4$ heterostructure for excitation in resonance with (d) the 2s exciton of WSe$_2$ (A$_{2\text{s}}^{\text{WSe}_2}$) and (e) the X$^{\text{CT}}$ transition. The excitonic resonances are indicated by arrows. (f) Schematic band alignment of the TMD-perovskite heterostructure. Spin-conserving charge transfer of the hole from WSe$_2$ to PbI$_4$ and formation of an interlayer exciton are schematically shown.}
\label{fig:PL_ChargeTransfer_PolarizationIX}
\end{figure*}

%\section*{Results and discussion}
%\subsection*{Reversed photoluminescence helicity}
The heterostructure used in our investigation was fabricated by vertically stacking a (BA)$_2$PbI$_4$ flake and a WSe$_2$ monolayer and is fully encapsulated in hBN to improve optical quality \cite{cadiz2017excitonic} and minimize degradation under illumination or exposure to ambient conditions \cite{seitz2019long,ziegler2020fast}. A micrograph of the sample is shown in Fig.~\ref{SI_fig:MicrographPLmaps}(a) of the Supplementary Information. The micro-PL ($\upmu$PL) and reflectivity spectra of the individual materials %consist of multiple resonances, attributed to localized, neutral, and charged excitonic species. These assignments 
are discussed in Fig.~\ref{SI_fig:ReflectivityPL}. In the heterostructure, the PL peaks related to the recombination of the intralayer exciton species of WSe$_2$ are strongly quenched, as shown in Fig.~\ref{fig:PL_ChargeTransfer_PolarizationIX}(a) (see also the PL map in Fig.~\ref{SI_fig:MicrographPLmaps}(c)), due to charge and energy transfer across the heterostructure. The PL spectrum displays an additional broad peak at \SI{1.55}{\eV}, which is only observed in the heterostructure region, as confirmed by the PL map in Fig.~\ref{SI_fig:MicrographPLmaps}(b). We attribute this feature to the recombination of an IX, with the electron and the hole spatially separated in the lead-halide octahedral plane and the TMD monolayer, respectively, enabled by the type II band alignment \cite{zhang2020excitonic,chen2020robust,karpinska2022interlayer,singh2023valley}. This band alignment, obtained by the band structure calculated using DFT (see Fig.~\ref{SI_fig:BandStructure_DOS}), and the IX complex are shown schematically in the left panel of Fig.\ \ref{fig:PL_ChargeTransfer_PolarizationIX}(c). Additional characteristic features of the IX are illustrated in the power and temperature dependence of the PL spectrum (Figs.~\ref{SI_fig:PowerDependence} and \ref{SI_fig:TemperatureDependence}) \cite{yang2023organic}. In the reflectivity spectrum of the heterostructure, shown in Fig.\ \ref{fig:PL_ChargeTransfer_PolarizationIX}(b), we identify the excitonic resonances of the individual materials and an additional resonance at $\simeq \SI{2.1}{\eV}$. We attribute this feature to an interlayer hybridized charge transfer exciton X$^{\text{CT}}$ (see discussion of DFT calculations for more details), in which the hole occupies a hybridized state involving primarily the organic spacer and the TMD, as shown schematically in the right panel of Fig.\ \ref{fig:PL_ChargeTransfer_PolarizationIX}(c). %The large energy separation of $\approx \SI{80}{\milli\eV}$ from the B exciton of WSe$_2$ allows us to rule out a charged exciton state as the origin of this resonance.

To demonstrate control over the circular polarization of the IX emission, we perform helicity-resolved PL measurements. We initially excite with a circularly polarized laser tuned in resonance with the 2s exciton of WSe$_2$. The IX emission displays a strong co-circular polarization, inherited from the WSe$_2$ intralayer exciton \cite{jasinski2024control}, as shown in Fig.~\ref{fig:PL_ChargeTransfer_PolarizationIX}(d). The degree of circular polarization $P_{\text{c}}$, defined as $P_{\text{c}} = (I_{\text{co}} - I_{\text{cr}})/(I_{\text{co}} + I_{\text{cr}})$, where $I_{\text{co/cr}}$ denotes the intensity of the co-polarized and cross-polarized emission, respectively, amounts to $\simeq 32\%$ for the IX and to $\simeq 26\%$ for the WSe$_2$ intralayer exciton. For intralayer excitons, co-polarized PL arises from the spin-valley locking exhibited by TMD monolayers \cite{xiao2012coupled}. The co-polarized PL of the IX indicates that the spin of the holes transferred from WSe$_2$ to (BA)$_2$PbI$_4$ is preserved during charge transfer. After the holes transfer to the 2D perovskite, IXs preferentially spin-polarized as the excitation laser recombine, which results in co-polarized PL \cite{singh2023valley} (see schematic in Fig.~\ref{fig:PL_ChargeTransfer_PolarizationIX}(f)).

Intriguingly, we can invert the sign of the degree of circular polarization of the IX emission by tuning the excitation energy of the laser. The polarization-resolved PL spectrum of the heterostructure excited in proximity of the B exciton of the WSe$_2$ monolayer is shown in Fig.~\ref{fig:PL_ChargeTransfer_PolarizationIX}(e). For this excitation condition, the PL of WSe$_2$ monolayers is negligible or slightly negative \cite{jasinski2024control}. Unexpectedly, the IX PL exhibits opposite helicity with respect to the excitation laser. This effect is maximized when the laser is tuned to resonance with X$^{\text{CT}}$. 
\begin{figure*}[!ht]
\centering
\includegraphics[width=1.0\linewidth]{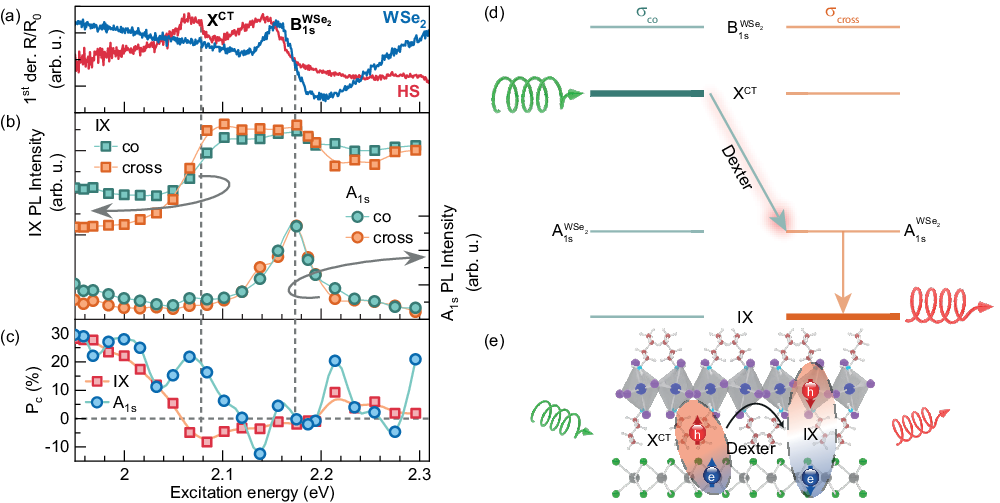}
\caption{(a) First derivative of the reflectivity contrast spectrum measured on the WSe$_2$ monolayer and on the heterostructure. (b) Helicity resolved PL intensity of the interlayer exciton and of the 1s exciton of WSe$_2$ monolayer and (c) degree of circular polarization $P_{\text{c}}$ as a function of the excitation energy. The vertical dashed lines indicate excitonic resonances. (d) Schematic of transfer of the exciton population towards states not driven optically but with the same spin configuration as the optically driven state mediated by Dexter-like coupling. (e) Pictorial view of the reversed circular polarization emission of the interlayer exciton.}
\label{fig:PolarizationResolvedPLE}
\end{figure*} 

We measure the degree of polarization of the PL spectrum as a function of the photon energy of the excitation laser in the spectral region corresponding to the reflectivity spectrum in Fig.\ \ref{fig:PolarizationResolvedPLE}(a). The PL intensity of WSe$_2$ has a maximum for excitation energies corresponding to the B exciton of the WSe$_2$ monolayer, as shown in Fig.\,\ref{fig:PolarizationResolvedPLE}(b), while its degree of polarization exhibits a minimum at this resonance \cite{jasinski2024control}. Subsequently, we consider the helicity-resolved PL spectrum of the IX as a function of the excitation energy. Its intensity, displayed in Fig.~\ref{fig:PolarizationResolvedPLE}(b), reaches its maximum over a relatively broad energy range, which includes both the intralayer B exciton and the interlayer X$^{\text{CT}}$. %This points to an efficient carrier injection towards the IX when we excite in resonance with these transitions. 
From the data shown in Fig.~\ref{fig:PolarizationResolvedPLE}(b), we estimate the corresponding degree of circular polarization, reported in Fig.~\ref{fig:PolarizationResolvedPLE}(c) as a function of the excitation energy. Crucially, when we excite the heterostructure close to the resonance of X$^{\text{CT}}$, the degree of circular polarization of the IX emission becomes increasingly negative, reaching $\simeq -10\%$. In contrast, in this spectral region the degree of circular polarization of the WSe$_2$ intralayer exciton increases with decreasing excitation energy. Otherwise, the trend of the degree of polarization follows that of the WSe$_2$ intralayer exciton, which stems from the charge transfer between WSe$_2$ monolayer and (BA)$_2$PbI$_4$.

For excitation close to resonance with the B exciton, the degree of circular polarization of the WSe$_2$ monolayer is determined by a Dexter-like coupling between exciton states with the same spin configuration \cite{berghauser2018inverted,jasinski2024control}. A similar process could explain the negative polarization of the IX. X$^{\text{CT}}$ is likely formed from electron states at the K point originating from the conduction band of WSe$_2$, while hole states are mainly formed from those of the organic spacer of the 2D perovskite at the K point, which hybridize with states from the spin-orbit split valence band of WSe$_2$. Thus, the negative polarization of the IX could be explained if the interlayer X$^{\text{CT}}$ inherits the same spin configuration of the WSe$_2$ B exciton following this hybridization. This would enable Dexter-like coupling between same-spin states following the resonant excitation of the X$^{\text{CT}}$. This coupling leads to the transfer of the carrier population to IX states that couple to the polarization not directly driven optically, possibly via relaxation from the A exciton of WSe$_2$, as schematically shown in Fig.\ \ref{fig:PolarizationResolvedPLE}(d). This process leads to the emission of counter-polarized PL from the IX, as illustrated in Fig.\ \ref{fig:PolarizationResolvedPLE}(e). The presence of electronic coupling and charge transfer between the (BA)$_4$PbI$_4$ and the WSe$_2$ monolayer is therefore pivotal in sensitizing the IX with the optical selection rules of WSe$_2$, which enables a fully optical control over the spin transferred from WSe$_2$ to (BA)$_2$PbI$_4$ and the helicity of the IX emission.

\begin{figure}[!ht]
\centering
\includegraphics[width=1.0\linewidth]{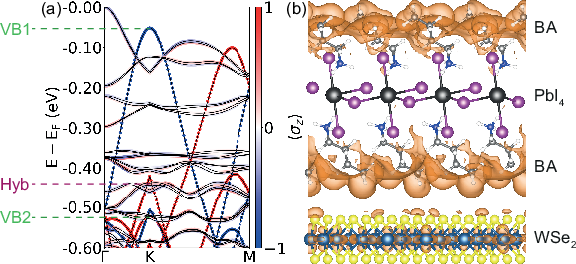}
\caption{(a) Zoom-in to the valence band of the (BA)$_2$PbI$_4$/WSe$_2$ heterostructure from Fig.~\ref{SI_fig:BandStructure_DOS} together with the spin expectation values, $\sigma_z$. Bands plotted with full dots originate from the WSe$_2$ monolayer, while states plotted with a coloured shading originate from (BA)$_2$PbI$_4$. VB1 and VB2 indicate the spin-orbit split valence bands of WSe$_2$, while Hyb designates the hybridized states involved in the X$^{\text{CT}}$ transition. (b) Eigenstate of one of the hybridized states involved in the X$^{\text{CT}}$ transition. An isosurface value of \SI{1e-6}{\angstrom^{-3}}, which is the probability density of this state, was used. This corresponds to about \SI{2.8e-8}{electron / \angstrom^{3}}. For the valence band maximum of WSe$_2$, these values are \SI{1e-3}{\angstrom^{-3}} and \SI{2.8e-5}{electron / \angstrom^{3}}, respectively}
\label{fig:SpinPolarizedVB_ElectronIsosurface}
\end{figure}

%\subsection*{Microscopic model}
To understand the origin of X$^{\text{CT}}$ and identify possible direct interlayer charge transfer transitions at energies between those of the A and B excitons of WSe$_2$, we calculated the electronic structure of the interface (see Fig.~\ref{SI_fig:BandStructure_DOS}).
The calculation %proves that the heterostructure is characterized by a type-II band alignment. It also 
reveals the presence of states that originate from the 2D perovskite at the K point, which are energetically located between the spin-orbit split valence bands of the WSe$_2$ monolayer, labelled VB1 and VB2 in Fig.\ \ref{fig:SpinPolarizedVB_ElectronIsosurface}(a). For the region around the K point, we calculated the spin expectation values, $\sigma_z$, to reveal the spin components in the $z$ direction (perpendicular to the layers), shown in Fig.~\ref{fig:SpinPolarizedVB_ElectronIsosurface}(a). In Fig.\ \ref{fig:SpinPolarizedVB_ElectronIsosurface}(b), we plot the isosurface of one of the eigenstates found between the spin-orbit split valence bands of the WSe$_2$ monolayer, labelled Hyb in Fig.\ \ref{fig:SpinPolarizedVB_ElectronIsosurface}(a). Although this state originates from the (BA)$_2$PbI$_4$, the isosurface demonstrates that in the heterostructure it exhibits some hybridization with d-orbital-like states of the transition metal atom in the WSe$_2$ monolayer \cite{liu2013three}. T%hese hybridized states span an energy range of \SIrange{70}{160}{\milli\eV} above the maximum of the spin-orbit split valence band VB2 of WSe$_2$, which is in good agreement with the experimental results. In particular, t
he Hyb state found \SI{75}{\milli\eV} above the valence band maximum could be responsible for the interlayer X$^{\text{CT}}$ resonance. This energy difference is compatible with the $\simeq \SI{80}{\milli\eV}$ shift of the X$^{\text{CT}}$ transition observed in the differential reflectivity spectrum of the heterostructure with respect to the B exciton of the WSe$_2$ monolayer (see Fig.~\ref{fig:PolarizationResolvedPLE}(a)). The observation of this state in reflectivity measurements suggests that X$^{\text{CT}}$ has a non-negligible oscillator strength, which can be attributed to its partial intralayer character due to the hybridization shown in Fig.\ \ref{fig:SpinPolarizedVB_ElectronIsosurface}(b). Moreover, the %valence band states responsible for the formation of X$^{\text{CT}}$ arise from the hybridization of the TMD and the organic spacer states. Their 
spin polarization of the hybridized states supports the role of Dexter-like coupling between same-spin states as the mechanism behind the observed negative circular polarization of the IX PL, which stems from the radiative recombination of electrons confined in the lead halide slab of the 2D perovskites and holes in the TMD monolayer.

\begin{figure*}[!ht]
\centering
\includegraphics[width=1.0\linewidth]{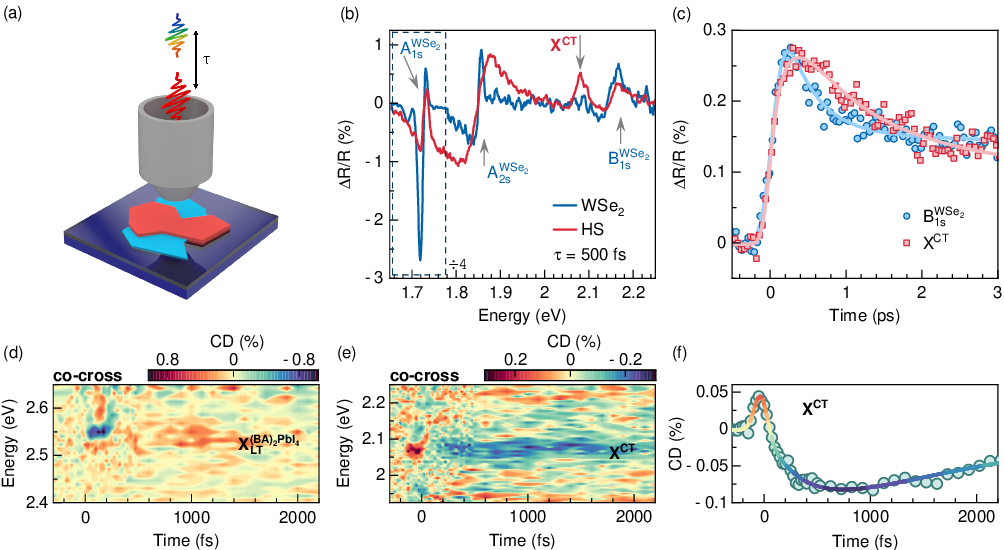}
\caption{(a) Schematic of broadband pump-probe measurements performed on the (BA)$_2$PbI$_4$/WSe$_2$ heterostructure. (b) Transient reflectivity spectrum of the heterostructure and of the isolated WSe$_2$ monolayer excited at \SI{2.64}{\eV} (above the quasi-particle band gap of both materials) and extracted at a delay $\tau = \SI{500}{\femto\second}$. The excitonic resonances are indicated. The resonance corresponding to the WSe$_2$ A exciton (A$_{1\text{s}}^{\text{WSe}_2}$) has been rescaled for increased clarity. (c) Transient differential reflectivity of the B exciton (B$_{1\text{s}}^{\text{WSe}_2}$) and of the interlayer charge transfer exciton (X$^{\text{CT}}$) measured as a function of the pump-probe delay. The line is the fit of an exponential rise and a bi-exponential decay model to the experimental data. Transient circular dichroism $(CD = (\Delta R / R)_{\text{co}} - (\Delta R / R)_{\text{cr}})$ maps of the (d) exciton transition of the low temperature phase of (BA)$_2$PbI$_4$ (X$_{\text{LT}}^{\text{(BA)}_2\text{PbI}_4}$) and (e) X$^{\text{CT}}$ excited in resonance with the WSe$_2$ A exciton obtained by subtracting the cross-polarized transient absorption from the co-polarized transient absorption. (f) Dynamics of the transient circular dichroism of X$^{\text{CT}}$ as a function of the pump-probe delay. The line is the fit to an exponential rise and an exponential decay model.}
\label{fig:IX_pump_probe}
\end{figure*}

%\subsection*{Charge and spin dynamics}
To investigate the dynamics of charge and spin transfer and exciton dynamics in the heterostructure, we performed broadband optical pump-probe spectroscopy (see schematic in Fig.\ \ref{fig:IX_pump_probe}(a)). We initially tuned the energy of the pump beam above the quasi-particle bandgap of both constituents of the heterostructure. The complete transient reflectivity maps of the WSe$_2$ monolayer and of the heterostructure as a function of the pump-probe delay and the probe energy are shown in Fig.\ \ref{SI_fig:PumpProbeMapsSpectra}. From these maps, we extract the transient reflectivity spectra $\Delta R/R$ ($\Delta R$ is the transient variation of reflectivity after pump excitation, while $R$ indicates the static reflectivity) at a fixed delay, which we show in Fig.\ \ref{fig:IX_pump_probe}(b). The transient spectrum of the WSe$_2$ monolayer is characterized by a transient signal at the energies of the 1s and 2s states of the A exciton, as well as the B exciton. The transient reflectivity spectrum of the heterostructure displays all excitonic resonances of WSe$_2$, broadened due to the presence of additional non-radiative recombination channels in the heterostructure \cite{hill2017exciton}, together with an additional transient signal at \SI{2.08}{\eV}, assigned to the X$^{\text{CT}}$ resonance (see Fig.\ \ref{fig:IX_pump_probe}(b)). %Based the line shape analysis performed in Ref.\ \cite{trovatello2022disentangling}, w
We attribute the transient signals measured at the energies of the A and B exciton resonances of WSe$_2$ under non-resonant photo-excitation primarily to photobleaching and broadening of the excitonic peaks \cite{trovatello2022disentangling}. The energy renormalization of the excitonic peaks is almost negligible under these excitation conditions \cite{trovatello2022disentangling}. The positive transient signal at the energy of X$^{\text{CT}}$ is attributed to a Pauli blocking process \cite{yuan2018photocarrier}. Coulomb screening induced by electron–hole pairs generated after pump excitation would result in an energy shift of the X$^{\text{CT}}$ peak and consequently in a derivative-like transient signal. We show the temporal dynamics of the X$^{\text{CT}}$ resonance in Fig.\ \ref{fig:IX_pump_probe}(c). This signal exhibits an instantaneous rise time, similar to the bleaching dynamics of the B exciton resonance of WSe$_2$, also displayed in Fig.\ \ref{fig:IX_pump_probe}(c), and limited by the instrument response function of the setup. Based on the results of DFT calculations, a possible mechanism that leads to instantaneous formation of X$^{\text{CT}}$ could be the direct excitation of an interlayer charge transfer transition, as already observed in WS$_2$/graphene heterostructures \cite{yuan2018photocarrier}. In this scenario, electrons from hybridized electronic states located between the spin-orbit split valence bands of the TMD are directly promoted by light excitation to the conduction band of WSe$_2$.

Polarization-resolved pump-probe measurements provide access to transient spin-dependent dynamics. Using linearly polarized pump pulses resonant with the 1s exciton of WSe$_2$, we observe a distinct bleaching signal at the energy of the exciton of (BA)$_2$PbI$_4$ as a consequence of an interlayer hole transfer process from the WSe$_2$ layer to the 2D halide perovskite (see Fig.\ \ref{SI_fig:ChargeTransfer}). We investigate the interlayer transfer dynamics of spin-polarized carriers by performing helicity-resolved optical pump-probe measurements. Due to spin-valley locking in monolayer TMDs \cite{xiao2012coupled}, we selectively photo-excite a population of spin-polarized electron-hole pairs in the TMD layer upon resonant excitation of the A exciton 1s transition of WSe$_2$ with circularly polarized pump pulses. The temporal evolution of the spin-polarized carriers is measured using co- and cross-polarized broadband probe pulses. We show in Fig.\ \ref{fig:IX_pump_probe}(d) the transient circular dichroism (CD) as a function of the probe photon energy in the spectral region of (BA)$_2$PbI$_4$ exciton and of the pump-probe delay. The CD is defined as the difference between the transient reflectivity signals co- and cross-polarized to the pump laser $(CD = (\Delta R / R)_{\text{co}} - (\Delta R / R)_{\text{cr}})$ and is proportional to the spin polarization of the exciton population. The spectrally resolved transient response initially displays a sharp feature at zero delay, which can be attributed to a coherent effect related to the spin/valley-dependent optical Stark shift of excitonic resonances \cite{kim2014ultrafast}. At later delays, a weak but non-negligible positive transient signal develops at the energy of (BA)$_2$PbI$_4$ exciton. This results from the photo-generation of spin-polarized holes only in WSe$_2$ upon its resonant excitation, followed by their directional transfer to (BA)$_2$PbI$_4$, similar to previous observations in heterobilayers based on TMDs \cite{schaibley2016directional}. The positive CD is in agreement with the degree of circular polarization of the IX observed in PL measurements under similar excitation conditions. The hole transfer preserves the optically injected spin polarization. %The majority of holes that undergo an interlayer transfer process preserve their spin polarization, maintaining the same orientation as the electrons left behind in the TMD layer. 
The radiative recombination of the resulting IX leads to the emission of a photon co-polarized with respect to the excitation laser.

The transient CD in the spectral range of X$^{\text{CT}}$ is shown in Fig.\ \ref{fig:IX_pump_probe}(e). Crucially, we notice a pronounced negative characteristic at energies corresponding to X$^{\text{CT}}$. By integrating the CD spectrum across this energy interval, we obtain the dynamics of its spin polarization \cite{kim2017observation,dal2015ultrafast,mai2014many}, which we show in Fig.~\ref{fig:IX_pump_probe}(f). Also in this case, the early dynamics are dominated by a sharp positive feature at zero delay related to the spin/valley dependent optical Stark shift of excitonic resonances \cite{kim2014ultrafast}. The opposite sign of the CD is consistent with the reversed optical selection rules characteristic of X$^{\text{CT}}$. A possible mechanism leading to this is related to the Dexter-like coupling between the WSe$_2$ A exciton resonantly excited and the B exciton, from which the holes could relax to hybridized states that contribute to the X$^{\text{CT}}$ resonance. The rise of the negative CD might reflect this relaxation process of spin-polarized holes. The decay of the CD is mainly driven by the depolarization via electron-hole exchange interaction \cite{bourelle2020exciton,zhao2020transient,chen2021tuning,singh2023valley}, which is expected to be relatively inefficient in this case, due to the reduced overlap of the electron-hole wave functions located in different layers \cite{rivera2016valley}.

%\subsection*{Discussion}
In conclusion, we have engineered a (BA)$_2$PbI$_4$/WSe$_2$ monolayer heterostructure with the goal of controlling the helicity of circularly polarized light emitted from a 2D perovskite with an excitation energy-tunable helicity. To reach this goal, we optically initialize the spin polarization of charge carriers in the WSe$_2$ monolayer and exploit the spin-conserving charge transfer of holes towards (BA)$_2$PbI$_4$, which enables the emission of circularly polarized light via the recombination of the IX. In the differential reflectivity spectrum of the heterostructure, we observe a new excitonic resonance at a slightly lower energy than that of the B exciton of WSe$_2$. When the circularly polarized excitation is resonant with this transition, the IX PL strikingly exhibits a circular polarization opposite to that of the excitation laser. The calculated band structure demonstrates the emergence of spin-polarized hybrid states between the A and B excitons of WSe$_2$, which participate in the formation of this interlayer X$^{\text{CT}}$. Helicity-resolved pump-probe measurements demonstrate that X$^{\text{CT}}$ exhibits a negative circular dichroism when the heterostructure is pumped in resonance with the A exciton of the WSe$_2$ monolayer. In general, we show that the WSe$_2$ monolayer can act as a spin filter to efficiently initialize a spin population in 2D perovskites, whose amount and sign can be conveniently controlled with a fully optical approach. These results can pave the way for the flexible design of opto-spintronic devices with an optically controllable circular dichroism.

\begin{acknowledgments}
A.S.\ gratefully acknowledges the support of the National Science Centre, Poland, grant no.\ 2023/50/E/ST3/00531. This study has been partially supported through the EUR grant NanoX no.\ ANR-17-EURE-0009 in the framework of the ``Programme des Investissements d'Avenir''. A.K.\ and A.C.\ thank Deutsche Forschungsgemeinschaft (DFG) within the CRC 1415 project. T.B.\ and A.K.\ gratefully acknowledge the computing time provided to them on the high-performance computers Beaker and Noctua 2 at the NHR Centers NHR@TUD (ZIH) and NHR@Padeborn (PC2). T.B.\ thanks the DFG for funding through the DIP Grant HE 3543/42-1 (Project No.\ 471289011). K.W.\ and T.T.\ acknowledge support from the JSPS KAKENHI (Grant Numbers 21H05233 and 23H02052) , the CREST (JPMJCR24A5), JST and World Premier International Research Center Initiative (WPI), MEXT, Japan. This project has received funding from the European Union's Horizon 2020 research and innovation programme under grant  agreement no.\ 871124 Laserlab-Europe (PID: 27173, ``Probing Ultrafast Dynamics in TMD/Perovskite Heterostructures Unraveling Spin Polarized Charge Transfer''). This work received funding from the European Union's Horizon 2020 research and innovation program under grant agreement 956813 (2Exciting).

\end{acknowledgments}

\bibliography{Bibliography}

%apsrev4-2.bst 2019-01-14 (MD) hand-edited version of apsrev4-1.bst
%Control: key (0)
%Control: author (8) initials jnrlst
%Control: editor formatted (1) identically to author
%Control: production of article title (0) allowed
%Control: page (0) single
%Control: year (1) truncated
%Control: production of eprint (0) enabled
\begin{thebibliography}{49}%
\makeatletter
\providecommand \@ifxundefined [1]{%
 \@ifx{#1\undefined}
}%
\providecommand \@ifnum [1]{%
 \ifnum #1\expandafter \@firstoftwo
 \else \expandafter \@secondoftwo
 \fi
}%
\providecommand \@ifx [1]{%
 \ifx #1\expandafter \@firstoftwo
 \else \expandafter \@secondoftwo
 \fi
}%
\providecommand \natexlab [1]{#1}%
\providecommand \enquote  [1]{``#1''}%
\providecommand \bibnamefont  [1]{#1}%
\providecommand \bibfnamefont [1]{#1}%
\providecommand \citenamefont [1]{#1}%
\providecommand \href@noop [0]{\@secondoftwo}%
\providecommand \href [0]{\begingroup \@sanitize@url \@href}%
\providecommand \@href[1]{\@@startlink{#1}\@@href}%
\providecommand \@@href[1]{\endgroup#1\@@endlink}%
\providecommand \@sanitize@url [0]{\catcode `\\12\catcode `\$12\catcode
  `\&12\catcode `\#12\catcode `\^12\catcode `\_12\catcode `\%12\relax}%
\providecommand \@@startlink[1]{}%
\providecommand \@@endlink[0]{}%
\providecommand \url  [0]{\begingroup\@sanitize@url \@url }%
\providecommand \@url [1]{\endgroup\@href {#1}{\urlprefix }}%
\providecommand \urlprefix  [0]{URL }%
\providecommand \Eprint [0]{\href }%
\providecommand \doibase [0]{https://doi.org/}%
\providecommand \selectlanguage [0]{\@gobble}%
\providecommand \bibinfo  [0]{\@secondoftwo}%
\providecommand \bibfield  [0]{\@secondoftwo}%
\providecommand \translation [1]{[#1]}%
\providecommand \BibitemOpen [0]{}%
\providecommand \bibitemStop [0]{}%
\providecommand \bibitemNoStop [0]{.\EOS\space}%
\providecommand \EOS [0]{\spacefactor3000\relax}%
\providecommand \BibitemShut  [1]{\csname bibitem#1\endcsname}%
\let\auto@bib@innerbib\@empty
%</preamble>
\bibitem [{\citenamefont {Castellanos-Gomez}\ \emph {et~al.}(2014)\citenamefont
  {Castellanos-Gomez}, \citenamefont {Buscema}, \citenamefont {Molenaar},
  \citenamefont {Singh}, \citenamefont {Janssen}, \citenamefont {Van
  Der~Zant},\ and\ \citenamefont {Steele}}]{castellanos2014deterministic}%
  \BibitemOpen
  \bibfield  {author} {\bibinfo {author} {\bibfnamefont {A.}~\bibnamefont
  {Castellanos-Gomez}}, \bibinfo {author} {\bibfnamefont {M.}~\bibnamefont
  {Buscema}}, \bibinfo {author} {\bibfnamefont {R.}~\bibnamefont {Molenaar}},
  \bibinfo {author} {\bibfnamefont {V.}~\bibnamefont {Singh}}, \bibinfo
  {author} {\bibfnamefont {L.}~\bibnamefont {Janssen}}, \bibinfo {author}
  {\bibfnamefont {H.~S.}\ \bibnamefont {Van Der~Zant}},\ and\ \bibinfo {author}
  {\bibfnamefont {G.~A.}\ \bibnamefont {Steele}},\ }\bibfield  {title}
  {\bibinfo {title} {Deterministic transfer of two-dimensional materials by
  all-dry viscoelastic stamping},\ }\href@noop {} {\bibfield  {journal}
  {\bibinfo  {journal} {2D Materials}\ }\textbf {\bibinfo {volume} {1}},\
  \bibinfo {pages} {011002} (\bibinfo {year} {2014})}\BibitemShut {NoStop}%
\bibitem [{\citenamefont {Kempt}(2021)}]{kempt_romankempthetbuilder_2021}%
  \BibitemOpen
  \bibfield  {author} {\bibinfo {author} {\bibfnamefont {R.}~\bibnamefont
  {Kempt}},\ }\href {https://doi.org/10.5281/ZENODO.4721346} {\bibinfo {title}
  {romankempt/hetbuilder: {Zenodo} {Release}}} (\bibinfo {year}
  {2021})\BibitemShut {NoStop}%
\bibitem [{\citenamefont {Blum}\ \emph {et~al.}(2009)\citenamefont {Blum},
  \citenamefont {Gehrke}, \citenamefont {Hanke}, \citenamefont {Havu},
  \citenamefont {Havu}, \citenamefont {Ren}, \citenamefont {Reuter},\ and\
  \citenamefont {Scheffler}}]{Blum2009}%
  \BibitemOpen
  \bibfield  {author} {\bibinfo {author} {\bibfnamefont {V.}~\bibnamefont
  {Blum}}, \bibinfo {author} {\bibfnamefont {R.}~\bibnamefont {Gehrke}},
  \bibinfo {author} {\bibfnamefont {F.}~\bibnamefont {Hanke}}, \bibinfo
  {author} {\bibfnamefont {P.}~\bibnamefont {Havu}}, \bibinfo {author}
  {\bibfnamefont {V.}~\bibnamefont {Havu}}, \bibinfo {author} {\bibfnamefont
  {X.}~\bibnamefont {Ren}}, \bibinfo {author} {\bibfnamefont {K.}~\bibnamefont
  {Reuter}},\ and\ \bibinfo {author} {\bibfnamefont {M.}~\bibnamefont
  {Scheffler}},\ }\bibfield  {title} {\bibinfo {title} {Ab initio molecular
  simulations with numeric atom-centered orbitals},\ }\href
  {https://doi.org/10.1016/j.cpc.2009.06.022} {\bibfield  {journal} {\bibinfo
  {journal} {Computer Physics Communications}\ }\textbf {\bibinfo {volume}
  {180}},\ \bibinfo {pages} {2175} (\bibinfo {year} {2009})}\BibitemShut
  {NoStop}%
\bibitem [{\citenamefont {Perdew}\ \emph {et~al.}(1996)\citenamefont {Perdew},
  \citenamefont {Burke},\ and\ \citenamefont {Ernzerhof}}]{Perdew1996}%
  \BibitemOpen
  \bibfield  {author} {\bibinfo {author} {\bibfnamefont {J.~P.}\ \bibnamefont
  {Perdew}}, \bibinfo {author} {\bibfnamefont {K.}~\bibnamefont {Burke}},\ and\
  \bibinfo {author} {\bibfnamefont {M.}~\bibnamefont {Ernzerhof}},\ }\bibfield
  {title} {\bibinfo {title} {Generalized gradient approximation made simple},\
  }\href {https://doi.org/10.1103/physrevlett.77.3865} {\bibfield  {journal}
  {\bibinfo  {journal} {Physical Review Letters}\ }\textbf {\bibinfo {volume}
  {77}},\ \bibinfo {pages} {3865} (\bibinfo {year} {1996})}\BibitemShut
  {NoStop}%
\bibitem [{\citenamefont {Tkatchenko}\ \emph {et~al.}(2013)\citenamefont
  {Tkatchenko}, \citenamefont {Ambrosetti},\ and\ \citenamefont
  {DiStasio}}]{Tkatchenko2013}%
  \BibitemOpen
  \bibfield  {author} {\bibinfo {author} {\bibfnamefont {A.}~\bibnamefont
  {Tkatchenko}}, \bibinfo {author} {\bibfnamefont {A.}~\bibnamefont
  {Ambrosetti}},\ and\ \bibinfo {author} {\bibfnamefont {R.~A.}\ \bibnamefont
  {DiStasio}},\ }\bibfield  {title} {\bibinfo {title} {Interatomic methods for
  the dispersion energy derived from the adiabatic connection
  fluctuation-dissipation theorem},\ }\href {https://doi.org/10.1063/1.4789814}
  {\bibfield  {journal} {\bibinfo  {journal} {The Journal of Chemical Physics}\
  }\textbf {\bibinfo {volume} {138}},\ \bibinfo {pages} {074106} (\bibinfo
  {year} {2013})}\BibitemShut {NoStop}%
\bibitem [{\citenamefont {Hermann}\ and\ \citenamefont
  {Tkatchenko}(2020)}]{Hermann2020}%
  \BibitemOpen
  \bibfield  {author} {\bibinfo {author} {\bibfnamefont {J.}~\bibnamefont
  {Hermann}}\ and\ \bibinfo {author} {\bibfnamefont {A.}~\bibnamefont
  {Tkatchenko}},\ }\bibfield  {title} {\bibinfo {title} {Density functional
  model for van der waals interactions: Unifying many-body atomic approaches
  with nonlocal functionals},\ }\href
  {https://doi.org/10.1103/physrevlett.124.146401} {\bibfield  {journal}
  {\bibinfo  {journal} {Physical Review Letters}\ }\textbf {\bibinfo {volume}
  {124}},\ \bibinfo {pages} {146401} (\bibinfo {year} {2020})}\BibitemShut
  {NoStop}%
\bibitem [{\citenamefont {Genco}\ \emph {et~al.}(2023)\citenamefont {Genco},
  \citenamefont {Trovatello}, \citenamefont {Louca}, \citenamefont {Watanabe},
  \citenamefont {Taniguchi}, \citenamefont {Tartakovskii}, \citenamefont
  {Cerullo},\ and\ \citenamefont {Dal~Conte}}]{Genco2023}%
  \BibitemOpen
  \bibfield  {author} {\bibinfo {author} {\bibfnamefont {A.}~\bibnamefont
  {Genco}}, \bibinfo {author} {\bibfnamefont {C.}~\bibnamefont {Trovatello}},
  \bibinfo {author} {\bibfnamefont {C.}~\bibnamefont {Louca}}, \bibinfo
  {author} {\bibfnamefont {K.}~\bibnamefont {Watanabe}}, \bibinfo {author}
  {\bibfnamefont {T.}~\bibnamefont {Taniguchi}}, \bibinfo {author}
  {\bibfnamefont {A.~I.}\ \bibnamefont {Tartakovskii}}, \bibinfo {author}
  {\bibfnamefont {G.}~\bibnamefont {Cerullo}},\ and\ \bibinfo {author}
  {\bibfnamefont {S.}~\bibnamefont {Dal~Conte}},\ }\bibfield  {title} {\bibinfo
  {title} {Ultrafast exciton and trion dynamics in high-quality encapsulated
  {MoS}$_2$ monolayers},\ }\href
  {https://doi.org/https://doi.org/10.1002/pssb.202200376} {\bibfield
  {journal} {\bibinfo  {journal} {physica status solidi (b)}\ }\textbf
  {\bibinfo {volume} {260}},\ \bibinfo {pages} {2200376} (\bibinfo {year}
  {2023})}\BibitemShut {NoStop}%
\bibitem [{\citenamefont {Manca}\ \emph {et~al.}(2017)\citenamefont {Manca},
  \citenamefont {Glazov}, \citenamefont {Robert}, \citenamefont {Cadiz},
  \citenamefont {Taniguchi}, \citenamefont {Watanabe}, \citenamefont
  {Courtade}, \citenamefont {Amand}, \citenamefont {Renucci}, \citenamefont
  {Marie}, \citenamefont {Wang},\ and\ \citenamefont
  {Urbaszek}}]{manca2017enabling}%
  \BibitemOpen
  \bibfield  {author} {\bibinfo {author} {\bibfnamefont {M.}~\bibnamefont
  {Manca}}, \bibinfo {author} {\bibfnamefont {M.~M.}\ \bibnamefont {Glazov}},
  \bibinfo {author} {\bibfnamefont {C.}~\bibnamefont {Robert}}, \bibinfo
  {author} {\bibfnamefont {F.}~\bibnamefont {Cadiz}}, \bibinfo {author}
  {\bibfnamefont {T.}~\bibnamefont {Taniguchi}}, \bibinfo {author}
  {\bibfnamefont {K.}~\bibnamefont {Watanabe}}, \bibinfo {author}
  {\bibfnamefont {E.}~\bibnamefont {Courtade}}, \bibinfo {author}
  {\bibfnamefont {T.}~\bibnamefont {Amand}}, \bibinfo {author} {\bibfnamefont
  {P.}~\bibnamefont {Renucci}}, \bibinfo {author} {\bibfnamefont
  {X.}~\bibnamefont {Marie}}, \bibinfo {author} {\bibfnamefont
  {G.}~\bibnamefont {Wang}},\ and\ \bibinfo {author} {\bibfnamefont
  {B.}~\bibnamefont {Urbaszek}},\ }\bibfield  {title} {\bibinfo {title}
  {Enabling valley selective exciton scattering in monolayer {WSe}$_2$ through
  upconversion},\ }\href@noop {} {\bibfield  {journal} {\bibinfo  {journal}
  {Nature Communications}\ }\textbf {\bibinfo {volume} {8}},\ \bibinfo {pages}
  {14927} (\bibinfo {year} {2017})}\BibitemShut {NoStop}%
\bibitem [{\citenamefont {Stier}\ \emph {et~al.}(2018)\citenamefont {Stier},
  \citenamefont {Wilson}, \citenamefont {Velizhanin}, \citenamefont {Kono},
  \citenamefont {Xu},\ and\ \citenamefont {Crooker}}]{stier2018magnetooptics}%
  \BibitemOpen
  \bibfield  {author} {\bibinfo {author} {\bibfnamefont {A.~V.}\ \bibnamefont
  {Stier}}, \bibinfo {author} {\bibfnamefont {N.~P.}\ \bibnamefont {Wilson}},
  \bibinfo {author} {\bibfnamefont {K.~A.}\ \bibnamefont {Velizhanin}},
  \bibinfo {author} {\bibfnamefont {J.}~\bibnamefont {Kono}}, \bibinfo {author}
  {\bibfnamefont {X.}~\bibnamefont {Xu}},\ and\ \bibinfo {author}
  {\bibfnamefont {S.~A.}\ \bibnamefont {Crooker}},\ }\bibfield  {title}
  {\bibinfo {title} {Magnetooptics of exciton {Rydberg} states in a monolayer
  semiconductor},\ }\href@noop {} {\bibfield  {journal} {\bibinfo  {journal}
  {Physical Review Letters}\ }\textbf {\bibinfo {volume} {120}},\ \bibinfo
  {pages} {057405} (\bibinfo {year} {2018})}\BibitemShut {NoStop}%
\bibitem [{\citenamefont {Chen}\ \emph
  {et~al.}(2018{\natexlab{a}})\citenamefont {Chen}, \citenamefont {Goldstein},
  \citenamefont {Tong}, \citenamefont {Taniguchi}, \citenamefont {Watanabe},\
  and\ \citenamefont {Yan}}]{chen2018superior}%
  \BibitemOpen
  \bibfield  {author} {\bibinfo {author} {\bibfnamefont {S.-Y.}\ \bibnamefont
  {Chen}}, \bibinfo {author} {\bibfnamefont {T.}~\bibnamefont {Goldstein}},
  \bibinfo {author} {\bibfnamefont {J.}~\bibnamefont {Tong}}, \bibinfo {author}
  {\bibfnamefont {T.}~\bibnamefont {Taniguchi}}, \bibinfo {author}
  {\bibfnamefont {K.}~\bibnamefont {Watanabe}},\ and\ \bibinfo {author}
  {\bibfnamefont {J.}~\bibnamefont {Yan}},\ }\bibfield  {title} {\bibinfo
  {title} {Superior valley polarization and coherence of 2s excitons in
  monolayer {WSe$_2$}},\ }\href@noop {} {\bibfield  {journal} {\bibinfo
  {journal} {Physical Review Letters}\ }\textbf {\bibinfo {volume} {120}},\
  \bibinfo {pages} {046402} (\bibinfo {year} {2018}{\natexlab{a}})}\BibitemShut
  {NoStop}%
\bibitem [{\citenamefont {Molas}\ \emph {et~al.}(2019)\citenamefont {Molas},
  \citenamefont {Slobodeniuk}, \citenamefont {Nogajewski}, \citenamefont
  {Bartos}, \citenamefont {Bala}, \citenamefont {Babi{\'n}ski}, \citenamefont
  {Watanabe}, \citenamefont {Taniguchi}, \citenamefont {Faugeras},\ and\
  \citenamefont {Potemski}}]{molas2019energy}%
  \BibitemOpen
  \bibfield  {author} {\bibinfo {author} {\bibfnamefont {M.}~\bibnamefont
  {Molas}}, \bibinfo {author} {\bibfnamefont {A.}~\bibnamefont {Slobodeniuk}},
  \bibinfo {author} {\bibfnamefont {K.}~\bibnamefont {Nogajewski}}, \bibinfo
  {author} {\bibfnamefont {M.}~\bibnamefont {Bartos}}, \bibinfo {author}
  {\bibfnamefont {{\L}.}~\bibnamefont {Bala}}, \bibinfo {author} {\bibfnamefont
  {A.}~\bibnamefont {Babi{\'n}ski}}, \bibinfo {author} {\bibfnamefont
  {K.}~\bibnamefont {Watanabe}}, \bibinfo {author} {\bibfnamefont
  {T.}~\bibnamefont {Taniguchi}}, \bibinfo {author} {\bibfnamefont
  {C.}~\bibnamefont {Faugeras}},\ and\ \bibinfo {author} {\bibfnamefont
  {M.}~\bibnamefont {Potemski}},\ }\bibfield  {title} {\bibinfo {title} {Energy
  spectrum of two-dimensional excitons in a nonuniform dielectric medium},\
  }\href@noop {} {\bibfield  {journal} {\bibinfo  {journal} {Physical Review
  Letters}\ }\textbf {\bibinfo {volume} {123}},\ \bibinfo {pages} {136801}
  (\bibinfo {year} {2019})}\BibitemShut {NoStop}%
\bibitem [{\citenamefont {Arora}\ \emph {et~al.}(2015)\citenamefont {Arora},
  \citenamefont {Koperski}, \citenamefont {Nogajewski}, \citenamefont {Marcus},
  \citenamefont {Faugeras},\ and\ \citenamefont
  {Potemski}}]{arora2015excitonic}%
  \BibitemOpen
  \bibfield  {author} {\bibinfo {author} {\bibfnamefont {A.}~\bibnamefont
  {Arora}}, \bibinfo {author} {\bibfnamefont {M.}~\bibnamefont {Koperski}},
  \bibinfo {author} {\bibfnamefont {K.}~\bibnamefont {Nogajewski}}, \bibinfo
  {author} {\bibfnamefont {J.}~\bibnamefont {Marcus}}, \bibinfo {author}
  {\bibfnamefont {C.}~\bibnamefont {Faugeras}},\ and\ \bibinfo {author}
  {\bibfnamefont {M.}~\bibnamefont {Potemski}},\ }\bibfield  {title} {\bibinfo
  {title} {Excitonic resonances in thin films of {WSe}$_2$: from monolayer to
  bulk material},\ }\href@noop {} {\bibfield  {journal} {\bibinfo  {journal}
  {Nanoscale}\ }\textbf {\bibinfo {volume} {7}},\ \bibinfo {pages} {10421}
  (\bibinfo {year} {2015})}\BibitemShut {NoStop}%
\bibitem [{\citenamefont {Hanbicki}\ \emph {et~al.}(2015)\citenamefont
  {Hanbicki}, \citenamefont {Currie}, \citenamefont {Kioseoglou}, \citenamefont
  {Friedman},\ and\ \citenamefont {Jonker}}]{hanbicki2015measurement}%
  \BibitemOpen
  \bibfield  {author} {\bibinfo {author} {\bibfnamefont {A.}~\bibnamefont
  {Hanbicki}}, \bibinfo {author} {\bibfnamefont {M.}~\bibnamefont {Currie}},
  \bibinfo {author} {\bibfnamefont {G.}~\bibnamefont {Kioseoglou}}, \bibinfo
  {author} {\bibfnamefont {A.}~\bibnamefont {Friedman}},\ and\ \bibinfo
  {author} {\bibfnamefont {B.}~\bibnamefont {Jonker}},\ }\bibfield  {title}
  {\bibinfo {title} {Measurement of high exciton binding energy in the
  monolayer transition-metal dichalcogenides {WS}$_2$ and {WSe}$_2$},\
  }\href@noop {} {\bibfield  {journal} {\bibinfo  {journal} {Solid State
  Communications}\ }\textbf {\bibinfo {volume} {203}},\ \bibinfo {pages} {16}
  (\bibinfo {year} {2015})}\BibitemShut {NoStop}%
\bibitem [{\citenamefont {Frisenda}\ \emph {et~al.}(2017)\citenamefont
  {Frisenda}, \citenamefont {Niu}, \citenamefont {Gant}, \citenamefont
  {Molina-Mendoza}, \citenamefont {Schmidt}, \citenamefont {Bratschitsch},
  \citenamefont {Liu}, \citenamefont {Fu}, \citenamefont {Dumcenco},
  \citenamefont {Kis}, \citenamefont {Perez~de Lara},\ and\ \citenamefont
  {Castellanos-Gomez}}]{frisenda2017micro}%
  \BibitemOpen
  \bibfield  {author} {\bibinfo {author} {\bibfnamefont {R.}~\bibnamefont
  {Frisenda}}, \bibinfo {author} {\bibfnamefont {Y.}~\bibnamefont {Niu}},
  \bibinfo {author} {\bibfnamefont {P.}~\bibnamefont {Gant}}, \bibinfo {author}
  {\bibfnamefont {A.~J.}\ \bibnamefont {Molina-Mendoza}}, \bibinfo {author}
  {\bibfnamefont {R.}~\bibnamefont {Schmidt}}, \bibinfo {author} {\bibfnamefont
  {R.}~\bibnamefont {Bratschitsch}}, \bibinfo {author} {\bibfnamefont
  {J.}~\bibnamefont {Liu}}, \bibinfo {author} {\bibfnamefont {L.}~\bibnamefont
  {Fu}}, \bibinfo {author} {\bibfnamefont {D.}~\bibnamefont {Dumcenco}},
  \bibinfo {author} {\bibfnamefont {A.}~\bibnamefont {Kis}}, \bibinfo {author}
  {\bibfnamefont {D.}~\bibnamefont {Perez~de Lara}},\ and\ \bibinfo {author}
  {\bibfnamefont {A.}~\bibnamefont {Castellanos-Gomez}},\ }\bibfield  {title}
  {\bibinfo {title} {Micro-reflectance and transmittance spectroscopy: a
  versatile and powerful tool to characterize {2D} materials},\ }\href@noop {}
  {\bibfield  {journal} {\bibinfo  {journal} {Journal of Physics D: Applied
  Physics}\ }\textbf {\bibinfo {volume} {50}},\ \bibinfo {pages} {074002}
  (\bibinfo {year} {2017})}\BibitemShut {NoStop}%
\bibitem [{\citenamefont {Baranowski}\ \emph {et~al.}(2019)\citenamefont
  {Baranowski}, \citenamefont {Zelewski}, \citenamefont {Kepenekian},
  \citenamefont {Traor{\'e}}, \citenamefont {Urban}, \citenamefont {Surrente},
  \citenamefont {Galkowski}, \citenamefont {Maude}, \citenamefont {Kuc},
  \citenamefont {Booker}, \citenamefont {Kudrawiec}, \citenamefont {Stranks},\
  and\ \citenamefont {Plochocka}}]{baranowski2019phase}%
  \BibitemOpen
  \bibfield  {author} {\bibinfo {author} {\bibfnamefont {M.}~\bibnamefont
  {Baranowski}}, \bibinfo {author} {\bibfnamefont {S.~J.}\ \bibnamefont
  {Zelewski}}, \bibinfo {author} {\bibfnamefont {M.}~\bibnamefont
  {Kepenekian}}, \bibinfo {author} {\bibfnamefont {B.}~\bibnamefont
  {Traor{\'e}}}, \bibinfo {author} {\bibfnamefont {J.~M.}\ \bibnamefont
  {Urban}}, \bibinfo {author} {\bibfnamefont {A.}~\bibnamefont {Surrente}},
  \bibinfo {author} {\bibfnamefont {K.}~\bibnamefont {Galkowski}}, \bibinfo
  {author} {\bibfnamefont {D.~K.}\ \bibnamefont {Maude}}, \bibinfo {author}
  {\bibfnamefont {A.}~\bibnamefont {Kuc}}, \bibinfo {author} {\bibfnamefont
  {E.~P.}\ \bibnamefont {Booker}}, \bibinfo {author} {\bibfnamefont
  {R.}~\bibnamefont {Kudrawiec}}, \bibinfo {author} {\bibfnamefont {S.~D.}\
  \bibnamefont {Stranks}},\ and\ \bibinfo {author} {\bibfnamefont
  {P.}~\bibnamefont {Plochocka}},\ }\bibfield  {title} {\bibinfo {title}
  {Phase-transition-induced carrier mass enhancement in {2D Ruddlesden--Popper}
  perovskites},\ }\href@noop {} {\bibfield  {journal} {\bibinfo  {journal} {ACS
  Energy Letters}\ }\textbf {\bibinfo {volume} {4}},\ \bibinfo {pages} {2386}
  (\bibinfo {year} {2019})}\BibitemShut {NoStop}%
\bibitem [{\citenamefont {Yaffe}\ \emph {et~al.}(2015)\citenamefont {Yaffe},
  \citenamefont {Chernikov}, \citenamefont {Norman}, \citenamefont {Zhong},
  \citenamefont {Velauthapillai}, \citenamefont {Van Der~Zande}, \citenamefont
  {Owen},\ and\ \citenamefont {Heinz}}]{yaffe2015excitons}%
  \BibitemOpen
  \bibfield  {author} {\bibinfo {author} {\bibfnamefont {O.}~\bibnamefont
  {Yaffe}}, \bibinfo {author} {\bibfnamefont {A.}~\bibnamefont {Chernikov}},
  \bibinfo {author} {\bibfnamefont {Z.~M.}\ \bibnamefont {Norman}}, \bibinfo
  {author} {\bibfnamefont {Y.}~\bibnamefont {Zhong}}, \bibinfo {author}
  {\bibfnamefont {A.}~\bibnamefont {Velauthapillai}}, \bibinfo {author}
  {\bibfnamefont {A.}~\bibnamefont {Van Der~Zande}}, \bibinfo {author}
  {\bibfnamefont {J.~S.}\ \bibnamefont {Owen}},\ and\ \bibinfo {author}
  {\bibfnamefont {T.~F.}\ \bibnamefont {Heinz}},\ }\bibfield  {title} {\bibinfo
  {title} {Excitons in ultrathin organic-inorganic perovskite crystals},\
  }\href@noop {} {\bibfield  {journal} {\bibinfo  {journal} {Physical Review
  B}\ }\textbf {\bibinfo {volume} {92}},\ \bibinfo {pages} {045414} (\bibinfo
  {year} {2015})}\BibitemShut {NoStop}%
\bibitem [{\citenamefont {Ye}\ \emph {et~al.}(2018)\citenamefont {Ye},
  \citenamefont {Waldecker}, \citenamefont {Ma}, \citenamefont {Rhodes},
  \citenamefont {Antony}, \citenamefont {Kim}, \citenamefont {Zhang},
  \citenamefont {Deng}, \citenamefont {Jiang}, \citenamefont {Lu},
  \citenamefont {Smirnov}, \citenamefont {Watanabe}, \citenamefont {Taniguchi},
  \citenamefont {Hone},\ and\ \citenamefont {Heinz}}]{ye2018efficient}%
  \BibitemOpen
  \bibfield  {author} {\bibinfo {author} {\bibfnamefont {Z.}~\bibnamefont
  {Ye}}, \bibinfo {author} {\bibfnamefont {L.}~\bibnamefont {Waldecker}},
  \bibinfo {author} {\bibfnamefont {E.~Y.}\ \bibnamefont {Ma}}, \bibinfo
  {author} {\bibfnamefont {D.}~\bibnamefont {Rhodes}}, \bibinfo {author}
  {\bibfnamefont {A.}~\bibnamefont {Antony}}, \bibinfo {author} {\bibfnamefont
  {B.}~\bibnamefont {Kim}}, \bibinfo {author} {\bibfnamefont {X.-X.}\
  \bibnamefont {Zhang}}, \bibinfo {author} {\bibfnamefont {M.}~\bibnamefont
  {Deng}}, \bibinfo {author} {\bibfnamefont {Y.}~\bibnamefont {Jiang}},
  \bibinfo {author} {\bibfnamefont {Z.}~\bibnamefont {Lu}}, \bibinfo {author}
  {\bibfnamefont {D.}~\bibnamefont {Smirnov}}, \bibinfo {author} {\bibfnamefont
  {K.}~\bibnamefont {Watanabe}}, \bibinfo {author} {\bibfnamefont
  {T.}~\bibnamefont {Taniguchi}}, \bibinfo {author} {\bibfnamefont
  {J.}~\bibnamefont {Hone}},\ and\ \bibinfo {author} {\bibfnamefont {T.~F.}\
  \bibnamefont {Heinz}},\ }\bibfield  {title} {\bibinfo {title} {Efficient
  generation of neutral and charged biexcitons in encapsulated {WSe}$_2$
  monolayers},\ }\href@noop {} {\bibfield  {journal} {\bibinfo  {journal}
  {Nature Communications}\ }\textbf {\bibinfo {volume} {9}},\ \bibinfo {pages}
  {3718} (\bibinfo {year} {2018})}\BibitemShut {NoStop}%
\bibitem [{\citenamefont {Chen}\ \emph
  {et~al.}(2018{\natexlab{b}})\citenamefont {Chen}, \citenamefont {Goldstein},
  \citenamefont {Taniguchi}, \citenamefont {Watanabe},\ and\ \citenamefont
  {Yan}}]{chen2018coulomb}%
  \BibitemOpen
  \bibfield  {author} {\bibinfo {author} {\bibfnamefont {S.-Y.}\ \bibnamefont
  {Chen}}, \bibinfo {author} {\bibfnamefont {T.}~\bibnamefont {Goldstein}},
  \bibinfo {author} {\bibfnamefont {T.}~\bibnamefont {Taniguchi}}, \bibinfo
  {author} {\bibfnamefont {K.}~\bibnamefont {Watanabe}},\ and\ \bibinfo
  {author} {\bibfnamefont {J.}~\bibnamefont {Yan}},\ }\bibfield  {title}
  {\bibinfo {title} {Coulomb-bound four-and five-particle intervalley states in
  an atomically-thin semiconductor},\ }\href@noop {} {\bibfield  {journal}
  {\bibinfo  {journal} {Nature Communications}\ }\textbf {\bibinfo {volume}
  {9}},\ \bibinfo {pages} {3717} (\bibinfo {year}
  {2018}{\natexlab{b}})}\BibitemShut {NoStop}%
\bibitem [{\citenamefont {Li}\ \emph {et~al.}(2018)\citenamefont {Li},
  \citenamefont {Wang}, \citenamefont {Lu}, \citenamefont {Jin}, \citenamefont
  {Chen}, \citenamefont {Meng}, \citenamefont {Lian}, \citenamefont
  {Taniguchi}, \citenamefont {Watanabe}, \citenamefont {Zhang}, \citenamefont
  {Smirnov},\ and\ \citenamefont {Shi}}]{li2018revealing}%
  \BibitemOpen
  \bibfield  {author} {\bibinfo {author} {\bibfnamefont {Z.}~\bibnamefont
  {Li}}, \bibinfo {author} {\bibfnamefont {T.}~\bibnamefont {Wang}}, \bibinfo
  {author} {\bibfnamefont {Z.}~\bibnamefont {Lu}}, \bibinfo {author}
  {\bibfnamefont {C.}~\bibnamefont {Jin}}, \bibinfo {author} {\bibfnamefont
  {Y.}~\bibnamefont {Chen}}, \bibinfo {author} {\bibfnamefont {Y.}~\bibnamefont
  {Meng}}, \bibinfo {author} {\bibfnamefont {Z.}~\bibnamefont {Lian}}, \bibinfo
  {author} {\bibfnamefont {T.}~\bibnamefont {Taniguchi}}, \bibinfo {author}
  {\bibfnamefont {K.}~\bibnamefont {Watanabe}}, \bibinfo {author}
  {\bibfnamefont {S.}~\bibnamefont {Zhang}}, \bibinfo {author} {\bibfnamefont
  {D.}~\bibnamefont {Smirnov}},\ and\ \bibinfo {author} {\bibfnamefont {S.-F.}\
  \bibnamefont {Shi}},\ }\bibfield  {title} {\bibinfo {title} {Revealing the
  biexciton and trion-exciton complexes in {BN} encapsulated {WSe}$_2$},\
  }\href@noop {} {\bibfield  {journal} {\bibinfo  {journal} {Nature
  Communications}\ }\textbf {\bibinfo {volume} {9}},\ \bibinfo {pages} {3719}
  (\bibinfo {year} {2018})}\BibitemShut {NoStop}%
\bibitem [{\citenamefont {Barbone}\ \emph {et~al.}(2018)\citenamefont
  {Barbone}, \citenamefont {Montblanch}, \citenamefont {Kara}, \citenamefont
  {Palacios-Berraquero}, \citenamefont {Cadore}, \citenamefont {De~Fazio},
  \citenamefont {Pingault}, \citenamefont {Mostaani}, \citenamefont {Li},
  \citenamefont {Chen}, \citenamefont {Watanabe}, \citenamefont {Taniguchi},
  \citenamefont {Tongay}, \citenamefont {Wang}, \citenamefont {Ferrari},\ and\
  \citenamefont {Atat{\"u}re}}]{barbone2018charge}%
  \BibitemOpen
  \bibfield  {author} {\bibinfo {author} {\bibfnamefont {M.}~\bibnamefont
  {Barbone}}, \bibinfo {author} {\bibfnamefont {A.~R.-P.}\ \bibnamefont
  {Montblanch}}, \bibinfo {author} {\bibfnamefont {D.~M.}\ \bibnamefont
  {Kara}}, \bibinfo {author} {\bibfnamefont {C.}~\bibnamefont
  {Palacios-Berraquero}}, \bibinfo {author} {\bibfnamefont {A.~R.}\
  \bibnamefont {Cadore}}, \bibinfo {author} {\bibfnamefont {D.}~\bibnamefont
  {De~Fazio}}, \bibinfo {author} {\bibfnamefont {B.}~\bibnamefont {Pingault}},
  \bibinfo {author} {\bibfnamefont {E.}~\bibnamefont {Mostaani}}, \bibinfo
  {author} {\bibfnamefont {H.}~\bibnamefont {Li}}, \bibinfo {author}
  {\bibfnamefont {B.}~\bibnamefont {Chen}}, \bibinfo {author} {\bibfnamefont
  {K.}~\bibnamefont {Watanabe}}, \bibinfo {author} {\bibfnamefont
  {T.}~\bibnamefont {Taniguchi}}, \bibinfo {author} {\bibfnamefont
  {S.}~\bibnamefont {Tongay}}, \bibinfo {author} {\bibfnamefont
  {G.}~\bibnamefont {Wang}}, \bibinfo {author} {\bibfnamefont {A.~C.}\
  \bibnamefont {Ferrari}},\ and\ \bibinfo {author} {\bibfnamefont
  {M.}~\bibnamefont {Atat{\"u}re}},\ }\bibfield  {title} {\bibinfo {title}
  {Charge-tuneable biexciton complexes in monolayer {WSe}$_2$},\ }\href@noop {}
  {\bibfield  {journal} {\bibinfo  {journal} {Nature Communications}\ }\textbf
  {\bibinfo {volume} {9}},\ \bibinfo {pages} {3721} (\bibinfo {year}
  {2018})}\BibitemShut {NoStop}%
\bibitem [{\citenamefont {Courtade}\ \emph {et~al.}(2017)\citenamefont
  {Courtade}, \citenamefont {Semina}, \citenamefont {Manca}, \citenamefont
  {Glazov}, \citenamefont {Robert}, \citenamefont {Cadiz}, \citenamefont
  {Wang}, \citenamefont {Taniguchi}, \citenamefont {Watanabe}, \citenamefont
  {Pierre}, \citenamefont {Escoffier}, \citenamefont {Ivchenko}, \citenamefont
  {Renucci}, \citenamefont {Marie}, \citenamefont {Amand},\ and\ \citenamefont
  {Urbaszek}}]{courtade2017charged}%
  \BibitemOpen
  \bibfield  {author} {\bibinfo {author} {\bibfnamefont {E.}~\bibnamefont
  {Courtade}}, \bibinfo {author} {\bibfnamefont {M.}~\bibnamefont {Semina}},
  \bibinfo {author} {\bibfnamefont {M.}~\bibnamefont {Manca}}, \bibinfo
  {author} {\bibfnamefont {M.}~\bibnamefont {Glazov}}, \bibinfo {author}
  {\bibfnamefont {C.}~\bibnamefont {Robert}}, \bibinfo {author} {\bibfnamefont
  {F.}~\bibnamefont {Cadiz}}, \bibinfo {author} {\bibfnamefont
  {G.}~\bibnamefont {Wang}}, \bibinfo {author} {\bibfnamefont {T.}~\bibnamefont
  {Taniguchi}}, \bibinfo {author} {\bibfnamefont {K.}~\bibnamefont {Watanabe}},
  \bibinfo {author} {\bibfnamefont {M.}~\bibnamefont {Pierre}}, \bibinfo
  {author} {\bibfnamefont {W.}~\bibnamefont {Escoffier}}, \bibinfo {author}
  {\bibfnamefont {E.}~\bibnamefont {Ivchenko}}, \bibinfo {author}
  {\bibfnamefont {P.}~\bibnamefont {Renucci}}, \bibinfo {author} {\bibfnamefont
  {X.}~\bibnamefont {Marie}}, \bibinfo {author} {\bibfnamefont
  {T.}~\bibnamefont {Amand}},\ and\ \bibinfo {author} {\bibfnamefont
  {B.}~\bibnamefont {Urbaszek}},\ }\bibfield  {title} {\bibinfo {title}
  {Charged excitons in monolayer {WSe}$_2$: Experiment and theory},\
  }\href@noop {} {\bibfield  {journal} {\bibinfo  {journal} {Physical Review
  B}\ }\textbf {\bibinfo {volume} {96}},\ \bibinfo {pages} {085302} (\bibinfo
  {year} {2017})}\BibitemShut {NoStop}%
\bibitem [{\citenamefont {Li}\ \emph {et~al.}(2019)\citenamefont {Li},
  \citenamefont {Wang}, \citenamefont {Lu}, \citenamefont {Khatoniar},
  \citenamefont {Lian}, \citenamefont {Meng}, \citenamefont {Blei},
  \citenamefont {Taniguchi}, \citenamefont {Watanabe}, \citenamefont {McGill},
  \citenamefont {Tongay}, \citenamefont {Menon}, \citenamefont {Smirnov},\ and\
  \citenamefont {Shi}}]{li2019direct}%
  \BibitemOpen
  \bibfield  {author} {\bibinfo {author} {\bibfnamefont {Z.}~\bibnamefont
  {Li}}, \bibinfo {author} {\bibfnamefont {T.}~\bibnamefont {Wang}}, \bibinfo
  {author} {\bibfnamefont {Z.}~\bibnamefont {Lu}}, \bibinfo {author}
  {\bibfnamefont {M.}~\bibnamefont {Khatoniar}}, \bibinfo {author}
  {\bibfnamefont {Z.}~\bibnamefont {Lian}}, \bibinfo {author} {\bibfnamefont
  {Y.}~\bibnamefont {Meng}}, \bibinfo {author} {\bibfnamefont {M.}~\bibnamefont
  {Blei}}, \bibinfo {author} {\bibfnamefont {T.}~\bibnamefont {Taniguchi}},
  \bibinfo {author} {\bibfnamefont {K.}~\bibnamefont {Watanabe}}, \bibinfo
  {author} {\bibfnamefont {S.~A.}\ \bibnamefont {McGill}}, \bibinfo {author}
  {\bibfnamefont {S.}~\bibnamefont {Tongay}}, \bibinfo {author} {\bibfnamefont
  {V.~M.}\ \bibnamefont {Menon}}, \bibinfo {author} {\bibfnamefont
  {D.}~\bibnamefont {Smirnov}},\ and\ \bibinfo {author} {\bibfnamefont {S.-F.}\
  \bibnamefont {Shi}},\ }\bibfield  {title} {\bibinfo {title} {Direct
  observation of gate-tunable dark trions in monolayer {WSe}$_2$},\ }\href@noop
  {} {\bibfield  {journal} {\bibinfo  {journal} {Nano Letters}\ }\textbf
  {\bibinfo {volume} {19}},\ \bibinfo {pages} {6886} (\bibinfo {year}
  {2019})}\BibitemShut {NoStop}%
\bibitem [{\citenamefont {Cadiz}\ \emph {et~al.}(2017)\citenamefont {Cadiz},
  \citenamefont {Courtade}, \citenamefont {Robert}, \citenamefont {Wang},
  \citenamefont {Shen}, \citenamefont {Cai}, \citenamefont {Taniguchi},
  \citenamefont {Watanabe}, \citenamefont {Carrere}, \citenamefont {Lagarde},
  \citenamefont {Manca}, \citenamefont {Amand}, \citenamefont {Renucci},
  \citenamefont {Tongay}, \citenamefont {Marie},\ and\ \citenamefont
  {Urbaszek}}]{cadiz2017excitonic}%
  \BibitemOpen
  \bibfield  {author} {\bibinfo {author} {\bibfnamefont {F.}~\bibnamefont
  {Cadiz}}, \bibinfo {author} {\bibfnamefont {E.}~\bibnamefont {Courtade}},
  \bibinfo {author} {\bibfnamefont {C.}~\bibnamefont {Robert}}, \bibinfo
  {author} {\bibfnamefont {G.}~\bibnamefont {Wang}}, \bibinfo {author}
  {\bibfnamefont {Y.}~\bibnamefont {Shen}}, \bibinfo {author} {\bibfnamefont
  {H.}~\bibnamefont {Cai}}, \bibinfo {author} {\bibfnamefont {T.}~\bibnamefont
  {Taniguchi}}, \bibinfo {author} {\bibfnamefont {K.}~\bibnamefont {Watanabe}},
  \bibinfo {author} {\bibfnamefont {H.}~\bibnamefont {Carrere}}, \bibinfo
  {author} {\bibfnamefont {D.}~\bibnamefont {Lagarde}}, \bibinfo {author}
  {\bibfnamefont {M.}~\bibnamefont {Manca}}, \bibinfo {author} {\bibfnamefont
  {T.}~\bibnamefont {Amand}}, \bibinfo {author} {\bibfnamefont
  {P.}~\bibnamefont {Renucci}}, \bibinfo {author} {\bibfnamefont
  {S.}~\bibnamefont {Tongay}}, \bibinfo {author} {\bibfnamefont
  {X.}~\bibnamefont {Marie}},\ and\ \bibinfo {author} {\bibfnamefont
  {B.}~\bibnamefont {Urbaszek}},\ }\bibfield  {title} {\bibinfo {title}
  {Excitonic linewidth approaching the homogeneous limit in {MoS}$_2$-based van
  der {Waals} heterostructures},\ }\href@noop {} {\bibfield  {journal}
  {\bibinfo  {journal} {Physical Review X}\ }\textbf {\bibinfo {volume} {7}},\
  \bibinfo {pages} {021026} (\bibinfo {year} {2017})}\BibitemShut {NoStop}%
\bibitem [{\citenamefont {Wu}\ \emph {et~al.}(2019)\citenamefont {Wu},
  \citenamefont {Chen}, \citenamefont {Zhou},\ and\ \citenamefont
  {Zhu}}]{wu2019ultrafast}%
  \BibitemOpen
  \bibfield  {author} {\bibinfo {author} {\bibfnamefont {L.}~\bibnamefont
  {Wu}}, \bibinfo {author} {\bibfnamefont {Y.}~\bibnamefont {Chen}}, \bibinfo
  {author} {\bibfnamefont {H.}~\bibnamefont {Zhou}},\ and\ \bibinfo {author}
  {\bibfnamefont {H.}~\bibnamefont {Zhu}},\ }\bibfield  {title} {\bibinfo
  {title} {Ultrafast energy transfer of both bright and dark excitons in {2D}
  van der {Waals} heterostructures beyond dipolar coupling},\ }\href@noop {}
  {\bibfield  {journal} {\bibinfo  {journal} {ACS Nano}\ }\textbf {\bibinfo
  {volume} {13}},\ \bibinfo {pages} {2341} (\bibinfo {year}
  {2019})}\BibitemShut {NoStop}%
\bibitem [{\citenamefont {Zhou}\ \emph {et~al.}(2020)\citenamefont {Zhou},
  \citenamefont {Zhao}, \citenamefont {Tao}, \citenamefont {Li}, \citenamefont
  {Zhou},\ and\ \citenamefont {Zhu}}]{zhou2020controlling}%
  \BibitemOpen
  \bibfield  {author} {\bibinfo {author} {\bibfnamefont {H.}~\bibnamefont
  {Zhou}}, \bibinfo {author} {\bibfnamefont {Y.}~\bibnamefont {Zhao}}, \bibinfo
  {author} {\bibfnamefont {W.}~\bibnamefont {Tao}}, \bibinfo {author}
  {\bibfnamefont {Y.}~\bibnamefont {Li}}, \bibinfo {author} {\bibfnamefont
  {Q.}~\bibnamefont {Zhou}},\ and\ \bibinfo {author} {\bibfnamefont
  {H.}~\bibnamefont {Zhu}},\ }\bibfield  {title} {\bibinfo {title} {Controlling
  exciton and valley dynamics in two-dimensional heterostructures with
  atomically precise interlayer proximity},\ }\href@noop {} {\bibfield
  {journal} {\bibinfo  {journal} {ACS Nano}\ }\textbf {\bibinfo {volume}
  {14}},\ \bibinfo {pages} {4618} (\bibinfo {year} {2020})}\BibitemShut
  {NoStop}%
\bibitem [{\citenamefont {Zhang}\ \emph {et~al.}(2020)\citenamefont {Zhang},
  \citenamefont {Linardy}, \citenamefont {Wang},\ and\ \citenamefont
  {Eda}}]{zhang2020excitonic}%
  \BibitemOpen
  \bibfield  {author} {\bibinfo {author} {\bibfnamefont {Q.}~\bibnamefont
  {Zhang}}, \bibinfo {author} {\bibfnamefont {E.}~\bibnamefont {Linardy}},
  \bibinfo {author} {\bibfnamefont {X.}~\bibnamefont {Wang}},\ and\ \bibinfo
  {author} {\bibfnamefont {G.}~\bibnamefont {Eda}},\ }\bibfield  {title}
  {\bibinfo {title} {Excitonic energy transfer in heterostructures of
  quasi-2{D} perovskite and monolayer {WS}$_2$},\ }\href@noop {} {\bibfield
  {journal} {\bibinfo  {journal} {ACS Nano}\ }\textbf {\bibinfo {volume}
  {14}},\ \bibinfo {pages} {11482} (\bibinfo {year} {2020})}\BibitemShut
  {NoStop}%
\bibitem [{\citenamefont {Chen}\ \emph
  {et~al.}(2020{\natexlab{a}})\citenamefont {Chen}, \citenamefont {Liu},
  \citenamefont {Li}, \citenamefont {Cheng}, \citenamefont {Ma}, \citenamefont
  {Wang},\ and\ \citenamefont {Li}}]{chen2020robust}%
  \BibitemOpen
  \bibfield  {author} {\bibinfo {author} {\bibfnamefont {Y.}~\bibnamefont
  {Chen}}, \bibinfo {author} {\bibfnamefont {Z.}~\bibnamefont {Liu}}, \bibinfo
  {author} {\bibfnamefont {J.}~\bibnamefont {Li}}, \bibinfo {author}
  {\bibfnamefont {X.}~\bibnamefont {Cheng}}, \bibinfo {author} {\bibfnamefont
  {J.}~\bibnamefont {Ma}}, \bibinfo {author} {\bibfnamefont {H.}~\bibnamefont
  {Wang}},\ and\ \bibinfo {author} {\bibfnamefont {D.}~\bibnamefont {Li}},\
  }\bibfield  {title} {\bibinfo {title} {Robust interlayer coupling in
  two-dimensional perovskite/monolayer transition metal dichalcogenide
  heterostructures},\ }\href@noop {} {\bibfield  {journal} {\bibinfo  {journal}
  {ACS Nano}\ }\textbf {\bibinfo {volume} {14}},\ \bibinfo {pages} {10258}
  (\bibinfo {year} {2020}{\natexlab{a}})}\BibitemShut {NoStop}%
\bibitem [{\citenamefont {Chen}\ \emph
  {et~al.}(2020{\natexlab{b}})\citenamefont {Chen}, \citenamefont {Ma},
  \citenamefont {Liu}, \citenamefont {Li}, \citenamefont {Duan},\ and\
  \citenamefont {Li}}]{chen2020manipulation}%
  \BibitemOpen
  \bibfield  {author} {\bibinfo {author} {\bibfnamefont {Y.}~\bibnamefont
  {Chen}}, \bibinfo {author} {\bibfnamefont {J.}~\bibnamefont {Ma}}, \bibinfo
  {author} {\bibfnamefont {Z.}~\bibnamefont {Liu}}, \bibinfo {author}
  {\bibfnamefont {J.}~\bibnamefont {Li}}, \bibinfo {author} {\bibfnamefont
  {X.}~\bibnamefont {Duan}},\ and\ \bibinfo {author} {\bibfnamefont
  {D.}~\bibnamefont {Li}},\ }\bibfield  {title} {\bibinfo {title} {Manipulation
  of valley pseudospin by selective spin injection in chiral two-dimensional
  perovskite/monolayer transition metal dichalcogenide heterostructures},\
  }\href@noop {} {\bibfield  {journal} {\bibinfo  {journal} {ACS Nano}\
  }\textbf {\bibinfo {volume} {14}},\ \bibinfo {pages} {15154} (\bibinfo {year}
  {2020}{\natexlab{b}})}\BibitemShut {NoStop}%
\bibitem [{\citenamefont {Wang}\ \emph {et~al.}(2020)\citenamefont {Wang},
  \citenamefont {Zhang}, \citenamefont {Luo}, \citenamefont {Wang},
  \citenamefont {Zhu}, \citenamefont {Liang}, \citenamefont {Zhang},
  \citenamefont {Yong}, \citenamefont {Yu~Wong}, \citenamefont {Eda},
  \citenamefont {Smet},\ and\ \citenamefont {Wee}}]{wang2020optoelectronic}%
  \BibitemOpen
  \bibfield  {author} {\bibinfo {author} {\bibfnamefont {Q.}~\bibnamefont
  {Wang}}, \bibinfo {author} {\bibfnamefont {Q.}~\bibnamefont {Zhang}},
  \bibinfo {author} {\bibfnamefont {X.}~\bibnamefont {Luo}}, \bibinfo {author}
  {\bibfnamefont {J.}~\bibnamefont {Wang}}, \bibinfo {author} {\bibfnamefont
  {R.}~\bibnamefont {Zhu}}, \bibinfo {author} {\bibfnamefont {Q.}~\bibnamefont
  {Liang}}, \bibinfo {author} {\bibfnamefont {L.}~\bibnamefont {Zhang}},
  \bibinfo {author} {\bibfnamefont {J.~Z.}\ \bibnamefont {Yong}}, \bibinfo
  {author} {\bibfnamefont {C.~P.}\ \bibnamefont {Yu~Wong}}, \bibinfo {author}
  {\bibfnamefont {G.}~\bibnamefont {Eda}}, \bibinfo {author} {\bibfnamefont
  {J.~H.}\ \bibnamefont {Smet}},\ and\ \bibinfo {author} {\bibfnamefont
  {A.~T.}\ \bibnamefont {Wee}},\ }\bibfield  {title} {\bibinfo {title}
  {Optoelectronic properties of a van der {Waals} {WS}$_2$ monolayer/2{D}
  perovskite vertical heterostructure},\ }\href@noop {} {\bibfield  {journal}
  {\bibinfo  {journal} {ACS Applied Materials \& Interfaces}\ }\textbf
  {\bibinfo {volume} {12}},\ \bibinfo {pages} {45235} (\bibinfo {year}
  {2020})}\BibitemShut {NoStop}%
\bibitem [{\citenamefont {Zhou}\ \emph {et~al.}(2022)\citenamefont {Zhou},
  \citenamefont {Lai}, \citenamefont {Sun}, \citenamefont {Zhang},
  \citenamefont {Wang}, \citenamefont {Liu}, \citenamefont {Zhou},\ and\
  \citenamefont {Xie}}]{zhou2022van}%
  \BibitemOpen
  \bibfield  {author} {\bibinfo {author} {\bibfnamefont {H.}~\bibnamefont
  {Zhou}}, \bibinfo {author} {\bibfnamefont {H.}~\bibnamefont {Lai}}, \bibinfo
  {author} {\bibfnamefont {X.}~\bibnamefont {Sun}}, \bibinfo {author}
  {\bibfnamefont {N.}~\bibnamefont {Zhang}}, \bibinfo {author} {\bibfnamefont
  {Y.}~\bibnamefont {Wang}}, \bibinfo {author} {\bibfnamefont {P.}~\bibnamefont
  {Liu}}, \bibinfo {author} {\bibfnamefont {Y.}~\bibnamefont {Zhou}},\ and\
  \bibinfo {author} {\bibfnamefont {W.}~\bibnamefont {Xie}},\ }\bibfield
  {title} {\bibinfo {title} {Van der {Waals MoS}$_2$/two-dimensional perovskite
  heterostructure for sensitive and ultrafast sub-band-gap photodetection},\
  }\href@noop {} {\bibfield  {journal} {\bibinfo  {journal} {ACS Applied
  Materials \& Interfaces}\ }\textbf {\bibinfo {volume} {14}},\ \bibinfo
  {pages} {3356} (\bibinfo {year} {2022})}\BibitemShut {NoStop}%
\bibitem [{\citenamefont {Karpinska}\ \emph {et~al.}(2021)\citenamefont
  {Karpinska}, \citenamefont {Liang}, \citenamefont {Kempt}, \citenamefont
  {Finzel}, \citenamefont {Kamminga}, \citenamefont {Dyksik}, \citenamefont
  {Zhang}, \citenamefont {Knodlseder}, \citenamefont {Maude}, \citenamefont
  {Baranowski}, \citenamefont {K{\l}optowski}, \citenamefont {Ye},
  \citenamefont {Kuc},\ and\ \citenamefont
  {Plochocka}}]{karpinska2021nonradiative}%
  \BibitemOpen
  \bibfield  {author} {\bibinfo {author} {\bibfnamefont {M.}~\bibnamefont
  {Karpinska}}, \bibinfo {author} {\bibfnamefont {M.}~\bibnamefont {Liang}},
  \bibinfo {author} {\bibfnamefont {R.}~\bibnamefont {Kempt}}, \bibinfo
  {author} {\bibfnamefont {K.}~\bibnamefont {Finzel}}, \bibinfo {author}
  {\bibfnamefont {M.}~\bibnamefont {Kamminga}}, \bibinfo {author}
  {\bibfnamefont {M.}~\bibnamefont {Dyksik}}, \bibinfo {author} {\bibfnamefont
  {N.}~\bibnamefont {Zhang}}, \bibinfo {author} {\bibfnamefont
  {C.}~\bibnamefont {Knodlseder}}, \bibinfo {author} {\bibfnamefont {D.~K.}\
  \bibnamefont {Maude}}, \bibinfo {author} {\bibfnamefont {M.}~\bibnamefont
  {Baranowski}}, \bibinfo {author} {\bibfnamefont {{\L}.}~\bibnamefont
  {K{\l}optowski}}, \bibinfo {author} {\bibfnamefont {J.}~\bibnamefont {Ye}},
  \bibinfo {author} {\bibfnamefont {A.}~\bibnamefont {Kuc}},\ and\ \bibinfo
  {author} {\bibfnamefont {P.}~\bibnamefont {Plochocka}},\ }\bibfield  {title}
  {\bibinfo {title} {Nonradiative energy transfer and selective charge transfer
  in a {WS}$_2$/({PEA})$_2${PbI}$_4$ heterostructure},\ }\href@noop {}
  {\bibfield  {journal} {\bibinfo  {journal} {ACS Applied Materials \&
  Interfaces}\ }\textbf {\bibinfo {volume} {13}},\ \bibinfo {pages} {33677}
  (\bibinfo {year} {2021})}\BibitemShut {NoStop}%
\bibitem [{\citenamefont {Karpi{\'n}ska}\ \emph {et~al.}(2022)\citenamefont
  {Karpi{\'n}ska}, \citenamefont {Jasi{\'n}ski}, \citenamefont {Kempt},
  \citenamefont {Ziegler}, \citenamefont {Sansom}, \citenamefont {Taniguchi},
  \citenamefont {Watanabe}, \citenamefont {Snaith}, \citenamefont {Surrente},
  \citenamefont {Dyksik}, \citenamefont {Maude}, \citenamefont {K{\l}optowski},
  \citenamefont {Chernikov}, \citenamefont {Kuc}, \citenamefont {Baranowski},\
  and\ \citenamefont {Plochocka}}]{karpinska2022interlayer}%
  \BibitemOpen
  \bibfield  {author} {\bibinfo {author} {\bibfnamefont {M.}~\bibnamefont
  {Karpi{\'n}ska}}, \bibinfo {author} {\bibfnamefont {J.}~\bibnamefont
  {Jasi{\'n}ski}}, \bibinfo {author} {\bibfnamefont {R.}~\bibnamefont {Kempt}},
  \bibinfo {author} {\bibfnamefont {J.}~\bibnamefont {Ziegler}}, \bibinfo
  {author} {\bibfnamefont {H.}~\bibnamefont {Sansom}}, \bibinfo {author}
  {\bibfnamefont {T.}~\bibnamefont {Taniguchi}}, \bibinfo {author}
  {\bibfnamefont {K.}~\bibnamefont {Watanabe}}, \bibinfo {author}
  {\bibfnamefont {H.}~\bibnamefont {Snaith}}, \bibinfo {author} {\bibfnamefont
  {A.}~\bibnamefont {Surrente}}, \bibinfo {author} {\bibfnamefont
  {M.}~\bibnamefont {Dyksik}}, \bibinfo {author} {\bibfnamefont
  {D.}~\bibnamefont {Maude}}, \bibinfo {author} {\bibfnamefont
  {{\L}.}~\bibnamefont {K{\l}optowski}}, \bibinfo {author} {\bibfnamefont
  {A.}~\bibnamefont {Chernikov}}, \bibinfo {author} {\bibfnamefont
  {A.}~\bibnamefont {Kuc}}, \bibinfo {author} {\bibfnamefont {M.}~\bibnamefont
  {Baranowski}},\ and\ \bibinfo {author} {\bibfnamefont {P.}~\bibnamefont
  {Plochocka}},\ }\bibfield  {title} {\bibinfo {title} {Interlayer excitons in
  {MoSe}$_2$/2{D} perovskite hybrid heterostructures--the interplay between
  charge and energy transfer},\ }\href@noop {} {\bibfield  {journal} {\bibinfo
  {journal} {Nanoscale}\ }\textbf {\bibinfo {volume} {14}},\ \bibinfo {pages}
  {8085} (\bibinfo {year} {2022})}\BibitemShut {NoStop}%
\bibitem [{\citenamefont {Li}\ \emph {et~al.}(2020)\citenamefont {Li},
  \citenamefont {Lu}, \citenamefont {Dubey}, \citenamefont {Devenica},\ and\
  \citenamefont {Srivastava}}]{li2020dipolar}%
  \BibitemOpen
  \bibfield  {author} {\bibinfo {author} {\bibfnamefont {W.}~\bibnamefont
  {Li}}, \bibinfo {author} {\bibfnamefont {X.}~\bibnamefont {Lu}}, \bibinfo
  {author} {\bibfnamefont {S.}~\bibnamefont {Dubey}}, \bibinfo {author}
  {\bibfnamefont {L.}~\bibnamefont {Devenica}},\ and\ \bibinfo {author}
  {\bibfnamefont {A.}~\bibnamefont {Srivastava}},\ }\bibfield  {title}
  {\bibinfo {title} {Dipolar interactions between localized interlayer excitons
  in van der {Waals} heterostructures},\ }\href@noop {} {\bibfield  {journal}
  {\bibinfo  {journal} {Nature Materials}\ }\textbf {\bibinfo {volume} {19}},\
  \bibinfo {pages} {624} (\bibinfo {year} {2020})}\BibitemShut {NoStop}%
\bibitem [{\citenamefont {Rivera}\ \emph {et~al.}(2016)\citenamefont {Rivera},
  \citenamefont {Seyler}, \citenamefont {Yu}, \citenamefont {Schaibley},
  \citenamefont {Yan}, \citenamefont {Mandrus}, \citenamefont {Yao},\ and\
  \citenamefont {Xu}}]{rivera2016valley}%
  \BibitemOpen
  \bibfield  {author} {\bibinfo {author} {\bibfnamefont {P.}~\bibnamefont
  {Rivera}}, \bibinfo {author} {\bibfnamefont {K.~L.}\ \bibnamefont {Seyler}},
  \bibinfo {author} {\bibfnamefont {H.}~\bibnamefont {Yu}}, \bibinfo {author}
  {\bibfnamefont {J.~R.}\ \bibnamefont {Schaibley}}, \bibinfo {author}
  {\bibfnamefont {J.}~\bibnamefont {Yan}}, \bibinfo {author} {\bibfnamefont
  {D.~G.}\ \bibnamefont {Mandrus}}, \bibinfo {author} {\bibfnamefont
  {W.}~\bibnamefont {Yao}},\ and\ \bibinfo {author} {\bibfnamefont
  {X.}~\bibnamefont {Xu}},\ }\bibfield  {title} {\bibinfo {title}
  {Valley-polarized exciton dynamics in a {2D} semiconductor heterostructure},\
  }\href@noop {} {\bibfield  {journal} {\bibinfo  {journal} {Science}\ }\textbf
  {\bibinfo {volume} {351}},\ \bibinfo {pages} {688} (\bibinfo {year}
  {2016})}\BibitemShut {NoStop}%
\bibitem [{\citenamefont {Montblanch}\ \emph {et~al.}(2021)\citenamefont
  {Montblanch}, \citenamefont {Kara}, \citenamefont {Paradisanos},
  \citenamefont {Purser}, \citenamefont {Feuer}, \citenamefont {Alexeev},
  \citenamefont {Stefan}, \citenamefont {Qin}, \citenamefont {Blei},
  \citenamefont {Wang}, \citenamefont {Cadore}, \citenamefont {Latawiec},
  \citenamefont {Lon{\v c}ar}, \citenamefont {Tongay}, \citenamefont
  {Ferrari},\ and\ \citenamefont {Atat{\"u}re}}]{montblanch2021confinement}%
  \BibitemOpen
  \bibfield  {author} {\bibinfo {author} {\bibfnamefont {A.~R.-P.}\
  \bibnamefont {Montblanch}}, \bibinfo {author} {\bibfnamefont {D.~M.}\
  \bibnamefont {Kara}}, \bibinfo {author} {\bibfnamefont {I.}~\bibnamefont
  {Paradisanos}}, \bibinfo {author} {\bibfnamefont {C.~M.}\ \bibnamefont
  {Purser}}, \bibinfo {author} {\bibfnamefont {M.~S.}\ \bibnamefont {Feuer}},
  \bibinfo {author} {\bibfnamefont {E.~M.}\ \bibnamefont {Alexeev}}, \bibinfo
  {author} {\bibfnamefont {L.}~\bibnamefont {Stefan}}, \bibinfo {author}
  {\bibfnamefont {Y.}~\bibnamefont {Qin}}, \bibinfo {author} {\bibfnamefont
  {M.}~\bibnamefont {Blei}}, \bibinfo {author} {\bibfnamefont {G.}~\bibnamefont
  {Wang}}, \bibinfo {author} {\bibfnamefont {A.~R.}\ \bibnamefont {Cadore}},
  \bibinfo {author} {\bibfnamefont {P.}~\bibnamefont {Latawiec}}, \bibinfo
  {author} {\bibfnamefont {M.}~\bibnamefont {Lon{\v c}ar}}, \bibinfo {author}
  {\bibfnamefont {S.}~\bibnamefont {Tongay}}, \bibinfo {author} {\bibfnamefont
  {A.~C.}\ \bibnamefont {Ferrari}},\ and\ \bibinfo {author} {\bibfnamefont
  {M.}~\bibnamefont {Atat{\"u}re}},\ }\bibfield  {title} {\bibinfo {title}
  {Confinement of long-lived interlayer excitons in {WS$_2$/WSe$_2$}
  heterostructures},\ }\href@noop {} {\bibfield  {journal} {\bibinfo  {journal}
  {Communications Physics}\ }\textbf {\bibinfo {volume} {4}},\ \bibinfo {pages}
  {119} (\bibinfo {year} {2021})}\BibitemShut {NoStop}%
\bibitem [{\citenamefont {Butov}\ \emph {et~al.}(1999)\citenamefont {Butov},
  \citenamefont {Shashkin}, \citenamefont {Dolgopolov}, \citenamefont
  {Campman},\ and\ \citenamefont {Gossard}}]{butov1999magneto}%
  \BibitemOpen
  \bibfield  {author} {\bibinfo {author} {\bibfnamefont {L.}~\bibnamefont
  {Butov}}, \bibinfo {author} {\bibfnamefont {A.}~\bibnamefont {Shashkin}},
  \bibinfo {author} {\bibfnamefont {V.}~\bibnamefont {Dolgopolov}}, \bibinfo
  {author} {\bibfnamefont {K.}~\bibnamefont {Campman}},\ and\ \bibinfo {author}
  {\bibfnamefont {A.}~\bibnamefont {Gossard}},\ }\bibfield  {title} {\bibinfo
  {title} {Magneto-optics of the spatially separated electron and hole layers
  in {GaAs/Al$_x$Ga$_{1-x}$As} coupled quantum wells},\ }\href@noop {}
  {\bibfield  {journal} {\bibinfo  {journal} {Physical Review B}\ }\textbf
  {\bibinfo {volume} {60}},\ \bibinfo {pages} {8753} (\bibinfo {year}
  {1999})}\BibitemShut {NoStop}%
\bibitem [{\citenamefont {Jauregui}\ \emph {et~al.}(2019)\citenamefont
  {Jauregui}, \citenamefont {Joe}, \citenamefont {Pistunova}, \citenamefont
  {Wild}, \citenamefont {High}, \citenamefont {Zhou}, \citenamefont {Scuri},
  \citenamefont {De~Greve}, \citenamefont {Sushko}, \citenamefont {Yu},
  \citenamefont {Taniguchi}, \citenamefont {Watanabe}, \citenamefont
  {Needleman}, \citenamefont {Lukin}, \citenamefont {Park},\ and\ \citenamefont
  {Kim}}]{jauregui2019electrical}%
  \BibitemOpen
  \bibfield  {author} {\bibinfo {author} {\bibfnamefont {L.~A.}\ \bibnamefont
  {Jauregui}}, \bibinfo {author} {\bibfnamefont {A.~Y.}\ \bibnamefont {Joe}},
  \bibinfo {author} {\bibfnamefont {K.}~\bibnamefont {Pistunova}}, \bibinfo
  {author} {\bibfnamefont {D.~S.}\ \bibnamefont {Wild}}, \bibinfo {author}
  {\bibfnamefont {A.~A.}\ \bibnamefont {High}}, \bibinfo {author}
  {\bibfnamefont {Y.}~\bibnamefont {Zhou}}, \bibinfo {author} {\bibfnamefont
  {G.}~\bibnamefont {Scuri}}, \bibinfo {author} {\bibfnamefont
  {K.}~\bibnamefont {De~Greve}}, \bibinfo {author} {\bibfnamefont
  {A.}~\bibnamefont {Sushko}}, \bibinfo {author} {\bibfnamefont {C.-H.}\
  \bibnamefont {Yu}}, \bibinfo {author} {\bibfnamefont {T.}~\bibnamefont
  {Taniguchi}}, \bibinfo {author} {\bibfnamefont {K.}~\bibnamefont {Watanabe}},
  \bibinfo {author} {\bibfnamefont {D.~J.}\ \bibnamefont {Needleman}}, \bibinfo
  {author} {\bibfnamefont {M.~D.}\ \bibnamefont {Lukin}}, \bibinfo {author}
  {\bibfnamefont {H.}~\bibnamefont {Park}},\ and\ \bibinfo {author}
  {\bibfnamefont {P.}~\bibnamefont {Kim}},\ }\bibfield  {title} {\bibinfo
  {title} {Electrical control of interlayer exciton dynamics in atomically thin
  heterostructures},\ }\href@noop {} {\bibfield  {journal} {\bibinfo  {journal}
  {Science}\ }\textbf {\bibinfo {volume} {366}},\ \bibinfo {pages} {870}
  (\bibinfo {year} {2019})}\BibitemShut {NoStop}%
\bibitem [{\citenamefont {Laikhtman}\ and\ \citenamefont
  {Rapaport}(2009)}]{laikhtman2009exciton}%
  \BibitemOpen
  \bibfield  {author} {\bibinfo {author} {\bibfnamefont {B.}~\bibnamefont
  {Laikhtman}}\ and\ \bibinfo {author} {\bibfnamefont {R.}~\bibnamefont
  {Rapaport}},\ }\bibfield  {title} {\bibinfo {title} {Exciton correlations in
  coupled quantum wells and their luminescence blue shift},\ }\href@noop {}
  {\bibfield  {journal} {\bibinfo  {journal} {Physical Review B}\ }\textbf
  {\bibinfo {volume} {80}},\ \bibinfo {pages} {195313} (\bibinfo {year}
  {2009})}\BibitemShut {NoStop}%
\bibitem [{\citenamefont {O’Donnell}\ and\ \citenamefont
  {Chen}(1991)}]{o1991temperature}%
  \BibitemOpen
  \bibfield  {author} {\bibinfo {author} {\bibfnamefont {K.~P.}\ \bibnamefont
  {O’Donnell}}\ and\ \bibinfo {author} {\bibfnamefont {X.}~\bibnamefont
  {Chen}},\ }\bibfield  {title} {\bibinfo {title} {Temperature dependence of
  semiconductor band gaps},\ }\href@noop {} {\bibfield  {journal} {\bibinfo
  {journal} {Applied Physics Letters}\ }\textbf {\bibinfo {volume} {58}},\
  \bibinfo {pages} {2924} (\bibinfo {year} {1991})}\BibitemShut {NoStop}%
\bibitem [{\citenamefont {Tongay}\ \emph {et~al.}(2012)\citenamefont {Tongay},
  \citenamefont {Zhou}, \citenamefont {Ataca}, \citenamefont {Lo},
  \citenamefont {Matthews}, \citenamefont {Li}, \citenamefont {Grossman},\ and\
  \citenamefont {Wu}}]{tongay2012thermally}%
  \BibitemOpen
  \bibfield  {author} {\bibinfo {author} {\bibfnamefont {S.}~\bibnamefont
  {Tongay}}, \bibinfo {author} {\bibfnamefont {J.}~\bibnamefont {Zhou}},
  \bibinfo {author} {\bibfnamefont {C.}~\bibnamefont {Ataca}}, \bibinfo
  {author} {\bibfnamefont {K.}~\bibnamefont {Lo}}, \bibinfo {author}
  {\bibfnamefont {T.~S.}\ \bibnamefont {Matthews}}, \bibinfo {author}
  {\bibfnamefont {J.}~\bibnamefont {Li}}, \bibinfo {author} {\bibfnamefont
  {J.~C.}\ \bibnamefont {Grossman}},\ and\ \bibinfo {author} {\bibfnamefont
  {J.}~\bibnamefont {Wu}},\ }\bibfield  {title} {\bibinfo {title} {Thermally
  driven crossover from indirect toward direct bandgap in {2D} semiconductors:
  {MoSe}$_2$ versus {MoS}$_2$},\ }\href@noop {} {\bibfield  {journal} {\bibinfo
   {journal} {Nano Letters}\ }\textbf {\bibinfo {volume} {12}},\ \bibinfo
  {pages} {5576} (\bibinfo {year} {2012})}\BibitemShut {NoStop}%
\bibitem [{\citenamefont {Ross}\ \emph {et~al.}(2013)\citenamefont {Ross},
  \citenamefont {Wu}, \citenamefont {Yu}, \citenamefont {Ghimire},
  \citenamefont {Jones}, \citenamefont {Aivazian}, \citenamefont {Yan},
  \citenamefont {Mandrus}, \citenamefont {Xiao}, \citenamefont {Yao},\ and\
  \citenamefont {Xu}}]{ross2013electrical}%
  \BibitemOpen
  \bibfield  {author} {\bibinfo {author} {\bibfnamefont {J.~S.}\ \bibnamefont
  {Ross}}, \bibinfo {author} {\bibfnamefont {S.}~\bibnamefont {Wu}}, \bibinfo
  {author} {\bibfnamefont {H.}~\bibnamefont {Yu}}, \bibinfo {author}
  {\bibfnamefont {N.~J.}\ \bibnamefont {Ghimire}}, \bibinfo {author}
  {\bibfnamefont {A.~M.}\ \bibnamefont {Jones}}, \bibinfo {author}
  {\bibfnamefont {G.}~\bibnamefont {Aivazian}}, \bibinfo {author}
  {\bibfnamefont {J.}~\bibnamefont {Yan}}, \bibinfo {author} {\bibfnamefont
  {D.~G.}\ \bibnamefont {Mandrus}}, \bibinfo {author} {\bibfnamefont
  {D.}~\bibnamefont {Xiao}}, \bibinfo {author} {\bibfnamefont {W.}~\bibnamefont
  {Yao}},\ and\ \bibinfo {author} {\bibfnamefont {X.}~\bibnamefont {Xu}},\
  }\bibfield  {title} {\bibinfo {title} {Electrical control of neutral and
  charged excitons in a monolayer semiconductor},\ }\href@noop {} {\bibfield
  {journal} {\bibinfo  {journal} {Nature Communications}\ }\textbf {\bibinfo
  {volume} {4}},\ \bibinfo {pages} {1474} (\bibinfo {year} {2013})}\BibitemShut
  {NoStop}%
\bibitem [{\citenamefont {Huang}\ \emph {et~al.}(2016)\citenamefont {Huang},
  \citenamefont {Hoang},\ and\ \citenamefont {Mikkelsen}}]{huang2016probing}%
  \BibitemOpen
  \bibfield  {author} {\bibinfo {author} {\bibfnamefont {J.}~\bibnamefont
  {Huang}}, \bibinfo {author} {\bibfnamefont {T.~B.}\ \bibnamefont {Hoang}},\
  and\ \bibinfo {author} {\bibfnamefont {M.~H.}\ \bibnamefont {Mikkelsen}},\
  }\bibfield  {title} {\bibinfo {title} {Probing the origin of excitonic states
  in monolayer {WSe}$_2$},\ }\href@noop {} {\bibfield  {journal} {\bibinfo
  {journal} {Scientific Reports}\ }\textbf {\bibinfo {volume} {6}},\ \bibinfo
  {pages} {22414} (\bibinfo {year} {2016})}\BibitemShut {NoStop}%
\bibitem [{\citenamefont {Fu}\ \emph {et~al.}(2021)\citenamefont {Fu},
  \citenamefont {Li}, \citenamefont {Solanki}, \citenamefont {Xu},
  \citenamefont {Lekina}, \citenamefont {Ramesh}, \citenamefont {Shen},\ and\
  \citenamefont {Sum}}]{fu2021electronic}%
  \BibitemOpen
  \bibfield  {author} {\bibinfo {author} {\bibfnamefont {J.}~\bibnamefont
  {Fu}}, \bibinfo {author} {\bibfnamefont {M.}~\bibnamefont {Li}}, \bibinfo
  {author} {\bibfnamefont {A.}~\bibnamefont {Solanki}}, \bibinfo {author}
  {\bibfnamefont {Q.}~\bibnamefont {Xu}}, \bibinfo {author} {\bibfnamefont
  {Y.}~\bibnamefont {Lekina}}, \bibinfo {author} {\bibfnamefont
  {S.}~\bibnamefont {Ramesh}}, \bibinfo {author} {\bibfnamefont {Z.~X.}\
  \bibnamefont {Shen}},\ and\ \bibinfo {author} {\bibfnamefont {T.~C.}\
  \bibnamefont {Sum}},\ }\bibfield  {title} {\bibinfo {title} {Electronic
  states modulation by coherent optical phonons in {2D} halide perovskites},\
  }\href@noop {} {\bibfield  {journal} {\bibinfo  {journal} {Advanced
  Materials}\ }\textbf {\bibinfo {volume} {33}},\ \bibinfo {pages} {2006233}
  (\bibinfo {year} {2021})}\BibitemShut {NoStop}%
\bibitem [{\citenamefont {Wilson}\ \emph {et~al.}(2019)\citenamefont {Wilson},
  \citenamefont {Frost}, \citenamefont {Wallace},\ and\ \citenamefont
  {Walsh}}]{wilson2019dielectric}%
  \BibitemOpen
  \bibfield  {author} {\bibinfo {author} {\bibfnamefont {J.~N.}\ \bibnamefont
  {Wilson}}, \bibinfo {author} {\bibfnamefont {J.~M.}\ \bibnamefont {Frost}},
  \bibinfo {author} {\bibfnamefont {S.~K.}\ \bibnamefont {Wallace}},\ and\
  \bibinfo {author} {\bibfnamefont {A.}~\bibnamefont {Walsh}},\ }\bibfield
  {title} {\bibinfo {title} {Dielectric and ferroic properties of metal halide
  perovskites},\ }\href@noop {} {\bibfield  {journal} {\bibinfo  {journal} {APL
  Materials}\ }\textbf {\bibinfo {volume} {7}} (\bibinfo {year}
  {2019})}\BibitemShut {NoStop}%
\bibitem [{\citenamefont {Godde}\ \emph {et~al.}(2016)\citenamefont {Godde},
  \citenamefont {Schmidt}, \citenamefont {Schmutzler}, \citenamefont
  {A{\ss}mann}, \citenamefont {Debus}, \citenamefont {Withers}, \citenamefont
  {Alexeev}, \citenamefont {Del Pozo-Zamudio}, \citenamefont {Skrypka},
  \citenamefont {Novoselov}, \citenamefont {Bayer},\ and\ \citenamefont
  {Tartakovskii}}]{godde2016exciton}%
  \BibitemOpen
  \bibfield  {author} {\bibinfo {author} {\bibfnamefont {T.}~\bibnamefont
  {Godde}}, \bibinfo {author} {\bibfnamefont {D.}~\bibnamefont {Schmidt}},
  \bibinfo {author} {\bibfnamefont {J.}~\bibnamefont {Schmutzler}}, \bibinfo
  {author} {\bibfnamefont {M.}~\bibnamefont {A{\ss}mann}}, \bibinfo {author}
  {\bibfnamefont {J.}~\bibnamefont {Debus}}, \bibinfo {author} {\bibfnamefont
  {F.}~\bibnamefont {Withers}}, \bibinfo {author} {\bibfnamefont
  {E.}~\bibnamefont {Alexeev}}, \bibinfo {author} {\bibfnamefont
  {O.}~\bibnamefont {Del Pozo-Zamudio}}, \bibinfo {author} {\bibfnamefont
  {O.}~\bibnamefont {Skrypka}}, \bibinfo {author} {\bibfnamefont
  {K.}~\bibnamefont {Novoselov}}, \bibinfo {author} {\bibfnamefont
  {M.}~\bibnamefont {Bayer}},\ and\ \bibinfo {author} {\bibfnamefont
  {A.}~\bibnamefont {Tartakovskii}},\ }\bibfield  {title} {\bibinfo {title}
  {Exciton and trion dynamics in atomically thin {MoSe}$_2$ and {WSe}$_2$:
  Effect of localization},\ }\href@noop {} {\bibfield  {journal} {\bibinfo
  {journal} {Physical Review B}\ }\textbf {\bibinfo {volume} {94}},\ \bibinfo
  {pages} {165301} (\bibinfo {year} {2016})}\BibitemShut {NoStop}%
\bibitem [{\citenamefont {Zhang}\ \emph {et~al.}(2015)\citenamefont {Zhang},
  \citenamefont {You}, \citenamefont {Zhao},\ and\ \citenamefont
  {Heinz}}]{zhang2015experimental}%
  \BibitemOpen
  \bibfield  {author} {\bibinfo {author} {\bibfnamefont {X.-X.}\ \bibnamefont
  {Zhang}}, \bibinfo {author} {\bibfnamefont {Y.}~\bibnamefont {You}}, \bibinfo
  {author} {\bibfnamefont {S.~Y.~F.}\ \bibnamefont {Zhao}},\ and\ \bibinfo
  {author} {\bibfnamefont {T.~F.}\ \bibnamefont {Heinz}},\ }\bibfield  {title}
  {\bibinfo {title} {Experimental evidence for dark excitons in monolayer
  {WSe}$_2$},\ }\href@noop {} {\bibfield  {journal} {\bibinfo  {journal}
  {Physical Review Letters}\ }\textbf {\bibinfo {volume} {115}},\ \bibinfo
  {pages} {257403} (\bibinfo {year} {2015})}\BibitemShut {NoStop}%
\bibitem [{\citenamefont {Shibata}(1998)}]{shibata1998negative}%
  \BibitemOpen
  \bibfield  {author} {\bibinfo {author} {\bibfnamefont {H.}~\bibnamefont
  {Shibata}},\ }\bibfield  {title} {\bibinfo {title} {Negative thermal
  quenching curves in photoluminescence of solids},\ }\href@noop {} {\bibfield
  {journal} {\bibinfo  {journal} {Japanese Journal of Applied Physics}\
  }\textbf {\bibinfo {volume} {37}},\ \bibinfo {pages} {550} (\bibinfo {year}
  {1998})}\BibitemShut {NoStop}%
\bibitem [{\citenamefont {Fang}\ \emph {et~al.}(2015)\citenamefont {Fang},
  \citenamefont {Wang}, \citenamefont {Sun}, \citenamefont {Lu}, \citenamefont
  {Deng}, \citenamefont {Ma}, \citenamefont {Jiang}, \citenamefont {Jia},
  \citenamefont {Wang}, \citenamefont {Zhou},\ and\ \citenamefont
  {Chen}}]{fang2015investigation}%
  \BibitemOpen
  \bibfield  {author} {\bibinfo {author} {\bibfnamefont {Y.}~\bibnamefont
  {Fang}}, \bibinfo {author} {\bibfnamefont {L.}~\bibnamefont {Wang}}, \bibinfo
  {author} {\bibfnamefont {Q.}~\bibnamefont {Sun}}, \bibinfo {author}
  {\bibfnamefont {T.}~\bibnamefont {Lu}}, \bibinfo {author} {\bibfnamefont
  {Z.}~\bibnamefont {Deng}}, \bibinfo {author} {\bibfnamefont {Z.}~\bibnamefont
  {Ma}}, \bibinfo {author} {\bibfnamefont {Y.}~\bibnamefont {Jiang}}, \bibinfo
  {author} {\bibfnamefont {H.}~\bibnamefont {Jia}}, \bibinfo {author}
  {\bibfnamefont {W.}~\bibnamefont {Wang}}, \bibinfo {author} {\bibfnamefont
  {J.}~\bibnamefont {Zhou}},\ and\ \bibinfo {author} {\bibfnamefont
  {H.}~\bibnamefont {Chen}},\ }\bibfield  {title} {\bibinfo {title}
  {Investigation of temperature-dependent photoluminescence in multi-quantum
  wells},\ }\href@noop {} {\bibfield  {journal} {\bibinfo  {journal}
  {Scientific Reports}\ }\textbf {\bibinfo {volume} {5}},\ \bibinfo {pages}
  {12718} (\bibinfo {year} {2015})}\BibitemShut {NoStop}%
\bibitem [{\citenamefont {Hill}\ \emph {et~al.}(2017)\citenamefont {Hill},
  \citenamefont {Rigosi}, \citenamefont {Raja}, \citenamefont {Chernikov},
  \citenamefont {Roquelet},\ and\ \citenamefont {Heinz}}]{hill2017exciton}%
  \BibitemOpen
  \bibfield  {author} {\bibinfo {author} {\bibfnamefont {H.~M.}\ \bibnamefont
  {Hill}}, \bibinfo {author} {\bibfnamefont {A.~F.}\ \bibnamefont {Rigosi}},
  \bibinfo {author} {\bibfnamefont {A.}~\bibnamefont {Raja}}, \bibinfo {author}
  {\bibfnamefont {A.}~\bibnamefont {Chernikov}}, \bibinfo {author}
  {\bibfnamefont {C.}~\bibnamefont {Roquelet}},\ and\ \bibinfo {author}
  {\bibfnamefont {T.~F.}\ \bibnamefont {Heinz}},\ }\bibfield  {title} {\bibinfo
  {title} {Exciton broadening in {WS}$_2$/graphene heterostructures},\
  }\href@noop {} {\bibfield  {journal} {\bibinfo  {journal} {Physical Review
  B}\ }\textbf {\bibinfo {volume} {96}},\ \bibinfo {pages} {205401} (\bibinfo
  {year} {2017})}\BibitemShut {NoStop}%
\end{thebibliography}%


%apsrev4-2.bst 2019-01-14 (MD) hand-edited version of apsrev4-1.bst
%Control: key (0)
%Control: author (8) initials jnrlst
%Control: editor formatted (1) identically to author
%Control: production of article title (0) allowed
%Control: page (0) single
%Control: year (1) truncated
%Control: production of eprint (0) enabled
\begin{thebibliography}{65}%
\makeatletter
\providecommand \@ifxundefined [1]{%
 \@ifx{#1\undefined}
}%
\providecommand \@ifnum [1]{%
 \ifnum #1\expandafter \@firstoftwo
 \else \expandafter \@secondoftwo
 \fi
}%
\providecommand \@ifx [1]{%
 \ifx #1\expandafter \@firstoftwo
 \else \expandafter \@secondoftwo
 \fi
}%
\providecommand \natexlab [1]{#1}%
\providecommand \enquote  [1]{``#1''}%
\providecommand \bibnamefont  [1]{#1}%
\providecommand \bibfnamefont [1]{#1}%
\providecommand \citenamefont [1]{#1}%
\providecommand \href@noop [0]{\@secondoftwo}%
\providecommand \href [0]{\begingroup \@sanitize@url \@href}%
\providecommand \@href[1]{\@@startlink{#1}\@@href}%
\providecommand \@@href[1]{\endgroup#1\@@endlink}%
\providecommand \@sanitize@url [0]{\catcode `\\12\catcode `\$12\catcode
  `\&12\catcode `\#12\catcode `\^12\catcode `\_12\catcode `\%12\relax}%
\providecommand \@@startlink[1]{}%
\providecommand \@@endlink[0]{}%
\providecommand \url  [0]{\begingroup\@sanitize@url \@url }%
\providecommand \@url [1]{\endgroup\@href {#1}{\urlprefix }}%
\providecommand \urlprefix  [0]{URL }%
\providecommand \Eprint [0]{\href }%
\providecommand \doibase [0]{https://doi.org/}%
\providecommand \selectlanguage [0]{\@gobble}%
\providecommand \bibinfo  [0]{\@secondoftwo}%
\providecommand \bibfield  [0]{\@secondoftwo}%
\providecommand \translation [1]{[#1]}%
\providecommand \BibitemOpen [0]{}%
\providecommand \bibitemStop [0]{}%
\providecommand \bibitemNoStop [0]{.\EOS\space}%
\providecommand \EOS [0]{\spacefactor3000\relax}%
\providecommand \BibitemShut  [1]{\csname bibitem#1\endcsname}%
\let\auto@bib@innerbib\@empty
%</preamble>
\bibitem [{\citenamefont {{\v{Z}}uti{\'c}}\ \emph {et~al.}(2004)\citenamefont
  {{\v{Z}}uti{\'c}}, \citenamefont {Fabian},\ and\ \citenamefont
  {Sarma}}]{vzutic2004spintronics}%
  \BibitemOpen
  \bibfield  {author} {\bibinfo {author} {\bibfnamefont {I.}~\bibnamefont
  {{\v{Z}}uti{\'c}}}, \bibinfo {author} {\bibfnamefont {J.}~\bibnamefont
  {Fabian}},\ and\ \bibinfo {author} {\bibfnamefont {S.~D.}\ \bibnamefont
  {Sarma}},\ }\bibfield  {title} {\bibinfo {title} {Spintronics: Fundamentals
  and applications},\ }\href@noop {} {\bibfield  {journal} {\bibinfo  {journal}
  {Reviews of Modern Physics}\ }\textbf {\bibinfo {volume} {76}},\ \bibinfo
  {pages} {323} (\bibinfo {year} {2004})}\BibitemShut {NoStop}%
\bibitem [{\citenamefont {Hirohata}\ \emph {et~al.}(2020)\citenamefont
  {Hirohata}, \citenamefont {Yamada}, \citenamefont {Nakatani}, \citenamefont
  {Prejbeanu}, \citenamefont {Di{\'e}ny}, \citenamefont {Pirro},\ and\
  \citenamefont {Hillebrands}}]{hirohata2020review}%
  \BibitemOpen
  \bibfield  {author} {\bibinfo {author} {\bibfnamefont {A.}~\bibnamefont
  {Hirohata}}, \bibinfo {author} {\bibfnamefont {K.}~\bibnamefont {Yamada}},
  \bibinfo {author} {\bibfnamefont {Y.}~\bibnamefont {Nakatani}}, \bibinfo
  {author} {\bibfnamefont {I.-L.}\ \bibnamefont {Prejbeanu}}, \bibinfo {author}
  {\bibfnamefont {B.}~\bibnamefont {Di{\'e}ny}}, \bibinfo {author}
  {\bibfnamefont {P.}~\bibnamefont {Pirro}},\ and\ \bibinfo {author}
  {\bibfnamefont {B.}~\bibnamefont {Hillebrands}},\ }\bibfield  {title}
  {\bibinfo {title} {Review on spintronics: Principles and device
  applications},\ }\href@noop {} {\bibfield  {journal} {\bibinfo  {journal}
  {Journal of Magnetism and Magnetic Materials}\ }\textbf {\bibinfo {volume}
  {509}},\ \bibinfo {pages} {166711} (\bibinfo {year} {2020})}\BibitemShut
  {NoStop}%
\bibitem [{\citenamefont {Nishizawa}\ \emph {et~al.}(2017)\citenamefont
  {Nishizawa}, \citenamefont {Nishibayashi},\ and\ \citenamefont
  {Munekata}}]{nishizawa2017pure}%
  \BibitemOpen
  \bibfield  {author} {\bibinfo {author} {\bibfnamefont {N.}~\bibnamefont
  {Nishizawa}}, \bibinfo {author} {\bibfnamefont {K.}~\bibnamefont
  {Nishibayashi}},\ and\ \bibinfo {author} {\bibfnamefont {H.}~\bibnamefont
  {Munekata}},\ }\bibfield  {title} {\bibinfo {title} {Pure circular
  polarization electroluminescence at room temperature with spin-polarized
  light-emitting diodes},\ }\href@noop {} {\bibfield  {journal} {\bibinfo
  {journal} {Proceedings of the National Academy of Sciences}\ }\textbf
  {\bibinfo {volume} {114}},\ \bibinfo {pages} {1783} (\bibinfo {year}
  {2017})}\BibitemShut {NoStop}%
\bibitem [{\citenamefont {Holub}\ \emph {et~al.}(2007)\citenamefont {Holub},
  \citenamefont {Shin}, \citenamefont {Saha},\ and\ \citenamefont
  {Bhattacharya}}]{holub2007electrical}%
  \BibitemOpen
  \bibfield  {author} {\bibinfo {author} {\bibfnamefont {M.}~\bibnamefont
  {Holub}}, \bibinfo {author} {\bibfnamefont {J.}~\bibnamefont {Shin}},
  \bibinfo {author} {\bibfnamefont {D.}~\bibnamefont {Saha}},\ and\ \bibinfo
  {author} {\bibfnamefont {P.}~\bibnamefont {Bhattacharya}},\ }\bibfield
  {title} {\bibinfo {title} {Electrical spin injection and threshold reduction
  in a semiconductor laser},\ }\href@noop {} {\bibfield  {journal} {\bibinfo
  {journal} {Physical Review Letters}\ }\textbf {\bibinfo {volume} {98}},\
  \bibinfo {pages} {146603} (\bibinfo {year} {2007})}\BibitemShut {NoStop}%
\bibitem [{\citenamefont {Zhai}\ \emph {et~al.}(2017)\citenamefont {Zhai},
  \citenamefont {Baniya}, \citenamefont {Zhang}, \citenamefont {Li},
  \citenamefont {Haney}, \citenamefont {Sheng}, \citenamefont {Ehrenfreund},\
  and\ \citenamefont {Vardeny}}]{zhai2017giant}%
  \BibitemOpen
  \bibfield  {author} {\bibinfo {author} {\bibfnamefont {Y.}~\bibnamefont
  {Zhai}}, \bibinfo {author} {\bibfnamefont {S.}~\bibnamefont {Baniya}},
  \bibinfo {author} {\bibfnamefont {C.}~\bibnamefont {Zhang}}, \bibinfo
  {author} {\bibfnamefont {J.}~\bibnamefont {Li}}, \bibinfo {author}
  {\bibfnamefont {P.}~\bibnamefont {Haney}}, \bibinfo {author} {\bibfnamefont
  {C.-X.}\ \bibnamefont {Sheng}}, \bibinfo {author} {\bibfnamefont
  {E.}~\bibnamefont {Ehrenfreund}},\ and\ \bibinfo {author} {\bibfnamefont
  {Z.~V.}\ \bibnamefont {Vardeny}},\ }\bibfield  {title} {\bibinfo {title}
  {Giant {Rashba} splitting in {2D} organic-inorganic halide perovskites
  measured by transient spectroscopies},\ }\href@noop {} {\bibfield  {journal}
  {\bibinfo  {journal} {Science Advances}\ }\textbf {\bibinfo {volume} {3}},\
  \bibinfo {pages} {e1700704} (\bibinfo {year} {2017})}\BibitemShut {NoStop}%
\bibitem [{\citenamefont {Niesner}\ \emph {et~al.}(2018)\citenamefont
  {Niesner}, \citenamefont {Hauck}, \citenamefont {Shrestha}, \citenamefont
  {Levchuk}, \citenamefont {Matt}, \citenamefont {Osvet}, \citenamefont
  {Batentschuk}, \citenamefont {Brabec}, \citenamefont {Weber},\ and\
  \citenamefont {Fauster}}]{niesner2018structural}%
  \BibitemOpen
  \bibfield  {author} {\bibinfo {author} {\bibfnamefont {D.}~\bibnamefont
  {Niesner}}, \bibinfo {author} {\bibfnamefont {M.}~\bibnamefont {Hauck}},
  \bibinfo {author} {\bibfnamefont {S.}~\bibnamefont {Shrestha}}, \bibinfo
  {author} {\bibfnamefont {I.}~\bibnamefont {Levchuk}}, \bibinfo {author}
  {\bibfnamefont {G.~J.}\ \bibnamefont {Matt}}, \bibinfo {author}
  {\bibfnamefont {A.}~\bibnamefont {Osvet}}, \bibinfo {author} {\bibfnamefont
  {M.}~\bibnamefont {Batentschuk}}, \bibinfo {author} {\bibfnamefont
  {C.}~\bibnamefont {Brabec}}, \bibinfo {author} {\bibfnamefont {H.~B.}\
  \bibnamefont {Weber}},\ and\ \bibinfo {author} {\bibfnamefont
  {T.}~\bibnamefont {Fauster}},\ }\bibfield  {title} {\bibinfo {title}
  {Structural fluctuations cause spin-split states in tetragonal
  {(CH$_3$NH$_3$)PbI$_3$} as evidenced by the circular photogalvanic effect},\
  }\href@noop {} {\bibfield  {journal} {\bibinfo  {journal} {Proceedings of the
  National Academy of Sciences}\ }\textbf {\bibinfo {volume} {115}},\ \bibinfo
  {pages} {9509} (\bibinfo {year} {2018})}\BibitemShut {NoStop}%
\bibitem [{\citenamefont {Niesner}\ \emph {et~al.}(2016)\citenamefont
  {Niesner}, \citenamefont {Wilhelm}, \citenamefont {Levchuk}, \citenamefont
  {Osvet}, \citenamefont {Shrestha}, \citenamefont {Batentschuk}, \citenamefont
  {Brabec},\ and\ \citenamefont {Fauster}}]{niesner2016giant}%
  \BibitemOpen
  \bibfield  {author} {\bibinfo {author} {\bibfnamefont {D.}~\bibnamefont
  {Niesner}}, \bibinfo {author} {\bibfnamefont {M.}~\bibnamefont {Wilhelm}},
  \bibinfo {author} {\bibfnamefont {I.}~\bibnamefont {Levchuk}}, \bibinfo
  {author} {\bibfnamefont {A.}~\bibnamefont {Osvet}}, \bibinfo {author}
  {\bibfnamefont {S.}~\bibnamefont {Shrestha}}, \bibinfo {author}
  {\bibfnamefont {M.}~\bibnamefont {Batentschuk}}, \bibinfo {author}
  {\bibfnamefont {C.}~\bibnamefont {Brabec}},\ and\ \bibinfo {author}
  {\bibfnamefont {T.}~\bibnamefont {Fauster}},\ }\bibfield  {title} {\bibinfo
  {title} {Giant {Rashba} splitting in {CH$_3$NH$_3$PbBr$_3$} organic-inorganic
  perovskite},\ }\href@noop {} {\bibfield  {journal} {\bibinfo  {journal}
  {Physical Review Letters}\ }\textbf {\bibinfo {volume} {117}},\ \bibinfo
  {pages} {126401} (\bibinfo {year} {2016})}\BibitemShut {NoStop}%
\bibitem [{\citenamefont {Yin}\ \emph {et~al.}(2021)\citenamefont {Yin},
  \citenamefont {Naphade}, \citenamefont {Maity}, \citenamefont
  {Guti{\'e}rrez-Arzaluz}, \citenamefont {Almalawi}, \citenamefont {Roqan},
  \citenamefont {Br{\'e}das}, \citenamefont {Bakr},\ and\ \citenamefont
  {Mohammed}}]{yin2021manipulation}%
  \BibitemOpen
  \bibfield  {author} {\bibinfo {author} {\bibfnamefont {J.}~\bibnamefont
  {Yin}}, \bibinfo {author} {\bibfnamefont {R.}~\bibnamefont {Naphade}},
  \bibinfo {author} {\bibfnamefont {P.}~\bibnamefont {Maity}}, \bibinfo
  {author} {\bibfnamefont {L.}~\bibnamefont {Guti{\'e}rrez-Arzaluz}}, \bibinfo
  {author} {\bibfnamefont {D.}~\bibnamefont {Almalawi}}, \bibinfo {author}
  {\bibfnamefont {I.~S.}\ \bibnamefont {Roqan}}, \bibinfo {author}
  {\bibfnamefont {J.-L.}\ \bibnamefont {Br{\'e}das}}, \bibinfo {author}
  {\bibfnamefont {O.~M.}\ \bibnamefont {Bakr}},\ and\ \bibinfo {author}
  {\bibfnamefont {O.~F.}\ \bibnamefont {Mohammed}},\ }\bibfield  {title}
  {\bibinfo {title} {Manipulation of hot carrier cooling dynamics in
  two-dimensional {Dion--Jacobson} hybrid perovskites via {Rashba} band
  splitting},\ }\href@noop {} {\bibfield  {journal} {\bibinfo  {journal}
  {Nature Communications}\ }\textbf {\bibinfo {volume} {12}},\ \bibinfo {pages}
  {3995} (\bibinfo {year} {2021})}\BibitemShut {NoStop}%
\bibitem [{\citenamefont {Even}\ \emph {et~al.}(2013)\citenamefont {Even},
  \citenamefont {Pedesseau}, \citenamefont {Jancu},\ and\ \citenamefont
  {Katan}}]{even2013importance}%
  \BibitemOpen
  \bibfield  {author} {\bibinfo {author} {\bibfnamefont {J.}~\bibnamefont
  {Even}}, \bibinfo {author} {\bibfnamefont {L.}~\bibnamefont {Pedesseau}},
  \bibinfo {author} {\bibfnamefont {J.-M.}\ \bibnamefont {Jancu}},\ and\
  \bibinfo {author} {\bibfnamefont {C.}~\bibnamefont {Katan}},\ }\bibfield
  {title} {\bibinfo {title} {Importance of spin--orbit coupling in hybrid
  organic/inorganic perovskites for photovoltaic applications},\ }\href@noop {}
  {\bibfield  {journal} {\bibinfo  {journal} {The Journal of Physical Chemistry
  Letters}\ }\textbf {\bibinfo {volume} {4}},\ \bibinfo {pages} {2999}
  (\bibinfo {year} {2013})}\BibitemShut {NoStop}%
\bibitem [{\citenamefont {Odenthal}\ \emph {et~al.}(2017)\citenamefont
  {Odenthal}, \citenamefont {Talmadge}, \citenamefont {Gundlach}, \citenamefont
  {Wang}, \citenamefont {Zhang}, \citenamefont {Sun}, \citenamefont {Yu},
  \citenamefont {Valy~Vardeny},\ and\ \citenamefont {Li}}]{odenthal2017spin}%
  \BibitemOpen
  \bibfield  {author} {\bibinfo {author} {\bibfnamefont {P.}~\bibnamefont
  {Odenthal}}, \bibinfo {author} {\bibfnamefont {W.}~\bibnamefont {Talmadge}},
  \bibinfo {author} {\bibfnamefont {N.}~\bibnamefont {Gundlach}}, \bibinfo
  {author} {\bibfnamefont {R.}~\bibnamefont {Wang}}, \bibinfo {author}
  {\bibfnamefont {C.}~\bibnamefont {Zhang}}, \bibinfo {author} {\bibfnamefont
  {D.}~\bibnamefont {Sun}}, \bibinfo {author} {\bibfnamefont {Z.-G.}\
  \bibnamefont {Yu}}, \bibinfo {author} {\bibfnamefont {Z.}~\bibnamefont
  {Valy~Vardeny}},\ and\ \bibinfo {author} {\bibfnamefont {Y.~S.}\ \bibnamefont
  {Li}},\ }\bibfield  {title} {\bibinfo {title} {Spin-polarized exciton quantum
  beating in hybrid organic--inorganic perovskites},\ }\href@noop {} {\bibfield
   {journal} {\bibinfo  {journal} {Nature Physics}\ }\textbf {\bibinfo {volume}
  {13}},\ \bibinfo {pages} {894} (\bibinfo {year} {2017})}\BibitemShut
  {NoStop}%
\bibitem [{\citenamefont {Chen}\ \emph {et~al.}(2018)\citenamefont {Chen},
  \citenamefont {Lu}, \citenamefont {Li}, \citenamefont {Zhai}, \citenamefont
  {Ndione}, \citenamefont {Berry}, \citenamefont {Zhu}, \citenamefont {Yang},\
  and\ \citenamefont {Beard}}]{chen2018impact}%
  \BibitemOpen
  \bibfield  {author} {\bibinfo {author} {\bibfnamefont {X.}~\bibnamefont
  {Chen}}, \bibinfo {author} {\bibfnamefont {H.}~\bibnamefont {Lu}}, \bibinfo
  {author} {\bibfnamefont {Z.}~\bibnamefont {Li}}, \bibinfo {author}
  {\bibfnamefont {Y.}~\bibnamefont {Zhai}}, \bibinfo {author} {\bibfnamefont
  {P.~F.}\ \bibnamefont {Ndione}}, \bibinfo {author} {\bibfnamefont {J.~J.}\
  \bibnamefont {Berry}}, \bibinfo {author} {\bibfnamefont {K.}~\bibnamefont
  {Zhu}}, \bibinfo {author} {\bibfnamefont {Y.}~\bibnamefont {Yang}},\ and\
  \bibinfo {author} {\bibfnamefont {M.~C.}\ \bibnamefont {Beard}},\ }\bibfield
  {title} {\bibinfo {title} {Impact of layer thickness on the charge carrier
  and spin coherence lifetime in two-dimensional layered perovskite single
  crystals},\ }\href@noop {} {\bibfield  {journal} {\bibinfo  {journal} {ACS
  Energy Letters}\ }\textbf {\bibinfo {volume} {3}},\ \bibinfo {pages} {2273}
  (\bibinfo {year} {2018})}\BibitemShut {NoStop}%
\bibitem [{\citenamefont {Zhan}\ \emph {et~al.}(2022)\citenamefont {Zhan},
  \citenamefont {Zhang}, \citenamefont {Zhang}, \citenamefont {Ou},
  \citenamefont {Yang}, \citenamefont {Qian}, \citenamefont {Zhang},
  \citenamefont {Xing}, \citenamefont {Zhang}, \citenamefont {Li},
  \citenamefont {Zhong}, \citenamefont {Yuan}, \citenamefont {Cao},
  \citenamefont {Zhou}, \citenamefont {Chen}, \citenamefont {Ma}, \citenamefont
  {Song}, \citenamefont {Zha}, \citenamefont {Huang}, \citenamefont {Wang},
  \citenamefont {Wang}, \citenamefont {Huang},\ and\ \citenamefont
  {Wang}}]{zhan2022stimulating}%
  \BibitemOpen
  \bibfield  {author} {\bibinfo {author} {\bibfnamefont {G.}~\bibnamefont
  {Zhan}}, \bibinfo {author} {\bibfnamefont {J.}~\bibnamefont {Zhang}},
  \bibinfo {author} {\bibfnamefont {L.}~\bibnamefont {Zhang}}, \bibinfo
  {author} {\bibfnamefont {Z.}~\bibnamefont {Ou}}, \bibinfo {author}
  {\bibfnamefont {H.}~\bibnamefont {Yang}}, \bibinfo {author} {\bibfnamefont
  {Y.}~\bibnamefont {Qian}}, \bibinfo {author} {\bibfnamefont {X.}~\bibnamefont
  {Zhang}}, \bibinfo {author} {\bibfnamefont {Z.}~\bibnamefont {Xing}},
  \bibinfo {author} {\bibfnamefont {L.}~\bibnamefont {Zhang}}, \bibinfo
  {author} {\bibfnamefont {C.}~\bibnamefont {Li}}, \bibinfo {author}
  {\bibfnamefont {J.}~\bibnamefont {Zhong}}, \bibinfo {author} {\bibfnamefont
  {J.}~\bibnamefont {Yuan}}, \bibinfo {author} {\bibfnamefont {Y.}~\bibnamefont
  {Cao}}, \bibinfo {author} {\bibfnamefont {D.}~\bibnamefont {Zhou}}, \bibinfo
  {author} {\bibfnamefont {X.}~\bibnamefont {Chen}}, \bibinfo {author}
  {\bibfnamefont {H.}~\bibnamefont {Ma}}, \bibinfo {author} {\bibfnamefont
  {X.}~\bibnamefont {Song}}, \bibinfo {author} {\bibfnamefont {C.}~\bibnamefont
  {Zha}}, \bibinfo {author} {\bibfnamefont {X.}~\bibnamefont {Huang}}, \bibinfo
  {author} {\bibfnamefont {J.}~\bibnamefont {Wang}}, \bibinfo {author}
  {\bibfnamefont {T.}~\bibnamefont {Wang}}, \bibinfo {author} {\bibfnamefont
  {W.}~\bibnamefont {Huang}},\ and\ \bibinfo {author} {\bibfnamefont
  {L.}~\bibnamefont {Wang}},\ }\bibfield  {title} {\bibinfo {title}
  {Stimulating and manipulating robust circularly polarized photoluminescence
  in achiral hybrid perovskites},\ }\href@noop {} {\bibfield  {journal}
  {\bibinfo  {journal} {Nano Letters}\ }\textbf {\bibinfo {volume} {22}},\
  \bibinfo {pages} {3961} (\bibinfo {year} {2022})}\BibitemShut {NoStop}%
\bibitem [{\citenamefont {Long}\ \emph {et~al.}(2018)\citenamefont {Long},
  \citenamefont {Jiang}, \citenamefont {Sabatini}, \citenamefont {Yang},
  \citenamefont {Wei}, \citenamefont {Quan}, \citenamefont {Liang},
  \citenamefont {Rasmita}, \citenamefont {Askerka}, \citenamefont {Walters},
  \citenamefont {Gong}, \citenamefont {Xing}, \citenamefont {Wen},
  \citenamefont {Quinteiro-Bermudez}, \citenamefont {Yuan}, \citenamefont
  {Xing}, \citenamefont {Wang}, \citenamefont {Song}, \citenamefont {Voznyy},
  \citenamefont {Zhang}, \citenamefont {Hoogland}, \citenamefont {Gao},
  \citenamefont {Xiong},\ and\ \citenamefont {Sargent}}]{long2018spin}%
  \BibitemOpen
  \bibfield  {author} {\bibinfo {author} {\bibfnamefont {G.}~\bibnamefont
  {Long}}, \bibinfo {author} {\bibfnamefont {C.}~\bibnamefont {Jiang}},
  \bibinfo {author} {\bibfnamefont {R.}~\bibnamefont {Sabatini}}, \bibinfo
  {author} {\bibfnamefont {Z.}~\bibnamefont {Yang}}, \bibinfo {author}
  {\bibfnamefont {M.}~\bibnamefont {Wei}}, \bibinfo {author} {\bibfnamefont
  {L.~N.}\ \bibnamefont {Quan}}, \bibinfo {author} {\bibfnamefont
  {Q.}~\bibnamefont {Liang}}, \bibinfo {author} {\bibfnamefont
  {A.}~\bibnamefont {Rasmita}}, \bibinfo {author} {\bibfnamefont
  {M.}~\bibnamefont {Askerka}}, \bibinfo {author} {\bibfnamefont
  {G.}~\bibnamefont {Walters}}, \bibinfo {author} {\bibfnamefont
  {X.}~\bibnamefont {Gong}}, \bibinfo {author} {\bibfnamefont {J.}~\bibnamefont
  {Xing}}, \bibinfo {author} {\bibfnamefont {X.}~\bibnamefont {Wen}}, \bibinfo
  {author} {\bibfnamefont {R.}~\bibnamefont {Quinteiro-Bermudez}}, \bibinfo
  {author} {\bibfnamefont {H.}~\bibnamefont {Yuan}}, \bibinfo {author}
  {\bibfnamefont {G.}~\bibnamefont {Xing}}, \bibinfo {author} {\bibfnamefont
  {X.~R.}\ \bibnamefont {Wang}}, \bibinfo {author} {\bibfnamefont
  {D.}~\bibnamefont {Song}}, \bibinfo {author} {\bibfnamefont {O.}~\bibnamefont
  {Voznyy}}, \bibinfo {author} {\bibfnamefont {M.}~\bibnamefont {Zhang}},
  \bibinfo {author} {\bibfnamefont {S.}~\bibnamefont {Hoogland}}, \bibinfo
  {author} {\bibfnamefont {W.}~\bibnamefont {Gao}}, \bibinfo {author}
  {\bibfnamefont {Q.}~\bibnamefont {Xiong}},\ and\ \bibinfo {author}
  {\bibfnamefont {E.~H.}\ \bibnamefont {Sargent}},\ }\bibfield  {title}
  {\bibinfo {title} {Spin control in reduced-dimensional chiral perovskites},\
  }\href@noop {} {\bibfield  {journal} {\bibinfo  {journal} {Nature Photonics}\
  }\textbf {\bibinfo {volume} {12}},\ \bibinfo {pages} {528} (\bibinfo {year}
  {2018})}\BibitemShut {NoStop}%
\bibitem [{\citenamefont {Singh}\ \emph {et~al.}(2023)\citenamefont {Singh},
  \citenamefont {Gong}, \citenamefont {Stevens}, \citenamefont {Hou},
  \citenamefont {Singh}, \citenamefont {Zhang}, \citenamefont {Anantharaman},
  \citenamefont {Mohite}, \citenamefont {Hendrickson}, \citenamefont {Yan},\
  and\ \citenamefont {Jariwala}}]{singh2023valley}%
  \BibitemOpen
  \bibfield  {author} {\bibinfo {author} {\bibfnamefont {S.}~\bibnamefont
  {Singh}}, \bibinfo {author} {\bibfnamefont {W.}~\bibnamefont {Gong}},
  \bibinfo {author} {\bibfnamefont {C.~E.}\ \bibnamefont {Stevens}}, \bibinfo
  {author} {\bibfnamefont {J.}~\bibnamefont {Hou}}, \bibinfo {author}
  {\bibfnamefont {A.}~\bibnamefont {Singh}}, \bibinfo {author} {\bibfnamefont
  {H.}~\bibnamefont {Zhang}}, \bibinfo {author} {\bibfnamefont {S.~B.}\
  \bibnamefont {Anantharaman}}, \bibinfo {author} {\bibfnamefont {A.~D.}\
  \bibnamefont {Mohite}}, \bibinfo {author} {\bibfnamefont {J.~R.}\
  \bibnamefont {Hendrickson}}, \bibinfo {author} {\bibfnamefont
  {Q.}~\bibnamefont {Yan}},\ and\ \bibinfo {author} {\bibfnamefont
  {D.}~\bibnamefont {Jariwala}},\ }\bibfield  {title} {\bibinfo {title}
  {Valley-polarized interlayer excitons in {2D} chalcogenide--halide
  perovskite--van der {Waals} heterostructures},\ }\href@noop {} {\bibfield
  {journal} {\bibinfo  {journal} {ACS Nano}\ }\textbf {\bibinfo {volume}
  {17}},\ \bibinfo {pages} {7487} (\bibinfo {year} {2023})}\BibitemShut
  {NoStop}%
\bibitem [{\citenamefont {Giovanni}\ \emph {et~al.}(2015)\citenamefont
  {Giovanni}, \citenamefont {Ma}, \citenamefont {Chua}, \citenamefont
  {Gr{\"a}tzel}, \citenamefont {Ramesh}, \citenamefont {Mhaisalkar},
  \citenamefont {Mathews},\ and\ \citenamefont {Sum}}]{giovanni2015highly}%
  \BibitemOpen
  \bibfield  {author} {\bibinfo {author} {\bibfnamefont {D.}~\bibnamefont
  {Giovanni}}, \bibinfo {author} {\bibfnamefont {H.}~\bibnamefont {Ma}},
  \bibinfo {author} {\bibfnamefont {J.}~\bibnamefont {Chua}}, \bibinfo {author}
  {\bibfnamefont {M.}~\bibnamefont {Gr{\"a}tzel}}, \bibinfo {author}
  {\bibfnamefont {R.}~\bibnamefont {Ramesh}}, \bibinfo {author} {\bibfnamefont
  {S.}~\bibnamefont {Mhaisalkar}}, \bibinfo {author} {\bibfnamefont
  {N.}~\bibnamefont {Mathews}},\ and\ \bibinfo {author} {\bibfnamefont {T.~C.}\
  \bibnamefont {Sum}},\ }\bibfield  {title} {\bibinfo {title} {Highly
  spin-polarized carrier dynamics and ultralarge photoinduced magnetization in
  {CH$_3$NH$_3$PbI$_3$} perovskite thin films},\ }\href@noop {} {\bibfield
  {journal} {\bibinfo  {journal} {Nano Letters}\ }\textbf {\bibinfo {volume}
  {15}},\ \bibinfo {pages} {1553} (\bibinfo {year} {2015})}\BibitemShut
  {NoStop}%
\bibitem [{\citenamefont {Kopteva}\ \emph {et~al.}(2024)\citenamefont
  {Kopteva}, \citenamefont {Yakovlev}, \citenamefont {Yalcin}, \citenamefont
  {Akimov}, \citenamefont {Nestoklon}, \citenamefont {Glazov}, \citenamefont
  {Kotur}, \citenamefont {Kudlacik}, \citenamefont {Zhukov}, \citenamefont
  {Kirstein}, \citenamefont {Hordiichuk}, \citenamefont {Dirin}, \citenamefont
  {Kovalenko},\ and\ \citenamefont {Bayer}}]{kopteva2024highly}%
  \BibitemOpen
  \bibfield  {author} {\bibinfo {author} {\bibfnamefont {N.~E.}\ \bibnamefont
  {Kopteva}}, \bibinfo {author} {\bibfnamefont {D.~R.}\ \bibnamefont
  {Yakovlev}}, \bibinfo {author} {\bibfnamefont {E.}~\bibnamefont {Yalcin}},
  \bibinfo {author} {\bibfnamefont {I.~A.}\ \bibnamefont {Akimov}}, \bibinfo
  {author} {\bibfnamefont {M.~O.}\ \bibnamefont {Nestoklon}}, \bibinfo {author}
  {\bibfnamefont {M.~M.}\ \bibnamefont {Glazov}}, \bibinfo {author}
  {\bibfnamefont {M.}~\bibnamefont {Kotur}}, \bibinfo {author} {\bibfnamefont
  {D.}~\bibnamefont {Kudlacik}}, \bibinfo {author} {\bibfnamefont {E.~A.}\
  \bibnamefont {Zhukov}}, \bibinfo {author} {\bibfnamefont {E.}~\bibnamefont
  {Kirstein}}, \bibinfo {author} {\bibfnamefont {O.}~\bibnamefont
  {Hordiichuk}}, \bibinfo {author} {\bibfnamefont {D.~N.}\ \bibnamefont
  {Dirin}}, \bibinfo {author} {\bibfnamefont {M.~V.}\ \bibnamefont
  {Kovalenko}},\ and\ \bibinfo {author} {\bibfnamefont {M.}~\bibnamefont
  {Bayer}},\ }\bibfield  {title} {\bibinfo {title} {Highly-polarized emission
  provided by giant optical orientation of exciton spins in lead halide
  perovskite crystals},\ }\href@noop {} {\bibfield  {journal} {\bibinfo
  {journal} {Advanced Science}\ }\textbf {\bibinfo {volume} {11}},\ \bibinfo
  {pages} {2403691} (\bibinfo {year} {2024})}\BibitemShut {NoStop}%
\bibitem [{\citenamefont {Lu}\ \emph {et~al.}(2024)\citenamefont {Lu},
  \citenamefont {Wang}, \citenamefont {Han}, \citenamefont {Zhao},
  \citenamefont {He}, \citenamefont {Song}, \citenamefont {Song},\ and\
  \citenamefont {Miao}}]{lu2024spintronic}%
  \BibitemOpen
  \bibfield  {author} {\bibinfo {author} {\bibfnamefont {Y.}~\bibnamefont
  {Lu}}, \bibinfo {author} {\bibfnamefont {Q.}~\bibnamefont {Wang}}, \bibinfo
  {author} {\bibfnamefont {L.}~\bibnamefont {Han}}, \bibinfo {author}
  {\bibfnamefont {Y.}~\bibnamefont {Zhao}}, \bibinfo {author} {\bibfnamefont
  {Z.}~\bibnamefont {He}}, \bibinfo {author} {\bibfnamefont {W.}~\bibnamefont
  {Song}}, \bibinfo {author} {\bibfnamefont {C.}~\bibnamefont {Song}},\ and\
  \bibinfo {author} {\bibfnamefont {Z.}~\bibnamefont {Miao}},\ }\bibfield
  {title} {\bibinfo {title} {Spintronic phenomena and applications in hybrid
  organic--inorganic perovskites},\ }\href@noop {} {\bibfield  {journal}
  {\bibinfo  {journal} {Advanced Functional Materials}\ }\textbf {\bibinfo
  {volume} {15}},\ \bibinfo {pages} {2314427} (\bibinfo {year}
  {2024})}\BibitemShut {NoStop}%
\bibitem [{\citenamefont {Yumoto}\ \emph {et~al.}(2022)\citenamefont {Yumoto},
  \citenamefont {Sekiguchi}, \citenamefont {Hashimoto}, \citenamefont
  {Nakamura}, \citenamefont {Wakamiya},\ and\ \citenamefont
  {Kanemitsu}}]{yumoto2022rapidly}%
  \BibitemOpen
  \bibfield  {author} {\bibinfo {author} {\bibfnamefont {G.}~\bibnamefont
  {Yumoto}}, \bibinfo {author} {\bibfnamefont {F.}~\bibnamefont {Sekiguchi}},
  \bibinfo {author} {\bibfnamefont {R.}~\bibnamefont {Hashimoto}}, \bibinfo
  {author} {\bibfnamefont {T.}~\bibnamefont {Nakamura}}, \bibinfo {author}
  {\bibfnamefont {A.}~\bibnamefont {Wakamiya}},\ and\ \bibinfo {author}
  {\bibfnamefont {Y.}~\bibnamefont {Kanemitsu}},\ }\bibfield  {title} {\bibinfo
  {title} {Rapidly expanding spin-polarized exciton halo in a two-dimensional
  halide perovskite at room temperature},\ }\href@noop {} {\bibfield  {journal}
  {\bibinfo  {journal} {Science Advances}\ }\textbf {\bibinfo {volume} {8}},\
  \bibinfo {pages} {eabp8135} (\bibinfo {year} {2022})}\BibitemShut {NoStop}%
\bibitem [{\citenamefont {Bourelle}\ \emph {et~al.}(2020)\citenamefont
  {Bourelle}, \citenamefont {Shivanna}, \citenamefont {Camargo}, \citenamefont
  {Ghosh}, \citenamefont {Gillett}, \citenamefont {Senanayak}, \citenamefont
  {Feldmann}, \citenamefont {Eyre}, \citenamefont {Ashoka}, \citenamefont
  {van~de Goor}, \citenamefont {Abolins}, \citenamefont {Wilker}, \citenamefont
  {Cerullo}, \citenamefont {Friend},\ and\ \citenamefont
  {Deschler}}]{bourelle2020exciton}%
  \BibitemOpen
  \bibfield  {author} {\bibinfo {author} {\bibfnamefont {S.~A.}\ \bibnamefont
  {Bourelle}}, \bibinfo {author} {\bibfnamefont {R.}~\bibnamefont {Shivanna}},
  \bibinfo {author} {\bibfnamefont {F.~V.}\ \bibnamefont {Camargo}}, \bibinfo
  {author} {\bibfnamefont {S.}~\bibnamefont {Ghosh}}, \bibinfo {author}
  {\bibfnamefont {A.~J.}\ \bibnamefont {Gillett}}, \bibinfo {author}
  {\bibfnamefont {S.~P.}\ \bibnamefont {Senanayak}}, \bibinfo {author}
  {\bibfnamefont {S.}~\bibnamefont {Feldmann}}, \bibinfo {author}
  {\bibfnamefont {L.}~\bibnamefont {Eyre}}, \bibinfo {author} {\bibfnamefont
  {A.}~\bibnamefont {Ashoka}}, \bibinfo {author} {\bibfnamefont {T.~W.}\
  \bibnamefont {van~de Goor}}, \bibinfo {author} {\bibfnamefont
  {H.}~\bibnamefont {Abolins}}, \bibinfo {author} {\bibfnamefont
  {T.}~\bibnamefont {Wilker}}, \bibinfo {author} {\bibfnamefont
  {G.}~\bibnamefont {Cerullo}}, \bibinfo {author} {\bibfnamefont {R.~H.}\
  \bibnamefont {Friend}},\ and\ \bibinfo {author} {\bibfnamefont
  {F.}~\bibnamefont {Deschler}},\ }\bibfield  {title} {\bibinfo {title} {How
  exciton interactions control spin-depolarization in layered hybrid
  perovskites},\ }\href@noop {} {\bibfield  {journal} {\bibinfo  {journal}
  {Nano Letters}\ }\textbf {\bibinfo {volume} {20}},\ \bibinfo {pages} {5678}
  (\bibinfo {year} {2020})}\BibitemShut {NoStop}%
\bibitem [{\citenamefont {Chen}\ \emph {et~al.}(2021)\citenamefont {Chen},
  \citenamefont {Lu}, \citenamefont {Wang}, \citenamefont {Zhai}, \citenamefont
  {Lunin}, \citenamefont {Sercel},\ and\ \citenamefont
  {Beard}}]{chen2021tuning}%
  \BibitemOpen
  \bibfield  {author} {\bibinfo {author} {\bibfnamefont {X.}~\bibnamefont
  {Chen}}, \bibinfo {author} {\bibfnamefont {H.}~\bibnamefont {Lu}}, \bibinfo
  {author} {\bibfnamefont {K.}~\bibnamefont {Wang}}, \bibinfo {author}
  {\bibfnamefont {Y.}~\bibnamefont {Zhai}}, \bibinfo {author} {\bibfnamefont
  {V.}~\bibnamefont {Lunin}}, \bibinfo {author} {\bibfnamefont {P.~C.}\
  \bibnamefont {Sercel}},\ and\ \bibinfo {author} {\bibfnamefont {M.~C.}\
  \bibnamefont {Beard}},\ }\bibfield  {title} {\bibinfo {title} {Tuning
  spin-polarized lifetime in two-dimensional metal--halide perovskite through
  exciton binding energy},\ }\href@noop {} {\bibfield  {journal} {\bibinfo
  {journal} {Journal of the American Chemical Society}\ }\textbf {\bibinfo
  {volume} {143}},\ \bibinfo {pages} {19438} (\bibinfo {year}
  {2021})}\BibitemShut {NoStop}%
\bibitem [{\citenamefont {Bourelle}\ \emph {et~al.}(2022)\citenamefont
  {Bourelle}, \citenamefont {Camargo}, \citenamefont {Ghosh}, \citenamefont
  {Neumann}, \citenamefont {van~de Goor}, \citenamefont {Shivanna},
  \citenamefont {Winkler}, \citenamefont {Cerullo},\ and\ \citenamefont
  {Deschler}}]{bourelle2022optical}%
  \BibitemOpen
  \bibfield  {author} {\bibinfo {author} {\bibfnamefont {S.~A.}\ \bibnamefont
  {Bourelle}}, \bibinfo {author} {\bibfnamefont {F.~V.}\ \bibnamefont
  {Camargo}}, \bibinfo {author} {\bibfnamefont {S.}~\bibnamefont {Ghosh}},
  \bibinfo {author} {\bibfnamefont {T.}~\bibnamefont {Neumann}}, \bibinfo
  {author} {\bibfnamefont {T.~W.}\ \bibnamefont {van~de Goor}}, \bibinfo
  {author} {\bibfnamefont {R.}~\bibnamefont {Shivanna}}, \bibinfo {author}
  {\bibfnamefont {T.}~\bibnamefont {Winkler}}, \bibinfo {author} {\bibfnamefont
  {G.}~\bibnamefont {Cerullo}},\ and\ \bibinfo {author} {\bibfnamefont
  {F.}~\bibnamefont {Deschler}},\ }\bibfield  {title} {\bibinfo {title}
  {Optical control of exciton spin dynamics in layered metal halide perovskites
  via polaronic state formation},\ }\href@noop {} {\bibfield  {journal}
  {\bibinfo  {journal} {Nature Communications}\ }\textbf {\bibinfo {volume}
  {13}},\ \bibinfo {pages} {3320} (\bibinfo {year} {2022})}\BibitemShut
  {NoStop}%
\bibitem [{\citenamefont {Long}\ \emph {et~al.}(2020)\citenamefont {Long},
  \citenamefont {Sabatini}, \citenamefont {Saidaminov}, \citenamefont
  {Lakhwani}, \citenamefont {Rasmita}, \citenamefont {Liu}, \citenamefont
  {Sargent},\ and\ \citenamefont {Gao}}]{long2020chiral}%
  \BibitemOpen
  \bibfield  {author} {\bibinfo {author} {\bibfnamefont {G.}~\bibnamefont
  {Long}}, \bibinfo {author} {\bibfnamefont {R.}~\bibnamefont {Sabatini}},
  \bibinfo {author} {\bibfnamefont {M.~I.}\ \bibnamefont {Saidaminov}},
  \bibinfo {author} {\bibfnamefont {G.}~\bibnamefont {Lakhwani}}, \bibinfo
  {author} {\bibfnamefont {A.}~\bibnamefont {Rasmita}}, \bibinfo {author}
  {\bibfnamefont {X.}~\bibnamefont {Liu}}, \bibinfo {author} {\bibfnamefont
  {E.~H.}\ \bibnamefont {Sargent}},\ and\ \bibinfo {author} {\bibfnamefont
  {W.}~\bibnamefont {Gao}},\ }\bibfield  {title} {\bibinfo {title}
  {Chiral-perovskite optoelectronics},\ }\href@noop {} {\bibfield  {journal}
  {\bibinfo  {journal} {Nature Reviews Materials}\ }\textbf {\bibinfo {volume}
  {5}},\ \bibinfo {pages} {423} (\bibinfo {year} {2020})}\BibitemShut {NoStop}%
\bibitem [{\citenamefont {Ma}\ \emph {et~al.}(2019)\citenamefont {Ma},
  \citenamefont {Fang}, \citenamefont {Chen}, \citenamefont {Jin},
  \citenamefont {Wang}, \citenamefont {Wang}, \citenamefont {Tang},\ and\
  \citenamefont {Li}}]{ma2019chiral}%
  \BibitemOpen
  \bibfield  {author} {\bibinfo {author} {\bibfnamefont {J.}~\bibnamefont
  {Ma}}, \bibinfo {author} {\bibfnamefont {C.}~\bibnamefont {Fang}}, \bibinfo
  {author} {\bibfnamefont {C.}~\bibnamefont {Chen}}, \bibinfo {author}
  {\bibfnamefont {L.}~\bibnamefont {Jin}}, \bibinfo {author} {\bibfnamefont
  {J.}~\bibnamefont {Wang}}, \bibinfo {author} {\bibfnamefont {S.}~\bibnamefont
  {Wang}}, \bibinfo {author} {\bibfnamefont {J.}~\bibnamefont {Tang}},\ and\
  \bibinfo {author} {\bibfnamefont {D.}~\bibnamefont {Li}},\ }\bibfield
  {title} {\bibinfo {title} {Chiral {2D} perovskites with a high degree of
  circularly polarized photoluminescence},\ }\href@noop {} {\bibfield
  {journal} {\bibinfo  {journal} {ACS Nano}\ }\textbf {\bibinfo {volume}
  {13}},\ \bibinfo {pages} {3659} (\bibinfo {year} {2019})}\BibitemShut
  {NoStop}%
\bibitem [{\citenamefont {Lu}\ \emph {et~al.}(2020)\citenamefont {Lu},
  \citenamefont {Xiao}, \citenamefont {Song}, \citenamefont {Li}, \citenamefont
  {Maughan}, \citenamefont {Levin}, \citenamefont {Brunecky}, \citenamefont
  {Berry}, \citenamefont {Mitzi}, \citenamefont {Blum},\ and\ \citenamefont
  {Beard}}]{lu2020highly}%
  \BibitemOpen
  \bibfield  {author} {\bibinfo {author} {\bibfnamefont {H.}~\bibnamefont
  {Lu}}, \bibinfo {author} {\bibfnamefont {C.}~\bibnamefont {Xiao}}, \bibinfo
  {author} {\bibfnamefont {R.}~\bibnamefont {Song}}, \bibinfo {author}
  {\bibfnamefont {T.}~\bibnamefont {Li}}, \bibinfo {author} {\bibfnamefont
  {A.~E.}\ \bibnamefont {Maughan}}, \bibinfo {author} {\bibfnamefont
  {A.}~\bibnamefont {Levin}}, \bibinfo {author} {\bibfnamefont
  {R.}~\bibnamefont {Brunecky}}, \bibinfo {author} {\bibfnamefont {J.~J.}\
  \bibnamefont {Berry}}, \bibinfo {author} {\bibfnamefont {D.~B.}\ \bibnamefont
  {Mitzi}}, \bibinfo {author} {\bibfnamefont {V.}~\bibnamefont {Blum}},\ and\
  \bibinfo {author} {\bibfnamefont {M.~C.}\ \bibnamefont {Beard}},\ }\bibfield
  {title} {\bibinfo {title} {Highly distorted chiral two-dimensional tin iodide
  perovskites for spin polarized charge transport},\ }\href@noop {} {\bibfield
  {journal} {\bibinfo  {journal} {Journal of the American Chemical Society}\
  }\textbf {\bibinfo {volume} {142}},\ \bibinfo {pages} {13030} (\bibinfo
  {year} {2020})}\BibitemShut {NoStop}%
\bibitem [{\citenamefont {Jana}\ \emph {et~al.}(2020)\citenamefont {Jana},
  \citenamefont {Song}, \citenamefont {Liu}, \citenamefont {Khanal},
  \citenamefont {Janke}, \citenamefont {Zhao}, \citenamefont {Liu},
  \citenamefont {Valy~Vardeny}, \citenamefont {Blum},\ and\ \citenamefont
  {Mitzi}}]{jana2020organic}%
  \BibitemOpen
  \bibfield  {author} {\bibinfo {author} {\bibfnamefont {M.~K.}\ \bibnamefont
  {Jana}}, \bibinfo {author} {\bibfnamefont {R.}~\bibnamefont {Song}}, \bibinfo
  {author} {\bibfnamefont {H.}~\bibnamefont {Liu}}, \bibinfo {author}
  {\bibfnamefont {D.~R.}\ \bibnamefont {Khanal}}, \bibinfo {author}
  {\bibfnamefont {S.~M.}\ \bibnamefont {Janke}}, \bibinfo {author}
  {\bibfnamefont {R.}~\bibnamefont {Zhao}}, \bibinfo {author} {\bibfnamefont
  {C.}~\bibnamefont {Liu}}, \bibinfo {author} {\bibfnamefont {Z.}~\bibnamefont
  {Valy~Vardeny}}, \bibinfo {author} {\bibfnamefont {V.}~\bibnamefont {Blum}},\
  and\ \bibinfo {author} {\bibfnamefont {D.~B.}\ \bibnamefont {Mitzi}},\
  }\bibfield  {title} {\bibinfo {title} {Organic-to-inorganic structural
  chirality transfer in a {2D} hybrid perovskite and impact on
  {Rashba-Dresselhaus} spin-orbit coupling},\ }\href@noop {} {\bibfield
  {journal} {\bibinfo  {journal} {Nature Communications}\ }\textbf {\bibinfo
  {volume} {11}},\ \bibinfo {pages} {4699} (\bibinfo {year}
  {2020})}\BibitemShut {NoStop}%
\bibitem [{\citenamefont {Wang}\ \emph {et~al.}(2023)\citenamefont {Wang},
  \citenamefont {Mao},\ and\ \citenamefont {Vardeny}}]{wang2023chirality}%
  \BibitemOpen
  \bibfield  {author} {\bibinfo {author} {\bibfnamefont {J.}~\bibnamefont
  {Wang}}, \bibinfo {author} {\bibfnamefont {B.}~\bibnamefont {Mao}},\ and\
  \bibinfo {author} {\bibfnamefont {Z.~V.}\ \bibnamefont {Vardeny}},\
  }\bibfield  {title} {\bibinfo {title} {Chirality induced spin selectivity in
  chiral hybrid organic--inorganic perovskites},\ }\href@noop {} {\bibfield
  {journal} {\bibinfo  {journal} {The Journal of Chemical Physics}\ }\textbf
  {\bibinfo {volume} {159}},\ \bibinfo {pages} {091002} (\bibinfo {year}
  {2023})}\BibitemShut {NoStop}%
\bibitem [{\citenamefont {Chen}\ \emph {et~al.}(2019)\citenamefont {Chen},
  \citenamefont {Gao}, \citenamefont {Gao}, \citenamefont {Ge}, \citenamefont
  {Du}, \citenamefont {Li}, \citenamefont {Yang}, \citenamefont {Niu},\ and\
  \citenamefont {Tang}}]{chen2019circularly}%
  \BibitemOpen
  \bibfield  {author} {\bibinfo {author} {\bibfnamefont {C.}~\bibnamefont
  {Chen}}, \bibinfo {author} {\bibfnamefont {L.}~\bibnamefont {Gao}}, \bibinfo
  {author} {\bibfnamefont {W.}~\bibnamefont {Gao}}, \bibinfo {author}
  {\bibfnamefont {C.}~\bibnamefont {Ge}}, \bibinfo {author} {\bibfnamefont
  {X.}~\bibnamefont {Du}}, \bibinfo {author} {\bibfnamefont {Z.}~\bibnamefont
  {Li}}, \bibinfo {author} {\bibfnamefont {Y.}~\bibnamefont {Yang}}, \bibinfo
  {author} {\bibfnamefont {G.}~\bibnamefont {Niu}},\ and\ \bibinfo {author}
  {\bibfnamefont {J.}~\bibnamefont {Tang}},\ }\bibfield  {title} {\bibinfo
  {title} {Circularly polarized light detection using chiral hybrid
  perovskite},\ }\href@noop {} {\bibfield  {journal} {\bibinfo  {journal}
  {Nature Communications}\ }\textbf {\bibinfo {volume} {10}},\ \bibinfo {pages}
  {1927} (\bibinfo {year} {2019})}\BibitemShut {NoStop}%
\bibitem [{\citenamefont {Kim}\ \emph {et~al.}(2021)\citenamefont {Kim},
  \citenamefont {Zhai}, \citenamefont {Lu}, \citenamefont {Pan}, \citenamefont
  {Xiao}, \citenamefont {Gaulding}, \citenamefont {Harvey}, \citenamefont
  {Berry}, \citenamefont {Vardeny}, \citenamefont {Luther},\ and\ \citenamefont
  {Beard}}]{kim2021chiral}%
  \BibitemOpen
  \bibfield  {author} {\bibinfo {author} {\bibfnamefont {Y.-H.}\ \bibnamefont
  {Kim}}, \bibinfo {author} {\bibfnamefont {Y.}~\bibnamefont {Zhai}}, \bibinfo
  {author} {\bibfnamefont {H.}~\bibnamefont {Lu}}, \bibinfo {author}
  {\bibfnamefont {X.}~\bibnamefont {Pan}}, \bibinfo {author} {\bibfnamefont
  {C.}~\bibnamefont {Xiao}}, \bibinfo {author} {\bibfnamefont {E.~A.}\
  \bibnamefont {Gaulding}}, \bibinfo {author} {\bibfnamefont {S.~P.}\
  \bibnamefont {Harvey}}, \bibinfo {author} {\bibfnamefont {J.~J.}\
  \bibnamefont {Berry}}, \bibinfo {author} {\bibfnamefont {Z.~V.}\ \bibnamefont
  {Vardeny}}, \bibinfo {author} {\bibfnamefont {J.~M.}\ \bibnamefont
  {Luther}},\ and\ \bibinfo {author} {\bibfnamefont {M.~C.}\ \bibnamefont
  {Beard}},\ }\bibfield  {title} {\bibinfo {title} {Chiral-induced spin
  selectivity enables a room-temperature spin light-emitting diode},\
  }\href@noop {} {\bibfield  {journal} {\bibinfo  {journal} {Science}\ }\textbf
  {\bibinfo {volume} {371}},\ \bibinfo {pages} {1129} (\bibinfo {year}
  {2021})}\BibitemShut {NoStop}%
\bibitem [{\citenamefont {Ye}\ \emph {et~al.}(2022)\citenamefont {Ye},
  \citenamefont {Jiang}, \citenamefont {Zou}, \citenamefont {Mi},\ and\
  \citenamefont {Xiao}}]{ye2022core}%
  \BibitemOpen
  \bibfield  {author} {\bibinfo {author} {\bibfnamefont {C.}~\bibnamefont
  {Ye}}, \bibinfo {author} {\bibfnamefont {J.}~\bibnamefont {Jiang}}, \bibinfo
  {author} {\bibfnamefont {S.}~\bibnamefont {Zou}}, \bibinfo {author}
  {\bibfnamefont {W.}~\bibnamefont {Mi}},\ and\ \bibinfo {author}
  {\bibfnamefont {Y.}~\bibnamefont {Xiao}},\ }\bibfield  {title} {\bibinfo
  {title} {Core--shell three-dimensional perovskite nanocrystals with
  chiral-induced spin selectivity for room-temperature spin light-emitting
  diodes},\ }\href@noop {} {\bibfield  {journal} {\bibinfo  {journal} {Journal
  of the American Chemical Society}\ }\textbf {\bibinfo {volume} {144}},\
  \bibinfo {pages} {9707} (\bibinfo {year} {2022})}\BibitemShut {NoStop}%
\bibitem [{\citenamefont {Zhang}\ \emph {et~al.}(2024)\citenamefont {Zhang},
  \citenamefont {Tian}, \citenamefont {Ye}, \citenamefont {Wang}, \citenamefont
  {Mi}, \citenamefont {Dai}, \citenamefont {Zou}, \citenamefont {Cao},
  \citenamefont {Gao},\ and\ \citenamefont {Xiao}}]{zhang2024efficient}%
  \BibitemOpen
  \bibfield  {author} {\bibinfo {author} {\bibfnamefont {R.}~\bibnamefont
  {Zhang}}, \bibinfo {author} {\bibfnamefont {Y.}~\bibnamefont {Tian}},
  \bibinfo {author} {\bibfnamefont {C.}~\bibnamefont {Ye}}, \bibinfo {author}
  {\bibfnamefont {Y.}~\bibnamefont {Wang}}, \bibinfo {author} {\bibfnamefont
  {W.}~\bibnamefont {Mi}}, \bibinfo {author} {\bibfnamefont {H.}~\bibnamefont
  {Dai}}, \bibinfo {author} {\bibfnamefont {S.}~\bibnamefont {Zou}}, \bibinfo
  {author} {\bibfnamefont {R.}~\bibnamefont {Cao}}, \bibinfo {author}
  {\bibfnamefont {H.}~\bibnamefont {Gao}},\ and\ \bibinfo {author}
  {\bibfnamefont {Y.}~\bibnamefont {Xiao}},\ }\bibfield  {title} {\bibinfo
  {title} {Efficient quasi-{2D} perovskite spin light-emitting diodes based on
  chiral-induced spin selectivity},\ }\href@noop {} {\bibfield  {journal}
  {\bibinfo  {journal} {Chemistry of Materials}\ }\textbf {\bibinfo {volume}
  {36}},\ \bibinfo {pages} {3812} (\bibinfo {year} {2024})}\BibitemShut
  {NoStop}%
\bibitem [{\citenamefont {Chen}\ \emph
  {et~al.}(2020{\natexlab{a}})\citenamefont {Chen}, \citenamefont {Ma},
  \citenamefont {Liu}, \citenamefont {Li}, \citenamefont {Duan},\ and\
  \citenamefont {Li}}]{chen2020manipulation}%
  \BibitemOpen
  \bibfield  {author} {\bibinfo {author} {\bibfnamefont {Y.}~\bibnamefont
  {Chen}}, \bibinfo {author} {\bibfnamefont {J.}~\bibnamefont {Ma}}, \bibinfo
  {author} {\bibfnamefont {Z.}~\bibnamefont {Liu}}, \bibinfo {author}
  {\bibfnamefont {J.}~\bibnamefont {Li}}, \bibinfo {author} {\bibfnamefont
  {X.}~\bibnamefont {Duan}},\ and\ \bibinfo {author} {\bibfnamefont
  {D.}~\bibnamefont {Li}},\ }\bibfield  {title} {\bibinfo {title} {Manipulation
  of valley pseudospin by selective spin injection in chiral two-dimensional
  perovskite/monolayer transition metal dichalcogenide heterostructures},\
  }\href@noop {} {\bibfield  {journal} {\bibinfo  {journal} {ACS Nano}\
  }\textbf {\bibinfo {volume} {14}},\ \bibinfo {pages} {15154} (\bibinfo {year}
  {2020}{\natexlab{a}})}\BibitemShut {NoStop}%
\bibitem [{\citenamefont {Chen}\ \emph
  {et~al.}(2020{\natexlab{b}})\citenamefont {Chen}, \citenamefont {Liu},
  \citenamefont {Li}, \citenamefont {Cheng}, \citenamefont {Ma}, \citenamefont
  {Wang},\ and\ \citenamefont {Li}}]{chen2020robust}%
  \BibitemOpen
  \bibfield  {author} {\bibinfo {author} {\bibfnamefont {Y.}~\bibnamefont
  {Chen}}, \bibinfo {author} {\bibfnamefont {Z.}~\bibnamefont {Liu}}, \bibinfo
  {author} {\bibfnamefont {J.}~\bibnamefont {Li}}, \bibinfo {author}
  {\bibfnamefont {X.}~\bibnamefont {Cheng}}, \bibinfo {author} {\bibfnamefont
  {J.}~\bibnamefont {Ma}}, \bibinfo {author} {\bibfnamefont {H.}~\bibnamefont
  {Wang}},\ and\ \bibinfo {author} {\bibfnamefont {D.}~\bibnamefont {Li}},\
  }\bibfield  {title} {\bibinfo {title} {Robust interlayer coupling in
  two-dimensional perovskite/monolayer transition metal dichalcogenide
  heterostructures},\ }\href@noop {} {\bibfield  {journal} {\bibinfo  {journal}
  {ACS Nano}\ }\textbf {\bibinfo {volume} {14}},\ \bibinfo {pages} {10258}
  (\bibinfo {year} {2020}{\natexlab{b}})}\BibitemShut {NoStop}%
\bibitem [{\citenamefont {Huang}\ \emph {et~al.}(2021)\citenamefont {Huang},
  \citenamefont {Zhang}, \citenamefont {Ge}, \citenamefont {He}, \citenamefont
  {Zeng}, \citenamefont {Wang}, \citenamefont {Liu}, \citenamefont {Wang},\
  and\ \citenamefont {Pan}}]{huang2021enhancing}%
  \BibitemOpen
  \bibfield  {author} {\bibinfo {author} {\bibfnamefont {L.}~\bibnamefont
  {Huang}}, \bibinfo {author} {\bibfnamefont {D.}~\bibnamefont {Zhang}},
  \bibinfo {author} {\bibfnamefont {C.}~\bibnamefont {Ge}}, \bibinfo {author}
  {\bibfnamefont {M.}~\bibnamefont {He}}, \bibinfo {author} {\bibfnamefont
  {Z.}~\bibnamefont {Zeng}}, \bibinfo {author} {\bibfnamefont {Y.}~\bibnamefont
  {Wang}}, \bibinfo {author} {\bibfnamefont {S.}~\bibnamefont {Liu}}, \bibinfo
  {author} {\bibfnamefont {X.}~\bibnamefont {Wang}},\ and\ \bibinfo {author}
  {\bibfnamefont {A.}~\bibnamefont {Pan}},\ }\bibfield  {title} {\bibinfo
  {title} {Enhancing circular polarization of photoluminescence of
  two-dimensional {Ruddlesden--Popper} perovskites by constructing van der
  {Waals} heterostructures},\ }\href@noop {} {\bibfield  {journal} {\bibinfo
  {journal} {Applied Physics Letters}\ }\textbf {\bibinfo {volume} {119}},\
  \bibinfo {pages} {151101} (\bibinfo {year} {2021})}\BibitemShut {NoStop}%
\bibitem [{\citenamefont {Hautzinger}\ \emph {et~al.}(2024)\citenamefont
  {Hautzinger}, \citenamefont {Pan}, \citenamefont {Hayden}, \citenamefont
  {Ye}, \citenamefont {Jiang}, \citenamefont {Wilson}, \citenamefont
  {Phillips}, \citenamefont {Dong}, \citenamefont {Raulerson}, \citenamefont
  {Leahy}, \citenamefont {Jiang}, \citenamefont {Blackburn}, \citenamefont
  {Luther}, \citenamefont {Lu}, \citenamefont {Jungjohann}, \citenamefont
  {Vardeny}, \citenamefont {Berry}, \citenamefont {Alberi},\ and\ \citenamefont
  {Beard}}]{hautzinger2024room}%
  \BibitemOpen
  \bibfield  {author} {\bibinfo {author} {\bibfnamefont {M.~P.}\ \bibnamefont
  {Hautzinger}}, \bibinfo {author} {\bibfnamefont {X.}~\bibnamefont {Pan}},
  \bibinfo {author} {\bibfnamefont {S.~C.}\ \bibnamefont {Hayden}}, \bibinfo
  {author} {\bibfnamefont {J.~Y.}\ \bibnamefont {Ye}}, \bibinfo {author}
  {\bibfnamefont {Q.}~\bibnamefont {Jiang}}, \bibinfo {author} {\bibfnamefont
  {M.~J.}\ \bibnamefont {Wilson}}, \bibinfo {author} {\bibfnamefont {A.~J.}\
  \bibnamefont {Phillips}}, \bibinfo {author} {\bibfnamefont {Y.}~\bibnamefont
  {Dong}}, \bibinfo {author} {\bibfnamefont {E.~K.}\ \bibnamefont {Raulerson}},
  \bibinfo {author} {\bibfnamefont {I.~A.}\ \bibnamefont {Leahy}}, \bibinfo
  {author} {\bibfnamefont {C.-S.}\ \bibnamefont {Jiang}}, \bibinfo {author}
  {\bibfnamefont {J.~L.}\ \bibnamefont {Blackburn}}, \bibinfo {author}
  {\bibfnamefont {J.~M.}\ \bibnamefont {Luther}}, \bibinfo {author}
  {\bibfnamefont {Y.}~\bibnamefont {Lu}}, \bibinfo {author} {\bibfnamefont
  {K.}~\bibnamefont {Jungjohann}}, \bibinfo {author} {\bibfnamefont {Z.~V.}\
  \bibnamefont {Vardeny}}, \bibinfo {author} {\bibfnamefont {J.~J.}\
  \bibnamefont {Berry}}, \bibinfo {author} {\bibfnamefont {K.}~\bibnamefont
  {Alberi}},\ and\ \bibinfo {author} {\bibfnamefont {M.~C.}\ \bibnamefont
  {Beard}},\ }\bibfield  {title} {\bibinfo {title} {Room-temperature spin
  injection across a chiral perovskite/{III--V} interface},\ }\href@noop {}
  {\bibfield  {journal} {\bibinfo  {journal} {Nature}\ ,\ \bibinfo {pages}
  {307}} (\bibinfo {year} {2024})}\BibitemShut {NoStop}%
\bibitem [{\citenamefont {Ma}\ \emph {et~al.}(2021)\citenamefont {Ma},
  \citenamefont {Wang},\ and\ \citenamefont {Li}}]{ma2021recent}%
  \BibitemOpen
  \bibfield  {author} {\bibinfo {author} {\bibfnamefont {J.}~\bibnamefont
  {Ma}}, \bibinfo {author} {\bibfnamefont {H.}~\bibnamefont {Wang}},\ and\
  \bibinfo {author} {\bibfnamefont {D.}~\bibnamefont {Li}},\ }\bibfield
  {title} {\bibinfo {title} {Recent progress of chiral perovskites: materials,
  synthesis, and properties},\ }\href@noop {} {\bibfield  {journal} {\bibinfo
  {journal} {Advanced Materials}\ }\textbf {\bibinfo {volume} {33}},\ \bibinfo
  {pages} {2008785} (\bibinfo {year} {2021})}\BibitemShut {NoStop}%
\bibitem [{\citenamefont {Wang}\ \emph {et~al.}(2024)\citenamefont {Wang},
  \citenamefont {Zhu}, \citenamefont {Tan}, \citenamefont {Hao}, \citenamefont
  {Ye}, \citenamefont {Tang}, \citenamefont {Wang}, \citenamefont {Ma},
  \citenamefont {Sun}, \citenamefont {Zhang}, \citenamefont {Zheng},
  \citenamefont {Zhang}, \citenamefont {Choi}, \citenamefont {Choy},
  \citenamefont {Wu}, \citenamefont {Sun},\ and\ \citenamefont
  {Wang}}]{wang2024spin}%
  \BibitemOpen
  \bibfield  {author} {\bibinfo {author} {\bibfnamefont {Q.}~\bibnamefont
  {Wang}}, \bibinfo {author} {\bibfnamefont {H.}~\bibnamefont {Zhu}}, \bibinfo
  {author} {\bibfnamefont {Y.}~\bibnamefont {Tan}}, \bibinfo {author}
  {\bibfnamefont {J.}~\bibnamefont {Hao}}, \bibinfo {author} {\bibfnamefont
  {T.}~\bibnamefont {Ye}}, \bibinfo {author} {\bibfnamefont {H.}~\bibnamefont
  {Tang}}, \bibinfo {author} {\bibfnamefont {Z.}~\bibnamefont {Wang}}, \bibinfo
  {author} {\bibfnamefont {J.}~\bibnamefont {Ma}}, \bibinfo {author}
  {\bibfnamefont {J.}~\bibnamefont {Sun}}, \bibinfo {author} {\bibfnamefont
  {T.}~\bibnamefont {Zhang}}, \bibinfo {author} {\bibfnamefont
  {F.}~\bibnamefont {Zheng}}, \bibinfo {author} {\bibfnamefont
  {F.}~\bibnamefont {Zhang}}, \bibinfo {author} {\bibfnamefont {H.~W.}\
  \bibnamefont {Choi}}, \bibinfo {author} {\bibfnamefont {W.~C.}\ \bibnamefont
  {Choy}}, \bibinfo {author} {\bibfnamefont {D.}~\bibnamefont {Wu}}, \bibinfo
  {author} {\bibfnamefont {X.~W.}\ \bibnamefont {Sun}},\ and\ \bibinfo {author}
  {\bibfnamefont {K.}~\bibnamefont {Wang}},\ }\bibfield  {title} {\bibinfo
  {title} {Spin quantum dot light-emitting diodes enabled by {2D} chiral
  perovskite with spin-dependent carrier transport},\ }\href@noop {} {\bibfield
   {journal} {\bibinfo  {journal} {Advanced Materials}\ }\textbf {\bibinfo
  {volume} {36}},\ \bibinfo {pages} {2305604} (\bibinfo {year}
  {2024})}\BibitemShut {NoStop}%
\bibitem [{\citenamefont {Xiao}\ \emph {et~al.}(2012)\citenamefont {Xiao},
  \citenamefont {Liu}, \citenamefont {Feng}, \citenamefont {Xu},\ and\
  \citenamefont {Yao}}]{xiao2012coupled}%
  \BibitemOpen
  \bibfield  {author} {\bibinfo {author} {\bibfnamefont {D.}~\bibnamefont
  {Xiao}}, \bibinfo {author} {\bibfnamefont {G.-B.}\ \bibnamefont {Liu}},
  \bibinfo {author} {\bibfnamefont {W.}~\bibnamefont {Feng}}, \bibinfo {author}
  {\bibfnamefont {X.}~\bibnamefont {Xu}},\ and\ \bibinfo {author}
  {\bibfnamefont {W.}~\bibnamefont {Yao}},\ }\bibfield  {title} {\bibinfo
  {title} {Coupled spin and valley physics in monolayers of {MoS}$_2$ and other
  group-{VI} dichalcogenides},\ }\href@noop {} {\bibfield  {journal} {\bibinfo
  {journal} {Physical Review Letters}\ }\textbf {\bibinfo {volume} {108}},\
  \bibinfo {pages} {196802} (\bibinfo {year} {2012})}\BibitemShut {NoStop}%
\bibitem [{\citenamefont {Baranowski}\ \emph {et~al.}(2022)\citenamefont
  {Baranowski}, \citenamefont {Surrente},\ and\ \citenamefont
  {Plochocka}}]{baranowski2022two}%
  \BibitemOpen
  \bibfield  {author} {\bibinfo {author} {\bibfnamefont {M.}~\bibnamefont
  {Baranowski}}, \bibinfo {author} {\bibfnamefont {A.}~\bibnamefont
  {Surrente}},\ and\ \bibinfo {author} {\bibfnamefont {P.}~\bibnamefont
  {Plochocka}},\ }\bibfield  {title} {\bibinfo {title} {Two dimensional
  perovskites/transition metal dichalcogenides heterostructures: Puzzles and
  challenges},\ }\href@noop {} {\bibfield  {journal} {\bibinfo  {journal}
  {Israel Journal of Chemistry}\ }\textbf {\bibinfo {volume} {62}},\ \bibinfo
  {pages} {e202100120} (\bibinfo {year} {2022})}\BibitemShut {NoStop}%
\bibitem [{\citenamefont {Fu}\ \emph {et~al.}(2019)\citenamefont {Fu},
  \citenamefont {Wang}, \citenamefont {Liu}, \citenamefont {Dong},
  \citenamefont {Liu}, \citenamefont {Zheng}, \citenamefont {Chaturvedi},
  \citenamefont {Zhou}, \citenamefont {Hu}, \citenamefont {Zhu}, \citenamefont
  {Bo}, \citenamefont {Long},\ and\ \citenamefont {Liu}}]{fu2019ultrathin}%
  \BibitemOpen
  \bibfield  {author} {\bibinfo {author} {\bibfnamefont {Q.}~\bibnamefont
  {Fu}}, \bibinfo {author} {\bibfnamefont {X.}~\bibnamefont {Wang}}, \bibinfo
  {author} {\bibfnamefont {F.}~\bibnamefont {Liu}}, \bibinfo {author}
  {\bibfnamefont {Y.}~\bibnamefont {Dong}}, \bibinfo {author} {\bibfnamefont
  {Z.}~\bibnamefont {Liu}}, \bibinfo {author} {\bibfnamefont {S.}~\bibnamefont
  {Zheng}}, \bibinfo {author} {\bibfnamefont {A.}~\bibnamefont {Chaturvedi}},
  \bibinfo {author} {\bibfnamefont {J.}~\bibnamefont {Zhou}}, \bibinfo {author}
  {\bibfnamefont {P.}~\bibnamefont {Hu}}, \bibinfo {author} {\bibfnamefont
  {Z.}~\bibnamefont {Zhu}}, \bibinfo {author} {\bibfnamefont {F.}~\bibnamefont
  {Bo}}, \bibinfo {author} {\bibfnamefont {Y.}~\bibnamefont {Long}},\ and\
  \bibinfo {author} {\bibfnamefont {Z.}~\bibnamefont {Liu}},\ }\bibfield
  {title} {\bibinfo {title} {Ultrathin {Ruddlesden--Popper} perovskite
  heterojunction for sensitive photodetection},\ }\href@noop {} {\bibfield
  {journal} {\bibinfo  {journal} {Small}\ }\textbf {\bibinfo {volume} {15}},\
  \bibinfo {pages} {1902890} (\bibinfo {year} {2019})}\BibitemShut {NoStop}%
\bibitem [{\citenamefont {Fang}\ \emph {et~al.}(2019)\citenamefont {Fang},
  \citenamefont {Han}, \citenamefont {Robert}, \citenamefont {Semina},
  \citenamefont {Lagarde}, \citenamefont {Courtade}, \citenamefont {Taniguchi},
  \citenamefont {Watanabe}, \citenamefont {Amand}, \citenamefont {Urbaszek},
  \citenamefont {Glazov},\ and\ \citenamefont {Marie}}]{fang2019control}%
  \BibitemOpen
  \bibfield  {author} {\bibinfo {author} {\bibfnamefont {H.}~\bibnamefont
  {Fang}}, \bibinfo {author} {\bibfnamefont {B.}~\bibnamefont {Han}}, \bibinfo
  {author} {\bibfnamefont {C.}~\bibnamefont {Robert}}, \bibinfo {author}
  {\bibfnamefont {M.}~\bibnamefont {Semina}}, \bibinfo {author} {\bibfnamefont
  {D.}~\bibnamefont {Lagarde}}, \bibinfo {author} {\bibfnamefont
  {E.}~\bibnamefont {Courtade}}, \bibinfo {author} {\bibfnamefont
  {T.}~\bibnamefont {Taniguchi}}, \bibinfo {author} {\bibfnamefont
  {K.}~\bibnamefont {Watanabe}}, \bibinfo {author} {\bibfnamefont
  {T.}~\bibnamefont {Amand}}, \bibinfo {author} {\bibfnamefont
  {B.}~\bibnamefont {Urbaszek}}, \bibinfo {author} {\bibfnamefont
  {M.}~\bibnamefont {Glazov}},\ and\ \bibinfo {author} {\bibfnamefont
  {X.}~\bibnamefont {Marie}},\ }\bibfield  {title} {\bibinfo {title} {Control
  of the exciton radiative lifetime in van der waals heterostructures},\
  }\href@noop {} {\bibfield  {journal} {\bibinfo  {journal} {Physical Review
  Letters}\ }\textbf {\bibinfo {volume} {123}},\ \bibinfo {pages} {067401}
  (\bibinfo {year} {2019})}\BibitemShut {NoStop}%
\bibitem [{\citenamefont {Wang}\ \emph {et~al.}(2020)\citenamefont {Wang},
  \citenamefont {Zhang}, \citenamefont {Luo}, \citenamefont {Wang},
  \citenamefont {Zhu}, \citenamefont {Liang}, \citenamefont {Zhang},
  \citenamefont {Yong}, \citenamefont {Yu~Wong}, \citenamefont {Eda},
  \citenamefont {Smet},\ and\ \citenamefont {Wee}}]{wang2020optoelectronic}%
  \BibitemOpen
  \bibfield  {author} {\bibinfo {author} {\bibfnamefont {Q.}~\bibnamefont
  {Wang}}, \bibinfo {author} {\bibfnamefont {Q.}~\bibnamefont {Zhang}},
  \bibinfo {author} {\bibfnamefont {X.}~\bibnamefont {Luo}}, \bibinfo {author}
  {\bibfnamefont {J.}~\bibnamefont {Wang}}, \bibinfo {author} {\bibfnamefont
  {R.}~\bibnamefont {Zhu}}, \bibinfo {author} {\bibfnamefont {Q.}~\bibnamefont
  {Liang}}, \bibinfo {author} {\bibfnamefont {L.}~\bibnamefont {Zhang}},
  \bibinfo {author} {\bibfnamefont {J.~Z.}\ \bibnamefont {Yong}}, \bibinfo
  {author} {\bibfnamefont {C.~P.}\ \bibnamefont {Yu~Wong}}, \bibinfo {author}
  {\bibfnamefont {G.}~\bibnamefont {Eda}}, \bibinfo {author} {\bibfnamefont
  {J.~H.}\ \bibnamefont {Smet}},\ and\ \bibinfo {author} {\bibfnamefont
  {A.~T.}\ \bibnamefont {Wee}},\ }\bibfield  {title} {\bibinfo {title}
  {Optoelectronic properties of a van der {Waals} {WS}$_2$ monolayer/2{D}
  perovskite vertical heterostructure},\ }\href@noop {} {\bibfield  {journal}
  {\bibinfo  {journal} {ACS Applied Materials \& Interfaces}\ }\textbf
  {\bibinfo {volume} {12}},\ \bibinfo {pages} {45235} (\bibinfo {year}
  {2020})}\BibitemShut {NoStop}%
\bibitem [{\citenamefont {Zhou}\ \emph {et~al.}(2022)\citenamefont {Zhou},
  \citenamefont {Lai}, \citenamefont {Sun}, \citenamefont {Zhang},
  \citenamefont {Wang}, \citenamefont {Liu}, \citenamefont {Zhou},\ and\
  \citenamefont {Xie}}]{zhou2022van}%
  \BibitemOpen
  \bibfield  {author} {\bibinfo {author} {\bibfnamefont {H.}~\bibnamefont
  {Zhou}}, \bibinfo {author} {\bibfnamefont {H.}~\bibnamefont {Lai}}, \bibinfo
  {author} {\bibfnamefont {X.}~\bibnamefont {Sun}}, \bibinfo {author}
  {\bibfnamefont {N.}~\bibnamefont {Zhang}}, \bibinfo {author} {\bibfnamefont
  {Y.}~\bibnamefont {Wang}}, \bibinfo {author} {\bibfnamefont {P.}~\bibnamefont
  {Liu}}, \bibinfo {author} {\bibfnamefont {Y.}~\bibnamefont {Zhou}},\ and\
  \bibinfo {author} {\bibfnamefont {W.}~\bibnamefont {Xie}},\ }\bibfield
  {title} {\bibinfo {title} {Van der {Waals MoS}$_2$/two-dimensional perovskite
  heterostructure for sensitive and ultrafast sub-band-gap photodetection},\
  }\href@noop {} {\bibfield  {journal} {\bibinfo  {journal} {ACS Applied
  Materials \& Interfaces}\ }\textbf {\bibinfo {volume} {14}},\ \bibinfo
  {pages} {3356} (\bibinfo {year} {2022})}\BibitemShut {NoStop}%
\bibitem [{\citenamefont {Zhang}\ \emph {et~al.}(2020)\citenamefont {Zhang},
  \citenamefont {Linardy}, \citenamefont {Wang},\ and\ \citenamefont
  {Eda}}]{zhang2020excitonic}%
  \BibitemOpen
  \bibfield  {author} {\bibinfo {author} {\bibfnamefont {Q.}~\bibnamefont
  {Zhang}}, \bibinfo {author} {\bibfnamefont {E.}~\bibnamefont {Linardy}},
  \bibinfo {author} {\bibfnamefont {X.}~\bibnamefont {Wang}},\ and\ \bibinfo
  {author} {\bibfnamefont {G.}~\bibnamefont {Eda}},\ }\bibfield  {title}
  {\bibinfo {title} {Excitonic energy transfer in heterostructures of
  quasi-2{D} perovskite and monolayer {WS}$_2$},\ }\href@noop {} {\bibfield
  {journal} {\bibinfo  {journal} {ACS Nano}\ }\textbf {\bibinfo {volume}
  {14}},\ \bibinfo {pages} {11482} (\bibinfo {year} {2020})}\BibitemShut
  {NoStop}%
\bibitem [{\citenamefont {Karpinska}\ \emph {et~al.}(2021)\citenamefont
  {Karpinska}, \citenamefont {Liang}, \citenamefont {Kempt}, \citenamefont
  {Finzel}, \citenamefont {Kamminga}, \citenamefont {Dyksik}, \citenamefont
  {Zhang}, \citenamefont {Knodlseder}, \citenamefont {Maude}, \citenamefont
  {Baranowski}, \citenamefont {K{\l}optowski}, \citenamefont {Ye},
  \citenamefont {Kuc},\ and\ \citenamefont
  {Plochocka}}]{karpinska2021nonradiative}%
  \BibitemOpen
  \bibfield  {author} {\bibinfo {author} {\bibfnamefont {M.}~\bibnamefont
  {Karpinska}}, \bibinfo {author} {\bibfnamefont {M.}~\bibnamefont {Liang}},
  \bibinfo {author} {\bibfnamefont {R.}~\bibnamefont {Kempt}}, \bibinfo
  {author} {\bibfnamefont {K.}~\bibnamefont {Finzel}}, \bibinfo {author}
  {\bibfnamefont {M.}~\bibnamefont {Kamminga}}, \bibinfo {author}
  {\bibfnamefont {M.}~\bibnamefont {Dyksik}}, \bibinfo {author} {\bibfnamefont
  {N.}~\bibnamefont {Zhang}}, \bibinfo {author} {\bibfnamefont
  {C.}~\bibnamefont {Knodlseder}}, \bibinfo {author} {\bibfnamefont {D.~K.}\
  \bibnamefont {Maude}}, \bibinfo {author} {\bibfnamefont {M.}~\bibnamefont
  {Baranowski}}, \bibinfo {author} {\bibfnamefont {{\L}.}~\bibnamefont
  {K{\l}optowski}}, \bibinfo {author} {\bibfnamefont {J.}~\bibnamefont {Ye}},
  \bibinfo {author} {\bibfnamefont {A.}~\bibnamefont {Kuc}},\ and\ \bibinfo
  {author} {\bibfnamefont {P.}~\bibnamefont {Plochocka}},\ }\bibfield  {title}
  {\bibinfo {title} {Nonradiative energy transfer and selective charge transfer
  in a {WS}$_2$/({PEA})$_2${PbI}$_4$ heterostructure},\ }\href@noop {}
  {\bibfield  {journal} {\bibinfo  {journal} {ACS Applied Materials \&
  Interfaces}\ }\textbf {\bibinfo {volume} {13}},\ \bibinfo {pages} {33677}
  (\bibinfo {year} {2021})}\BibitemShut {NoStop}%
\bibitem [{\citenamefont {Karpi{\'n}ska}\ \emph {et~al.}(2022)\citenamefont
  {Karpi{\'n}ska}, \citenamefont {Jasi{\'n}ski}, \citenamefont {Kempt},
  \citenamefont {Ziegler}, \citenamefont {Sansom}, \citenamefont {Taniguchi},
  \citenamefont {Watanabe}, \citenamefont {Snaith}, \citenamefont {Surrente},
  \citenamefont {Dyksik}, \citenamefont {Maude}, \citenamefont {K{\l}optowski},
  \citenamefont {Chernikov}, \citenamefont {Kuc}, \citenamefont {Baranowski},\
  and\ \citenamefont {Plochocka}}]{karpinska2022interlayer}%
  \BibitemOpen
  \bibfield  {author} {\bibinfo {author} {\bibfnamefont {M.}~\bibnamefont
  {Karpi{\'n}ska}}, \bibinfo {author} {\bibfnamefont {J.}~\bibnamefont
  {Jasi{\'n}ski}}, \bibinfo {author} {\bibfnamefont {R.}~\bibnamefont {Kempt}},
  \bibinfo {author} {\bibfnamefont {J.}~\bibnamefont {Ziegler}}, \bibinfo
  {author} {\bibfnamefont {H.}~\bibnamefont {Sansom}}, \bibinfo {author}
  {\bibfnamefont {T.}~\bibnamefont {Taniguchi}}, \bibinfo {author}
  {\bibfnamefont {K.}~\bibnamefont {Watanabe}}, \bibinfo {author}
  {\bibfnamefont {H.}~\bibnamefont {Snaith}}, \bibinfo {author} {\bibfnamefont
  {A.}~\bibnamefont {Surrente}}, \bibinfo {author} {\bibfnamefont
  {M.}~\bibnamefont {Dyksik}}, \bibinfo {author} {\bibfnamefont
  {D.}~\bibnamefont {Maude}}, \bibinfo {author} {\bibfnamefont
  {{\L}.}~\bibnamefont {K{\l}optowski}}, \bibinfo {author} {\bibfnamefont
  {A.}~\bibnamefont {Chernikov}}, \bibinfo {author} {\bibfnamefont
  {A.}~\bibnamefont {Kuc}}, \bibinfo {author} {\bibfnamefont {M.}~\bibnamefont
  {Baranowski}},\ and\ \bibinfo {author} {\bibfnamefont {P.}~\bibnamefont
  {Plochocka}},\ }\bibfield  {title} {\bibinfo {title} {Interlayer excitons in
  {MoSe}$_2$/2{D} perovskite hybrid heterostructures--the interplay between
  charge and energy transfer},\ }\href@noop {} {\bibfield  {journal} {\bibinfo
  {journal} {Nanoscale}\ }\textbf {\bibinfo {volume} {14}},\ \bibinfo {pages}
  {8085} (\bibinfo {year} {2022})}\BibitemShut {NoStop}%
\bibitem [{\citenamefont {Soni}\ \emph {et~al.}(2024)\citenamefont {Soni},
  \citenamefont {Ghosal}, \citenamefont {Kundar}, \citenamefont {Pati},\ and\
  \citenamefont {Pal}}]{soni2024long}%
  \BibitemOpen
  \bibfield  {author} {\bibinfo {author} {\bibfnamefont {A.}~\bibnamefont
  {Soni}}, \bibinfo {author} {\bibfnamefont {S.}~\bibnamefont {Ghosal}},
  \bibinfo {author} {\bibfnamefont {M.}~\bibnamefont {Kundar}}, \bibinfo
  {author} {\bibfnamefont {S.~K.}\ \bibnamefont {Pati}},\ and\ \bibinfo
  {author} {\bibfnamefont {S.~K.}\ \bibnamefont {Pal}},\ }\bibfield  {title}
  {\bibinfo {title} {Long-lived interlayer excitons in
  {WS$_2$/Ruddlesden--Popper Perovskite van der Waals} heterostructures},\
  }\href@noop {} {\bibfield  {journal} {\bibinfo  {journal} {ACS Applied
  Materials \& Interfaces}\ }\textbf {\bibinfo {volume} {16}},\ \bibinfo
  {pages} {35841–35851} (\bibinfo {year} {2024})}\BibitemShut {NoStop}%
\bibitem [{\citenamefont {Ghosh}\ \emph {et~al.}(2025)\citenamefont {Ghosh},
  \citenamefont {Kamat},\ and\ \citenamefont {Hartland}}]{ghosh2025exciton}%
  \BibitemOpen
  \bibfield  {author} {\bibinfo {author} {\bibfnamefont {B.}~\bibnamefont
  {Ghosh}}, \bibinfo {author} {\bibfnamefont {P.~V.}\ \bibnamefont {Kamat}},\
  and\ \bibinfo {author} {\bibfnamefont {G.~V.}\ \bibnamefont {Hartland}},\
  }\bibfield  {title} {\bibinfo {title} {Exciton and hot charge carrier
  dynamics in a mos2/(pea) 2pbi4 2d heterostructure},\ }\href@noop {}
  {\bibfield  {journal} {\bibinfo  {journal} {ACS Nano}\ }\textbf {\bibinfo
  {volume} {19}},\ \bibinfo {pages} {25770} (\bibinfo {year}
  {2025})}\BibitemShut {NoStop}%
\bibitem [{\citenamefont {Qin}\ \emph {et~al.}(2025)\citenamefont {Qin},
  \citenamefont {Hu}, \citenamefont {Zheng}, \citenamefont {Zhou},
  \citenamefont {Song}, \citenamefont {Jiao}, \citenamefont {Ma}, \citenamefont
  {Jia},\ and\ \citenamefont {Jiang}}]{qin2025carrier}%
  \BibitemOpen
  \bibfield  {author} {\bibinfo {author} {\bibfnamefont {C.}~\bibnamefont
  {Qin}}, \bibinfo {author} {\bibfnamefont {Y.}~\bibnamefont {Hu}}, \bibinfo
  {author} {\bibfnamefont {S.}~\bibnamefont {Zheng}}, \bibinfo {author}
  {\bibfnamefont {Z.}~\bibnamefont {Zhou}}, \bibinfo {author} {\bibfnamefont
  {J.}~\bibnamefont {Song}}, \bibinfo {author} {\bibfnamefont {Z.}~\bibnamefont
  {Jiao}}, \bibinfo {author} {\bibfnamefont {S.}~\bibnamefont {Ma}}, \bibinfo
  {author} {\bibfnamefont {G.}~\bibnamefont {Jia}},\ and\ \bibinfo {author}
  {\bibfnamefont {Y.}~\bibnamefont {Jiang}},\ }\bibfield  {title} {\bibinfo
  {title} {Carrier transfer in {2D} perovskite/{WSe}$_2$ monolayer
  heterostructure},\ }\href@noop {} {\bibfield  {journal} {\bibinfo  {journal}
  {Journal of Luminescence}\ }\textbf {\bibinfo {volume} {286}},\ \bibinfo
  {pages} {121451} (\bibinfo {year} {2025})}\BibitemShut {NoStop}%
\bibitem [{\citenamefont {Rivera}\ \emph {et~al.}(2016)\citenamefont {Rivera},
  \citenamefont {Seyler}, \citenamefont {Yu}, \citenamefont {Schaibley},
  \citenamefont {Yan}, \citenamefont {Mandrus}, \citenamefont {Yao},\ and\
  \citenamefont {Xu}}]{rivera2016valley}%
  \BibitemOpen
  \bibfield  {author} {\bibinfo {author} {\bibfnamefont {P.}~\bibnamefont
  {Rivera}}, \bibinfo {author} {\bibfnamefont {K.~L.}\ \bibnamefont {Seyler}},
  \bibinfo {author} {\bibfnamefont {H.}~\bibnamefont {Yu}}, \bibinfo {author}
  {\bibfnamefont {J.~R.}\ \bibnamefont {Schaibley}}, \bibinfo {author}
  {\bibfnamefont {J.}~\bibnamefont {Yan}}, \bibinfo {author} {\bibfnamefont
  {D.~G.}\ \bibnamefont {Mandrus}}, \bibinfo {author} {\bibfnamefont
  {W.}~\bibnamefont {Yao}},\ and\ \bibinfo {author} {\bibfnamefont
  {X.}~\bibnamefont {Xu}},\ }\bibfield  {title} {\bibinfo {title}
  {Valley-polarized exciton dynamics in a {2D} semiconductor heterostructure},\
  }\href@noop {} {\bibfield  {journal} {\bibinfo  {journal} {Science}\ }\textbf
  {\bibinfo {volume} {351}},\ \bibinfo {pages} {688} (\bibinfo {year}
  {2016})}\BibitemShut {NoStop}%
\bibitem [{\citenamefont {Cadiz}\ \emph {et~al.}(2017)\citenamefont {Cadiz},
  \citenamefont {Courtade}, \citenamefont {Robert}, \citenamefont {Wang},
  \citenamefont {Shen}, \citenamefont {Cai}, \citenamefont {Taniguchi},
  \citenamefont {Watanabe}, \citenamefont {Carrere}, \citenamefont {Lagarde},
  \citenamefont {Manca}, \citenamefont {Amand}, \citenamefont {Renucci},
  \citenamefont {Tongay}, \citenamefont {Marie},\ and\ \citenamefont
  {Urbaszek}}]{cadiz2017excitonic}%
  \BibitemOpen
  \bibfield  {author} {\bibinfo {author} {\bibfnamefont {F.}~\bibnamefont
  {Cadiz}}, \bibinfo {author} {\bibfnamefont {E.}~\bibnamefont {Courtade}},
  \bibinfo {author} {\bibfnamefont {C.}~\bibnamefont {Robert}}, \bibinfo
  {author} {\bibfnamefont {G.}~\bibnamefont {Wang}}, \bibinfo {author}
  {\bibfnamefont {Y.}~\bibnamefont {Shen}}, \bibinfo {author} {\bibfnamefont
  {H.}~\bibnamefont {Cai}}, \bibinfo {author} {\bibfnamefont {T.}~\bibnamefont
  {Taniguchi}}, \bibinfo {author} {\bibfnamefont {K.}~\bibnamefont {Watanabe}},
  \bibinfo {author} {\bibfnamefont {H.}~\bibnamefont {Carrere}}, \bibinfo
  {author} {\bibfnamefont {D.}~\bibnamefont {Lagarde}}, \bibinfo {author}
  {\bibfnamefont {M.}~\bibnamefont {Manca}}, \bibinfo {author} {\bibfnamefont
  {T.}~\bibnamefont {Amand}}, \bibinfo {author} {\bibfnamefont
  {P.}~\bibnamefont {Renucci}}, \bibinfo {author} {\bibfnamefont
  {S.}~\bibnamefont {Tongay}}, \bibinfo {author} {\bibfnamefont
  {X.}~\bibnamefont {Marie}},\ and\ \bibinfo {author} {\bibfnamefont
  {B.}~\bibnamefont {Urbaszek}},\ }\bibfield  {title} {\bibinfo {title}
  {Excitonic linewidth approaching the homogeneous limit in {MoS}$_2$-based van
  der {Waals} heterostructures},\ }\href@noop {} {\bibfield  {journal}
  {\bibinfo  {journal} {Physical Review X}\ }\textbf {\bibinfo {volume} {7}},\
  \bibinfo {pages} {021026} (\bibinfo {year} {2017})}\BibitemShut {NoStop}%
\bibitem [{\citenamefont {Seitz}\ \emph {et~al.}(2019)\citenamefont {Seitz},
  \citenamefont {Gant}, \citenamefont {Castellanos-Gomez},\ and\ \citenamefont
  {Prins}}]{seitz2019long}%
  \BibitemOpen
  \bibfield  {author} {\bibinfo {author} {\bibfnamefont {M.}~\bibnamefont
  {Seitz}}, \bibinfo {author} {\bibfnamefont {P.}~\bibnamefont {Gant}},
  \bibinfo {author} {\bibfnamefont {A.}~\bibnamefont {Castellanos-Gomez}},\
  and\ \bibinfo {author} {\bibfnamefont {F.}~\bibnamefont {Prins}},\ }\bibfield
   {title} {\bibinfo {title} {Long-term stabilization of two-dimensional
  perovskites by encapsulation with hexagonal boron nitride},\ }\href@noop {}
  {\bibfield  {journal} {\bibinfo  {journal} {Nanomaterials}\ }\textbf
  {\bibinfo {volume} {9}},\ \bibinfo {pages} {1120} (\bibinfo {year}
  {2019})}\BibitemShut {NoStop}%
\bibitem [{\citenamefont {Ziegler}\ \emph {et~al.}(2020)\citenamefont
  {Ziegler}, \citenamefont {Zipfel}, \citenamefont {Meisinger}, \citenamefont
  {Menahem}, \citenamefont {Zhu}, \citenamefont {Taniguchi}, \citenamefont
  {Watanabe}, \citenamefont {Yaffe}, \citenamefont {Egger},\ and\ \citenamefont
  {Chernikov}}]{ziegler2020fast}%
  \BibitemOpen
  \bibfield  {author} {\bibinfo {author} {\bibfnamefont {J.~D.}\ \bibnamefont
  {Ziegler}}, \bibinfo {author} {\bibfnamefont {J.}~\bibnamefont {Zipfel}},
  \bibinfo {author} {\bibfnamefont {B.}~\bibnamefont {Meisinger}}, \bibinfo
  {author} {\bibfnamefont {M.}~\bibnamefont {Menahem}}, \bibinfo {author}
  {\bibfnamefont {X.}~\bibnamefont {Zhu}}, \bibinfo {author} {\bibfnamefont
  {T.}~\bibnamefont {Taniguchi}}, \bibinfo {author} {\bibfnamefont
  {K.}~\bibnamefont {Watanabe}}, \bibinfo {author} {\bibfnamefont
  {O.}~\bibnamefont {Yaffe}}, \bibinfo {author} {\bibfnamefont {D.~A.}\
  \bibnamefont {Egger}},\ and\ \bibinfo {author} {\bibfnamefont
  {A.}~\bibnamefont {Chernikov}},\ }\bibfield  {title} {\bibinfo {title} {Fast
  and anomalous exciton diffusion in two-dimensional hybrid perovskites},\
  }\href@noop {} {\bibfield  {journal} {\bibinfo  {journal} {Nano Letters}\
  }\textbf {\bibinfo {volume} {20}},\ \bibinfo {pages} {6674} (\bibinfo {year}
  {2020})}\BibitemShut {NoStop}%
\bibitem [{\citenamefont {Yang}\ \emph {et~al.}(2023)\citenamefont {Yang},
  \citenamefont {Hu}, \citenamefont {Chen},\ and\ \citenamefont
  {Li}}]{yang2023organic}%
  \BibitemOpen
  \bibfield  {author} {\bibinfo {author} {\bibfnamefont {D.}~\bibnamefont
  {Yang}}, \bibinfo {author} {\bibfnamefont {J.}~\bibnamefont {Hu}}, \bibinfo
  {author} {\bibfnamefont {Y.}~\bibnamefont {Chen}},\ and\ \bibinfo {author}
  {\bibfnamefont {D.}~\bibnamefont {Li}},\ }\bibfield  {title} {\bibinfo
  {title} {Organic cation modulation of interlayer exciton emission in
  two-dimensional perovskite/monolayer transition metal dichalcogenide
  heterostructures},\ }\href@noop {} {\bibfield  {journal} {\bibinfo  {journal}
  {Advanced Optical Materials}\ }\textbf {\bibinfo {volume} {11}},\ \bibinfo
  {pages} {2300398} (\bibinfo {year} {2023})}\BibitemShut {NoStop}%
\bibitem [{\citenamefont {Jasi{\'n}ski}\ \emph {et~al.}(2024)\citenamefont
  {Jasi{\'n}ski}, \citenamefont {Thompson}, \citenamefont {Palai},
  \citenamefont {{\'S}miertka}, \citenamefont {Dyksik}, \citenamefont
  {Taniguchi}, \citenamefont {Watanabe}, \citenamefont {Baranowski},
  \citenamefont {Maude}, \citenamefont {Surrente}, \citenamefont {Malic},\ and\
  \citenamefont {Plochocka}}]{jasinski2024control}%
  \BibitemOpen
  \bibfield  {author} {\bibinfo {author} {\bibfnamefont {J.}~\bibnamefont
  {Jasi{\'n}ski}}, \bibinfo {author} {\bibfnamefont {J.~J.}\ \bibnamefont
  {Thompson}}, \bibinfo {author} {\bibfnamefont {S.}~\bibnamefont {Palai}},
  \bibinfo {author} {\bibfnamefont {M.}~\bibnamefont {{\'S}miertka}}, \bibinfo
  {author} {\bibfnamefont {M.}~\bibnamefont {Dyksik}}, \bibinfo {author}
  {\bibfnamefont {T.}~\bibnamefont {Taniguchi}}, \bibinfo {author}
  {\bibfnamefont {K.}~\bibnamefont {Watanabe}}, \bibinfo {author}
  {\bibfnamefont {M.}~\bibnamefont {Baranowski}}, \bibinfo {author}
  {\bibfnamefont {D.~K.}\ \bibnamefont {Maude}}, \bibinfo {author}
  {\bibfnamefont {A.}~\bibnamefont {Surrente}}, \bibinfo {author}
  {\bibfnamefont {E.}~\bibnamefont {Malic}},\ and\ \bibinfo {author}
  {\bibfnamefont {P.}~\bibnamefont {Plochocka}},\ }\bibfield  {title} {\bibinfo
  {title} {Control of the valley polarization of monolayer {WSe$_2$ by
  Dexter}-like coupling},\ }\href@noop {} {\bibfield  {journal} {\bibinfo
  {journal} {2D Materials}\ }\textbf {\bibinfo {volume} {11}},\ \bibinfo
  {pages} {025007} (\bibinfo {year} {2024})}\BibitemShut {NoStop}%
\bibitem [{\citenamefont {Bergh{\"a}user}\ \emph {et~al.}(2018)\citenamefont
  {Bergh{\"a}user}, \citenamefont {Bernal-Villamil}, \citenamefont {Schmidt},
  \citenamefont {Schneider}, \citenamefont {Niehues}, \citenamefont {Erhart},
  \citenamefont {Michaelis~de Vasconcellos}, \citenamefont {Bratschitsch},
  \citenamefont {Knorr},\ and\ \citenamefont {Malic}}]{berghauser2018inverted}%
  \BibitemOpen
  \bibfield  {author} {\bibinfo {author} {\bibfnamefont {G.}~\bibnamefont
  {Bergh{\"a}user}}, \bibinfo {author} {\bibfnamefont {I.}~\bibnamefont
  {Bernal-Villamil}}, \bibinfo {author} {\bibfnamefont {R.}~\bibnamefont
  {Schmidt}}, \bibinfo {author} {\bibfnamefont {R.}~\bibnamefont {Schneider}},
  \bibinfo {author} {\bibfnamefont {I.}~\bibnamefont {Niehues}}, \bibinfo
  {author} {\bibfnamefont {P.}~\bibnamefont {Erhart}}, \bibinfo {author}
  {\bibfnamefont {S.}~\bibnamefont {Michaelis~de Vasconcellos}}, \bibinfo
  {author} {\bibfnamefont {R.}~\bibnamefont {Bratschitsch}}, \bibinfo {author}
  {\bibfnamefont {A.}~\bibnamefont {Knorr}},\ and\ \bibinfo {author}
  {\bibfnamefont {E.}~\bibnamefont {Malic}},\ }\bibfield  {title} {\bibinfo
  {title} {Inverted valley polarization in optically excited transition metal
  dichalcogenides},\ }\href@noop {} {\bibfield  {journal} {\bibinfo  {journal}
  {Nature Communications}\ }\textbf {\bibinfo {volume} {9}},\ \bibinfo {pages}
  {971} (\bibinfo {year} {2018})}\BibitemShut {NoStop}%
\bibitem [{\citenamefont {Liu}\ \emph {et~al.}(2013)\citenamefont {Liu},
  \citenamefont {Shan}, \citenamefont {Yao}, \citenamefont {Yao},\ and\
  \citenamefont {Xiao}}]{liu2013three}%
  \BibitemOpen
  \bibfield  {author} {\bibinfo {author} {\bibfnamefont {G.-B.}\ \bibnamefont
  {Liu}}, \bibinfo {author} {\bibfnamefont {W.-Y.}\ \bibnamefont {Shan}},
  \bibinfo {author} {\bibfnamefont {Y.}~\bibnamefont {Yao}}, \bibinfo {author}
  {\bibfnamefont {W.}~\bibnamefont {Yao}},\ and\ \bibinfo {author}
  {\bibfnamefont {D.}~\bibnamefont {Xiao}},\ }\bibfield  {title} {\bibinfo
  {title} {Three-band tight-binding model for monolayers of group-{VIB}
  transition metal dichalcogenides},\ }\href@noop {} {\bibfield  {journal}
  {\bibinfo  {journal} {Physical Review B—Condensed Matter and Materials
  Physics}\ }\textbf {\bibinfo {volume} {88}},\ \bibinfo {pages} {085433}
  (\bibinfo {year} {2013})}\BibitemShut {NoStop}%
\bibitem [{\citenamefont {Hill}\ \emph {et~al.}(2017)\citenamefont {Hill},
  \citenamefont {Rigosi}, \citenamefont {Raja}, \citenamefont {Chernikov},
  \citenamefont {Roquelet},\ and\ \citenamefont {Heinz}}]{hill2017exciton}%
  \BibitemOpen
  \bibfield  {author} {\bibinfo {author} {\bibfnamefont {H.~M.}\ \bibnamefont
  {Hill}}, \bibinfo {author} {\bibfnamefont {A.~F.}\ \bibnamefont {Rigosi}},
  \bibinfo {author} {\bibfnamefont {A.}~\bibnamefont {Raja}}, \bibinfo {author}
  {\bibfnamefont {A.}~\bibnamefont {Chernikov}}, \bibinfo {author}
  {\bibfnamefont {C.}~\bibnamefont {Roquelet}},\ and\ \bibinfo {author}
  {\bibfnamefont {T.~F.}\ \bibnamefont {Heinz}},\ }\bibfield  {title} {\bibinfo
  {title} {Exciton broadening in {WS}$_2$/graphene heterostructures},\
  }\href@noop {} {\bibfield  {journal} {\bibinfo  {journal} {Physical Review
  B}\ }\textbf {\bibinfo {volume} {96}},\ \bibinfo {pages} {205401} (\bibinfo
  {year} {2017})}\BibitemShut {NoStop}%
\bibitem [{\citenamefont {Trovatello}\ \emph {et~al.}(2022)\citenamefont
  {Trovatello}, \citenamefont {Katsch}, \citenamefont {Li}, \citenamefont
  {Zhu}, \citenamefont {Knorr}, \citenamefont {Cerullo},\ and\ \citenamefont
  {Dal~Conte}}]{trovatello2022disentangling}%
  \BibitemOpen
  \bibfield  {author} {\bibinfo {author} {\bibfnamefont {C.}~\bibnamefont
  {Trovatello}}, \bibinfo {author} {\bibfnamefont {F.}~\bibnamefont {Katsch}},
  \bibinfo {author} {\bibfnamefont {Q.}~\bibnamefont {Li}}, \bibinfo {author}
  {\bibfnamefont {X.}~\bibnamefont {Zhu}}, \bibinfo {author} {\bibfnamefont
  {A.}~\bibnamefont {Knorr}}, \bibinfo {author} {\bibfnamefont
  {G.}~\bibnamefont {Cerullo}},\ and\ \bibinfo {author} {\bibfnamefont
  {S.}~\bibnamefont {Dal~Conte}},\ }\bibfield  {title} {\bibinfo {title}
  {Disentangling many-body effects in the coherent optical response of {2D}
  semiconductors},\ }\href@noop {} {\bibfield  {journal} {\bibinfo  {journal}
  {Nano Letters}\ }\textbf {\bibinfo {volume} {22}},\ \bibinfo {pages} {5322}
  (\bibinfo {year} {2022})}\BibitemShut {NoStop}%
\bibitem [{\citenamefont {Yuan}\ \emph {et~al.}(2018)\citenamefont {Yuan},
  \citenamefont {Chung}, \citenamefont {Kuc}, \citenamefont {Wan},
  \citenamefont {Xu}, \citenamefont {Chen}, \citenamefont {Heine},\ and\
  \citenamefont {Huang}}]{yuan2018photocarrier}%
  \BibitemOpen
  \bibfield  {author} {\bibinfo {author} {\bibfnamefont {L.}~\bibnamefont
  {Yuan}}, \bibinfo {author} {\bibfnamefont {T.-F.}\ \bibnamefont {Chung}},
  \bibinfo {author} {\bibfnamefont {A.}~\bibnamefont {Kuc}}, \bibinfo {author}
  {\bibfnamefont {Y.}~\bibnamefont {Wan}}, \bibinfo {author} {\bibfnamefont
  {Y.}~\bibnamefont {Xu}}, \bibinfo {author} {\bibfnamefont {Y.~P.}\
  \bibnamefont {Chen}}, \bibinfo {author} {\bibfnamefont {T.}~\bibnamefont
  {Heine}},\ and\ \bibinfo {author} {\bibfnamefont {L.}~\bibnamefont {Huang}},\
  }\bibfield  {title} {\bibinfo {title} {Photocarrier generation from
  interlayer charge-transfer transitions in {WS}$_2$-graphene
  heterostructures},\ }\href@noop {} {\bibfield  {journal} {\bibinfo  {journal}
  {Science Advances}\ }\textbf {\bibinfo {volume} {4}},\ \bibinfo {pages}
  {e1700324} (\bibinfo {year} {2018})}\BibitemShut {NoStop}%
\bibitem [{\citenamefont {Kim}\ \emph {et~al.}(2014)\citenamefont {Kim},
  \citenamefont {Hong}, \citenamefont {Jin}, \citenamefont {Shi}, \citenamefont
  {Chang}, \citenamefont {Chiu}, \citenamefont {Li},\ and\ \citenamefont
  {Wang}}]{kim2014ultrafast}%
  \BibitemOpen
  \bibfield  {author} {\bibinfo {author} {\bibfnamefont {J.}~\bibnamefont
  {Kim}}, \bibinfo {author} {\bibfnamefont {X.}~\bibnamefont {Hong}}, \bibinfo
  {author} {\bibfnamefont {C.}~\bibnamefont {Jin}}, \bibinfo {author}
  {\bibfnamefont {S.-F.}\ \bibnamefont {Shi}}, \bibinfo {author} {\bibfnamefont
  {C.-Y.~S.}\ \bibnamefont {Chang}}, \bibinfo {author} {\bibfnamefont {M.-H.}\
  \bibnamefont {Chiu}}, \bibinfo {author} {\bibfnamefont {L.-J.}\ \bibnamefont
  {Li}},\ and\ \bibinfo {author} {\bibfnamefont {F.}~\bibnamefont {Wang}},\
  }\bibfield  {title} {\bibinfo {title} {Ultrafast generation of
  pseudo-magnetic field for valley excitons in {WSe}$_2$ monolayers},\
  }\href@noop {} {\bibfield  {journal} {\bibinfo  {journal} {Science}\ }\textbf
  {\bibinfo {volume} {346}},\ \bibinfo {pages} {1205} (\bibinfo {year}
  {2014})}\BibitemShut {NoStop}%
\bibitem [{\citenamefont {Schaibley}\ \emph {et~al.}(2016)\citenamefont
  {Schaibley}, \citenamefont {Rivera}, \citenamefont {Yu}, \citenamefont
  {Seyler}, \citenamefont {Yan}, \citenamefont {Mandrus}, \citenamefont
  {Taniguchi}, \citenamefont {Watanabe}, \citenamefont {Yao},\ and\
  \citenamefont {Xu}}]{schaibley2016directional}%
  \BibitemOpen
  \bibfield  {author} {\bibinfo {author} {\bibfnamefont {J.~R.}\ \bibnamefont
  {Schaibley}}, \bibinfo {author} {\bibfnamefont {P.}~\bibnamefont {Rivera}},
  \bibinfo {author} {\bibfnamefont {H.}~\bibnamefont {Yu}}, \bibinfo {author}
  {\bibfnamefont {K.~L.}\ \bibnamefont {Seyler}}, \bibinfo {author}
  {\bibfnamefont {J.}~\bibnamefont {Yan}}, \bibinfo {author} {\bibfnamefont
  {D.~G.}\ \bibnamefont {Mandrus}}, \bibinfo {author} {\bibfnamefont
  {T.}~\bibnamefont {Taniguchi}}, \bibinfo {author} {\bibfnamefont
  {K.}~\bibnamefont {Watanabe}}, \bibinfo {author} {\bibfnamefont
  {W.}~\bibnamefont {Yao}},\ and\ \bibinfo {author} {\bibfnamefont
  {X.}~\bibnamefont {Xu}},\ }\bibfield  {title} {\bibinfo {title} {Directional
  interlayer spin-valley transfer in two-dimensional heterostructures},\
  }\href@noop {} {\bibfield  {journal} {\bibinfo  {journal} {Nature
  Communications}\ }\textbf {\bibinfo {volume} {7}},\ \bibinfo {pages} {13747}
  (\bibinfo {year} {2016})}\BibitemShut {NoStop}%
\bibitem [{\citenamefont {Kim}\ \emph {et~al.}(2017)\citenamefont {Kim},
  \citenamefont {Jin}, \citenamefont {Chen}, \citenamefont {Cai}, \citenamefont
  {Zhao}, \citenamefont {Lee}, \citenamefont {Kahn}, \citenamefont {Watanabe},
  \citenamefont {Taniguchi}, \citenamefont {Tongay}, \citenamefont {Crommie},\
  and\ \citenamefont {Wang}}]{kim2017observation}%
  \BibitemOpen
  \bibfield  {author} {\bibinfo {author} {\bibfnamefont {J.}~\bibnamefont
  {Kim}}, \bibinfo {author} {\bibfnamefont {C.}~\bibnamefont {Jin}}, \bibinfo
  {author} {\bibfnamefont {B.}~\bibnamefont {Chen}}, \bibinfo {author}
  {\bibfnamefont {H.}~\bibnamefont {Cai}}, \bibinfo {author} {\bibfnamefont
  {T.}~\bibnamefont {Zhao}}, \bibinfo {author} {\bibfnamefont {P.}~\bibnamefont
  {Lee}}, \bibinfo {author} {\bibfnamefont {S.}~\bibnamefont {Kahn}}, \bibinfo
  {author} {\bibfnamefont {K.}~\bibnamefont {Watanabe}}, \bibinfo {author}
  {\bibfnamefont {T.}~\bibnamefont {Taniguchi}}, \bibinfo {author}
  {\bibfnamefont {S.}~\bibnamefont {Tongay}}, \bibinfo {author} {\bibfnamefont
  {M.~F.}\ \bibnamefont {Crommie}},\ and\ \bibinfo {author} {\bibfnamefont
  {F.}~\bibnamefont {Wang}},\ }\bibfield  {title} {\bibinfo {title}
  {Observation of ultralong valley lifetime in {WSe}$_2$/{MoS}$_2$
  heterostructures},\ }\href@noop {} {\bibfield  {journal} {\bibinfo  {journal}
  {Science Advances}\ }\textbf {\bibinfo {volume} {3}},\ \bibinfo {pages}
  {e1700518} (\bibinfo {year} {2017})}\BibitemShut {NoStop}%
\bibitem [{\citenamefont {Dal~Conte}\ \emph {et~al.}(2015)\citenamefont
  {Dal~Conte}, \citenamefont {Bottegoni}, \citenamefont {Pogna}, \citenamefont
  {De~Fazio}, \citenamefont {Ambrogio}, \citenamefont {Bargigia}, \citenamefont
  {D'Andrea}, \citenamefont {Lombardo}, \citenamefont {Bruna}, \citenamefont
  {Ciccacci}, \citenamefont {Ferrari}, \citenamefont {Cerullo},\ and\
  \citenamefont {Finazzi}}]{dal2015ultrafast}%
  \BibitemOpen
  \bibfield  {author} {\bibinfo {author} {\bibfnamefont {S.}~\bibnamefont
  {Dal~Conte}}, \bibinfo {author} {\bibfnamefont {F.}~\bibnamefont
  {Bottegoni}}, \bibinfo {author} {\bibfnamefont {E.}~\bibnamefont {Pogna}},
  \bibinfo {author} {\bibfnamefont {D.}~\bibnamefont {De~Fazio}}, \bibinfo
  {author} {\bibfnamefont {S.}~\bibnamefont {Ambrogio}}, \bibinfo {author}
  {\bibfnamefont {I.}~\bibnamefont {Bargigia}}, \bibinfo {author}
  {\bibfnamefont {C.}~\bibnamefont {D'Andrea}}, \bibinfo {author}
  {\bibfnamefont {A.}~\bibnamefont {Lombardo}}, \bibinfo {author}
  {\bibfnamefont {M.}~\bibnamefont {Bruna}}, \bibinfo {author} {\bibfnamefont
  {F.}~\bibnamefont {Ciccacci}}, \bibinfo {author} {\bibfnamefont {A.~C.}\
  \bibnamefont {Ferrari}}, \bibinfo {author} {\bibfnamefont {G.}~\bibnamefont
  {Cerullo}},\ and\ \bibinfo {author} {\bibfnamefont {M.}~\bibnamefont
  {Finazzi}},\ }\bibfield  {title} {\bibinfo {title} {Ultrafast valley
  relaxation dynamics in monolayer {MoS}$_2$ probed by nonequilibrium optical
  techniques},\ }\href@noop {} {\bibfield  {journal} {\bibinfo  {journal}
  {Physical Review B}\ }\textbf {\bibinfo {volume} {92}},\ \bibinfo {pages}
  {235425} (\bibinfo {year} {2015})}\BibitemShut {NoStop}%
\bibitem [{\citenamefont {Mai}\ \emph {et~al.}(2014)\citenamefont {Mai},
  \citenamefont {Barrette}, \citenamefont {Yu}, \citenamefont {Semenov},
  \citenamefont {Kim}, \citenamefont {Cao},\ and\ \citenamefont
  {Gundogdu}}]{mai2014many}%
  \BibitemOpen
  \bibfield  {author} {\bibinfo {author} {\bibfnamefont {C.}~\bibnamefont
  {Mai}}, \bibinfo {author} {\bibfnamefont {A.}~\bibnamefont {Barrette}},
  \bibinfo {author} {\bibfnamefont {Y.}~\bibnamefont {Yu}}, \bibinfo {author}
  {\bibfnamefont {Y.~G.}\ \bibnamefont {Semenov}}, \bibinfo {author}
  {\bibfnamefont {K.~W.}\ \bibnamefont {Kim}}, \bibinfo {author} {\bibfnamefont
  {L.}~\bibnamefont {Cao}},\ and\ \bibinfo {author} {\bibfnamefont
  {K.}~\bibnamefont {Gundogdu}},\ }\bibfield  {title} {\bibinfo {title}
  {Many-body effects in valleytronics: direct measurement of valley lifetimes
  in single-layer {MoS}$_2$},\ }\href@noop {} {\bibfield  {journal} {\bibinfo
  {journal} {Nano Letters}\ }\textbf {\bibinfo {volume} {14}},\ \bibinfo
  {pages} {202} (\bibinfo {year} {2014})}\BibitemShut {NoStop}%
\bibitem [{\citenamefont {Zhao}\ \emph {et~al.}(2020)\citenamefont {Zhao},
  \citenamefont {Su}, \citenamefont {Huang}, \citenamefont {Wu}, \citenamefont
  {Fong}, \citenamefont {Feng},\ and\ \citenamefont
  {Xiong}}]{zhao2020transient}%
  \BibitemOpen
  \bibfield  {author} {\bibinfo {author} {\bibfnamefont {W.}~\bibnamefont
  {Zhao}}, \bibinfo {author} {\bibfnamefont {R.}~\bibnamefont {Su}}, \bibinfo
  {author} {\bibfnamefont {Y.}~\bibnamefont {Huang}}, \bibinfo {author}
  {\bibfnamefont {J.}~\bibnamefont {Wu}}, \bibinfo {author} {\bibfnamefont
  {C.~F.}\ \bibnamefont {Fong}}, \bibinfo {author} {\bibfnamefont
  {J.}~\bibnamefont {Feng}},\ and\ \bibinfo {author} {\bibfnamefont
  {Q.}~\bibnamefont {Xiong}},\ }\bibfield  {title} {\bibinfo {title} {Transient
  circular dichroism and exciton spin dynamics in all-inorganic halide
  perovskites},\ }\href@noop {} {\bibfield  {journal} {\bibinfo  {journal}
  {Nature Communications}\ }\textbf {\bibinfo {volume} {11}},\ \bibinfo {pages}
  {5665} (\bibinfo {year} {2020})}\BibitemShut {NoStop}%
\end{thebibliography}%

\end{document}

%\title{Supplementary Information\\Optical polarization switching and spin injection in a (BA)$_2$PbI$_4$/WSe$_2$ heterostructure}
\title{Supplementary Information\\Spin injection and emission helicity switching in a 2D \texorpdfstring{perovskite/WSe$_2$}{perovskite/WSe2} heterostructure}

\author{Jakub Jasi{\'n}ski}
\altaffiliation{These authors contributed equally to the work}
\affiliation{Department of Experimental Physics, Faculty of Fundamental Problems of Technology, Wroclaw University of Science and Technology, 50-370 Wroclaw, Poland}
\affiliation{Laboratoire National des Champs Magn\'etiques Intenses, EMFL, CNRS UPR 3228, Universit{\'e} Grenoble Alpes, Universit{\'e} Toulouse, Universit{\'e} Toulouse 3, INSA-T, Grenoble and Toulouse, France}

\author{Francesco Gucci}
\altaffiliation{These authors contributed equally to the work}
\affiliation{Department of Physics, Politecnico di Milano, Piazza Leonardo da Vinci 32, 20133 Milan, Italy}

\author{Thomas Brumme}
\affiliation{Chair of Theoretical Chemistry, Technische Universit{\"a}t Dresden, Bergstra{\ss}e 66, 01069 Dresden, Germany}

\author{Swaroop Palai}
\affiliation{Laboratoire National des Champs Magn\'etiques Intenses, EMFL, CNRS UPR 3228, Universit{\'e} Grenoble Alpes, Universit{\'e} Toulouse, Universit{\'e} Toulouse 3, INSA-T, Grenoble and Toulouse, France}

\author{Armando Genco}
\affiliation{Department of Physics, Politecnico di Milano, Piazza Leonardo da Vinci 32, 20133 Milan, Italy}

\author{Alessandro Baserga}
\affiliation{Department of Physics, Politecnico di Milano, Piazza Leonardo da Vinci 32, 20133 Milan, Italy}

\author{Jonas D.\ Ziegler}
\affiliation{Institute of Applied Physics and W{\"u}rzburg-Dresden Cluster of Excellence ct.qmat, Technische Universit{\"a}t Dresden, 01062 Dresden, Germany}

\author{Takashi Taniguchi}
\affiliation{International Center for Materials Nanoarchitectonics, National Institute for Materials Science, Tsukuba, Ibaraki 305-004, Japan}

\author{Kenji Watanabe}
\affiliation{Research Center for Functional Materials, National Institute for Materials Science, Tsukuba, Ibaraki 305-004, Japan}

\author{Mateusz Dyksik}
\affiliation{Department of Experimental Physics, Faculty of Fundamental Problems of Technology, Wroclaw University of Science and Technology, 50-370 Wroclaw, Poland}

\author{Christoph Gadermaier}
\affiliation{Department of Physics, Politecnico di Milano, Piazza Leonardo da Vinci 32, 20133 Milan, Italy}

\author{Micha{\l} Baranowski}
\affiliation{Department of Experimental Physics, Faculty of Fundamental Problems of Technology, Wroclaw University of Science and Technology, 50-370 Wroclaw, Poland}

\author{Duncan K.\ Maude}
\affiliation{Laboratoire National des Champs Magn\'etiques Intenses, EMFL, CNRS UPR 3228, Universit{\'e} Grenoble Alpes, Universit{\'e} Toulouse, Universit{\'e} Toulouse 3, INSA-T, Grenoble and Toulouse, France}

\author{Alexey Chernikov}
\affiliation{Institute of Applied Physics and W{\"u}rzburg-Dresden Cluster of Excellence ct.qmat, Technische Universit{\"a}t Dresden, 01062 Dresden, Germany}

\author{Giulio Cerullo}
\affiliation{Department of Physics, Politecnico di Milano, Piazza Leonardo da Vinci 32, 20133 Milan, Italy}

\author{Agnieszka Kuc}
\affiliation{Helmholtz-Zentrum Dresden-Rossendorf, HZDR, Bautzner Landstra{\ss}e 400, 01328 Dresden, Germany}
\affiliation{Center for Advanced Systems Understanding, CASUS, Conrad-Schiedt-Stra{\ss}e 20, 02826 G\"orlitz, Germany}

\author{Stefano Dal Conte}\email{stefano.dalconte@polimi.it}
\affiliation{Department of Physics, Politecnico di Milano, Piazza Leonardo da Vinci 32, 20133 Milan, Italy}

\author{Paulina Plochocka}\email{paulina.plochocka@lncmi.cnrs.fr}
\affiliation{Department of Experimental Physics, Faculty of Fundamental Problems of Technology, Wroclaw University of Science and Technology, 50-370 Wroclaw, Poland}
\affiliation{Laboratoire National des Champs Magn\'etiques Intenses, EMFL, CNRS UPR 3228, Universit{\'e} Grenoble Alpes, Universit{\'e} Toulouse, Universit{\'e} Toulouse 3, INSA-T, Grenoble and Toulouse, France}

\author{Alessandro Surrente}\email{alessandro.surrente@pwr.edu.pl}
\affiliation{Department of Experimental Physics, Faculty of Fundamental Problems of Technology, Wroclaw University of Science and Technology, 50-370 Wroclaw, Poland}

\date{\today}

\renewcommand{\thefigure}{S\arabic{figure}} 
\renewcommand{\thepage}{S\arabic{page}}
\renewcommand{\thetable}{S\arabic{table}}
\renewcommand{\theequation}{S\arabic{equation}}

\maketitle

\section{Methods}\label{sec:methods}
\subsection*{Fabrication and optical spectroscopy}
The sample was fabricated using flakes mechanically exfoliated on PDMS and then assembled on the silicon substrate flake by flake via the dry transfer method \cite{castellanos2014deterministic}. Photoluminescence (PL) and reflectivity spectra have been obtained at cryogenic temperatures of $\sim\SI{4}{\kelvin}$, unless otherwise specified. For PL and PL excitation (PLE) measurements, the sample was mounted on the cold finger of a He flow cryostat. A pulsed Ti-sapphire laser (\SI{80}{\mega\hertz} repetition rate, \SI{150}{\femto\second} pulse width) was used to pump an optical parametric oscillator, which was used as the excitation source in PLE measurements, to achieve wavelength tuning in a wide range. For excitation power and temperature-dependent PL measurements, \SI{640}{\nano\metre} or \SI{630}{\nano\metre} wavelengths were used. A 50$\times$ microscope objective with a numerical aperture 0.55 was used to focus the excitation laser on the sample and collect the signal with a spatial resolution of $\sim\SI{1}{\micro\metre}$. The signal was directed to a spectrometer equipped with a liquid-nitrogen-cooled CCD camera. For reflectivity measurements, a white light source was used instead of the laser. Position dependent PL measurements were performed with a scanning step of \SI{1}{\micro\metre} along both horizontal and vertical directions. The scans were enabled by using an automated $xy$ translation stage on which the cryostat was mounted. For PL maps of the WSe$_2$ and the heterostructure region, a \SI{532}{\nano\metre} CW laser was used. For PL maps of the 2D perovskite flake, the wavelength of the Ti-sapphire laser was tuned to \SI{430}{\nano\metre} to achieve above band gap excitation conditions. %%%Time-resolved PL decays were measured by using a Supercontinuum white light laser by NKT Photonics as the excitation source with the repetition rate set to \SI{20}{\mega\hertz}. The broadband light pulse was spectrally filtered by a monochromator. The PL signal was spectrally filtered with a bandpass filter to select only the emission of the interlayer exciton. The spectrally filtered PL signal was fed into an avalanche photodiode. This detector was connected to a time-correlated single-photon counter module, which allowed us to reconstruct the time-resolved histogram of single-photon detection events.

\subsection*{Band structure calculation}
The model of the (BA)$_2$PbI$_4$/WSe$_2$ heterostructure was created using a coincidence lattice algorithm implemented in hetbuilder \cite{kempt_romankempthetbuilder_2021} with less than 0.2\% strain on the individual layers. The (BA)$_2$PbI$_4$/WSe$_2$ heterostructure in the low-temperature phase consists of 207 atoms and can be characterized by supercell vectors $m_1=(-1, -4)$, $m_2 = (4,7)$, $n_1 =(0,-1)$ and $n_2 = (1,2)$ with a rotation angle of $\theta = \SI{20.65}{\degree}$ and a vacuum spacing of \SI{100}{\angstrom}, see Ref.~\cite{kempt_romankempthetbuilder_2021} for details (other representation choices are possible). This model structure was relaxed such that the supercell matched the hexagonal symmetry of WSe$_2$. The difference from a fully relaxed lattice was below 0.1\%. To fully optimize the heterostructure model, we used FHI-aims \cite{Blum2009} employing the Perdew-Burke-Ernzerhof (PBE) functional \cite{Perdew1996} on tight tier 1 numeric atom-centered orbitals, including the nonlocal many-body dispersion correction (MBD-nl) \cite{Tkatchenko2013,Hermann2020}, and scalar relativistic corrections (ZORA) on a $6\times3\times1$ $\Gamma$-centered $k$-grid. The $\gamma$-angle was kept fixed, while forces and stresses were minimized until below \SI{0.01}{\eV\angstrom^{-1}}. The SCF parameters were automatically adjusted after three steps. The resulting lattice vectors of the fully optimized system are $a = \SI{8.759}{\angstrom}$, $b = \SI{18.433}{\angstrom}$, and $\gamma = \SI{91.94}{\degree}$. The electronic band structure, the Mulliken projections, and the density of states were calculated including spin-orbit coupling (SOC) and considering the dipole correction on a $12\times6\times1$ $\Gamma$-centered $k$-grid.

\subsection*{Transient reflectivity measurements}\label{sec:PumpProbe}
The experimental setup used for the transient reflectivity measurements has been described in detail in Ref.\ \cite{Genco2023}. It is based on an amplified Ti:sapphire laser, which generates $\sim\SI{100}{\femto\second}$ pulses at \SI{800}{\nano\metre} (\SI{1.55}{\eV}) with a repetition rate of \SI{2}{\kilo\hertz}. The output of the laser is split into two beams (pump and probe). The pump pulses can be continuously tuned in energy between \SI{470}{\nano\metre} (\SI{2.64}{\eV}) to \SI{780}{\nano\metre} (\SI{1.59}{\eV}) with a bandwidth of $\sim\SI{10}{\nano\metre}$, by a non-collinear optical parametric amplifier (NOPA). The probe pulses are focused onto a sapphire crystal to generate a broadband supercontinuum, whose spectrum can be appropriately filtered using band-pass filters. A mechanical chopper is used to modulate the pump beam at \SI{250}{\hertz}, while a mechanical delay line controls the pump-probe temporal delay. Both beams are collinearly focused onto the sample with an objective lens (\SI{8}{\milli\metre} focal length, numerical aperture 0.3). The diameter of the probe beam at the sample position is estimated to be $\sim\SI{3}{\micro\metre}$, while the pump is slightly defocused to obtain a larger spot size. The sample is housed in a closed-cycle helium cryostat. All pump-probe measurements are performed at a base temperature of \SI{8}{\kelvin}. The probe beam reflected from the sample is collected through the objective lens and directed to a dispersive spectrometer equipped with a CCD detector (Princeton Instruments PIXIS 100) triggered by the laser.

%%%Unfolding of the band structure was performed using the BandUp code \cite{unfold1, unfold2} on the heterostructure model from FHI-aims and by performing single-point calculations using the Vienna $ab-initio$ simulation package (VASP) \cite{Kresse1996} to obtain the wavefuction. The adjusted lattice parameters to match the hexagonal symmetry of WSe$_2$ are $a$ = 8.757~\AA, $b$ = 18.418~\AA, and $\gamma$ = 91.73$^\circ$.
%The projector-augmented wave (PAW) technique was used to describe the ionic potentials. We used the PBE functional \cite{Perdew1996}. The plane-wave cutoff energy and the total energy convergence criterion were set to 350~eV and 10$^{-6}$~eV, respectively. A vacuum region of at least 20~\AA\ was used to avoid spurious interactions between the periodically repeated layers. The BZ integrations were performed on a $\Gamma$-centered 3$\times$1$\times$1 Monkhorst-Pack $k$-grid. SOC was taken into account during the electronic structure calculations.
%The settings used in this calculation resulted in a 100 GB wavefunction file, thus, denser k-grid was not feasible.
%give structures as SI

\section{PL maps}\label{sec:PLmaps}
\begin{figure*}[!ht]
\centering
\includegraphics[width=1.0\linewidth]{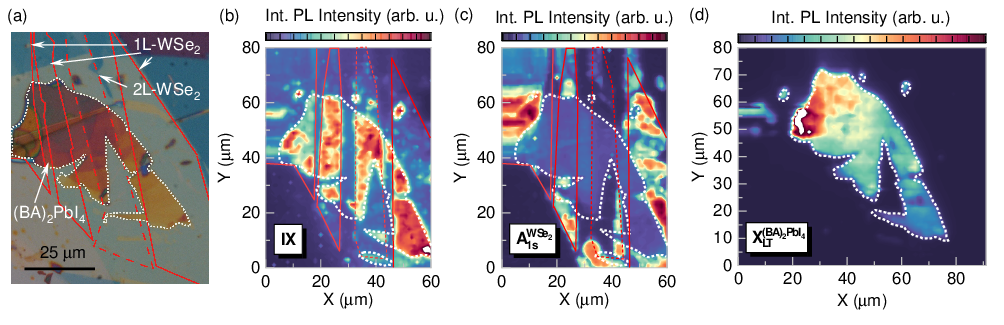}
\caption{(a) Micrograph of the investigated (BA)$_2$PbI$_4$/WSe$_2$ heterostructure. The contour of monolayer WSe$_2$ is highlighted by continuous red lines, that of bilayer WSe$_2$ by dashed red lines. The (BA)$_2$PbI$_4$ flake is traced in a dashed white line. PL intensity map of (b) the IX, (c) the intralayer WSe$_2$ exciton and (c) the intralayer exciton of the (BA)$_2$PbI$_4$ low temperature phase. %In the maps, the TMD flake is identified by red lines, while the perovskite flake contour is defined by a white dashed line.
}
\label{fig:MicrographPLmaps}
\end{figure*}

In Fig.\ \ref{fig:MicrographPLmaps}(a), we show a micrograph of the investigated sample. Mechanical exfoliation of a WSe$_2$ bulk crystal yields monolayer and bilayer areas, identified on the micrograph by a continuous and dashed contour, respectively. The heterostructure is finalized by transferring a mechanically exfoliated (BA)$_2$PbI$_4$ flake and encapsulating the structure in hBN.

To reveal the impact of the heterostructure on the PL spectrum, we measured PL maps. Fig.\ \ref{fig:MicrographPLmaps}(b) summarizes the spatial dependence of the intensity of the low energy peak attributed to the interlayer exciton (IX). As confirmed by comparing the PL map with the micrograph of Fig.\ \ref{fig:MicrographPLmaps}(a), this peak is only observed on areas where WSe$_2$ and (BA)$_2$PbI$_4$ overlap. This backs the assignment of the low energy peak in the PL spectrum of the heterostructure to the IX. In Fig.\ \ref{fig:MicrographPLmaps}(c), we show the PL map obtained by integrating the peak corresponding to the intralayer WSe$_2$ exciton. The intensity of this signal is considerably decreased in areas corresponding to the heterostructure, as expected for a type II band alignment. The spatial dependence of the exciton peak related to the low-temperature phase of (BA)$_2$PbI$_4$ is depicted in Fig.\ \ref{fig:MicrographPLmaps}(d). Although the 2D perovskite flake almost completely overlaps with WSe$_2$, there is a thin part of it that is not in direct contact. However, the PL intensity of (BA)$_2$PbI$_4$ seems to be fairly unaffected by the presence of a different material and is mainly influenced by the local thickness of the flake. The weak dependence of the intensity of the 2D perovskite PL suggests that only emission from regions close to WSe$_2$ is affected by the charge transfer process, while the effect on the global flake scale is relatively small.

\section{Detailed assignment of PL and reflectivity peaks}\label{sec:PL_reflectivity}
\begin{figure*}[!ht]
\centering
\includegraphics[width=1.0\linewidth]{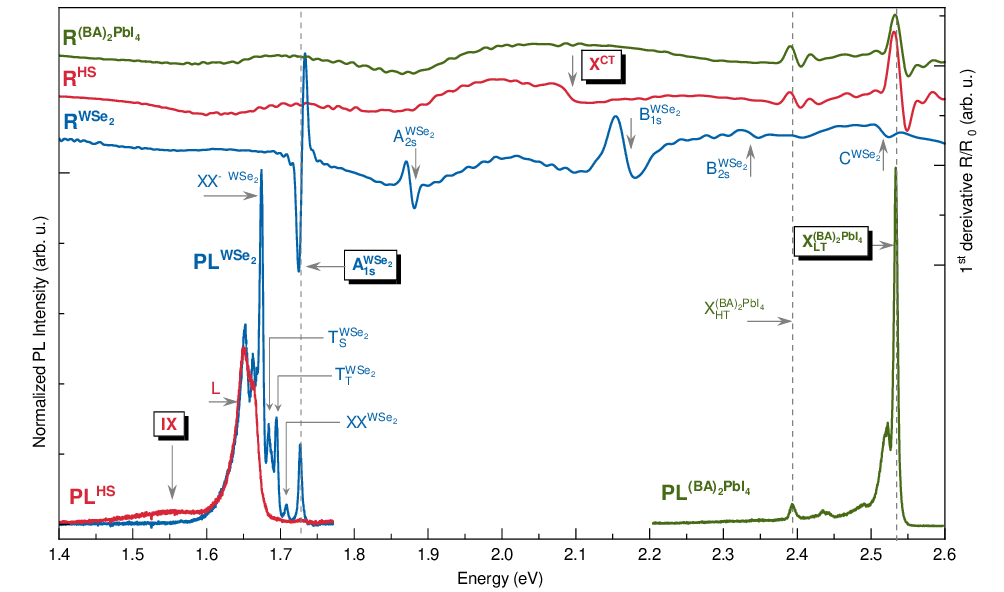}
\caption{Long range $\upmu$PL and first derivative of the reflectivity contrast spectra of isolated flakes and of the heterostructure (HS). The excitonic resonances are indicated by arrows.}
\label{fig:ReflectivityPL}
\end{figure*}

The differential reflectivity spectrum of WSe$_2$, shown in Fig.\ \ref{fig:ReflectivityPL}, consists of multiple resonances, assigned to 1s and 2s transitions of A (A$_{n\text{s}}^{\text{WSe}_2}$) and B (B$_{n\text{s}}^{\text{WSe}_2}$) excitons \cite{manca2017enabling,stier2018magnetooptics,chen2018superior,molas2019energy}, as well as to C exciton (C$^{\text{WSe}_2}$) \cite{arora2015excitonic,hanbicki2015measurement,frisenda2017micro}. The differential reflectivity of the (BA)$_2$PbI$_4$ flake presents a prominent resonance, which corresponds to the excitonic transition of the low temperature phase (X$_{\text{LT}}^{\text{(BA)}_2\text{PbI}_4}$) at $\sim\SI{2.54}{\eV}$. A weaker transition at $\sim\SI{2.38}{\eV}$, attributed to the exciton resonance of the high temperature phase (X$_{\text{HT}}^{\text{(BA)}_2\text{PbI}_4}$), can also be noted. The simultaneous observation of features related to both the low and high temperature phases has been reported for 2D perovskites deposited as thin films \cite{baranowski2019phase} or as exfoliated flakes \cite{yaffe2015excitons}. It is explained by a partial inhibition of the phase transitions of small domains as a consequence of the strong contact between the perovskite flake and the substrate \cite{yaffe2015excitons}. The PL spectrum of the WSe$_2$ monolayer consists of the neutral exciton peak and a series of lower energy features attributed to the biexciton (XX$^{\text{WSe}_2}$) \cite{ye2018efficient,chen2018coulomb,li2018revealing,barbone2018charge}, triplet (T$_{\text{T}}^{\text{WSe}_2}$) and singlet (T$_{\text{S}}^{\text{WSe}_2}$) charged excitons \cite{courtade2017charged,li2019direct,barbone2018charge}, and charged biexciton (XX$^{\text{-WSe}_2}$) \cite{ye2018efficient,chen2018coulomb,li2018revealing,barbone2018charge}, based on the energy separation with respect to the neutral exciton peak. The lowest energy peak originates from the recombination of localized excitons (L$^{\text{WSe}_2}$), which can contribute to the PL spectrum even after hBN encapsulation \cite{cadiz2017excitonic,ye2018efficient}. The PL spectrum of (BA)$_2$PbI$_4$ is dominated by the recombination of the exciton in the low temperature phase, with a weak, low energy peak, which testifies the presence of inclusions of high temperature domains within the excitation spot. In the differential reflectivity spectrum of the heterostructure of Fig.\ \ref{fig:ReflectivityPL}, the resonances of (BA)$_2$PbI$_4$ are visible, while those of WSe$_2$ are strongly weakened and broadened. This occurs due to the presence of strong electronic interactions and non-radiative charge or energy transfer across the heterostructure \cite{wu2019ultrafast,zhou2020controlling}. The most noticeable difference in the reflectivity spectrum of the heterostructure is a new feature, which we label X$^{\text{CT}}$, red-shifted with respect to the B exciton of WSe$_2$ of the isolated monolayer by $\sim \SI{80}{\milli\eV}$. This resonance is observed only on the heterostructure area.

\section{Band structure calculations}\label{sec:BandStructure} 
\begin{figure*}[!ht]
\centering
\includegraphics[width=1.0\linewidth]{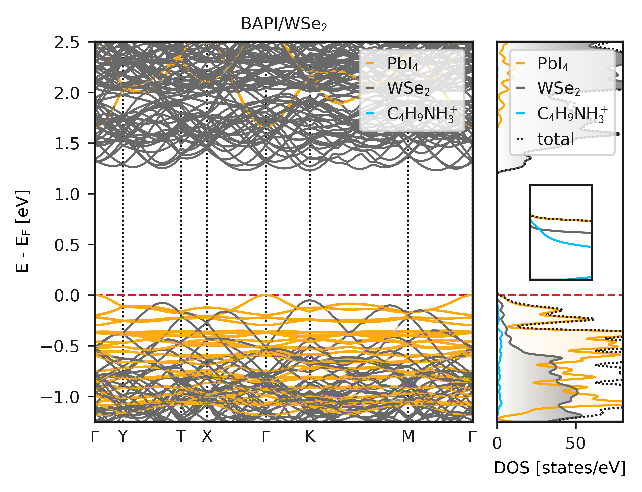}
\caption{(Left) Mulliken-projected band structure and (right) projected density of state (PDOS) of the (BA)$_2$PbI$_4$/WSe$_2$ monolayer heterostructure. The majority contributions to the electronic states are shown by colour-coding the specific building block mainly responsible for that state. The Fermi level is shifted to the top of valence band and set at zero (horizontal red dashed line). The top of the valence band is dominated by PbI$_4$ states, while the bottom of the conduction band by WSe$_2$ states. The inset in the PDOS shows a close-up view of the region close to the Fermi level, highlighting the states from the organic spacer between those of PbI$_4$ and WSe$_2$.}
\label{fig:BandStructure_DOS}
\end{figure*}

Prior calculations have demonstrated that 2D perovskite/TMD heterostructures are characterized by a type II band alignment \cite{zhang2020excitonic,chen2020robust,chen2020manipulation,wang2020optoelectronic,zhou2022van,karpinska2021nonradiative,karpinska2022interlayer}. However, to fully describe electronic properties, the heterostructure as a whole should be simulated \cite{karpinska2021nonradiative,karpinska2022interlayer}. The calculated band structure of our heterostructure is shown in Fig.~\ref{fig:BandStructure_DOS}. This simulation reveals that the edge of the conduction band is dominated by states belonging to the TMD monolayer. In the conduction band, the states of the organic barriers are energetically higher than those of both the TMD and the PbI$_4$ octahedral units, as schematically illustrated in Fig.~\ref{fig:PL_ChargeTransfer_PolarizationIX}(a,c). Thus, the organic spacer acts as a barrier for electron transfer in the conduction band. In contrast, the edge of the valence band is mainly contributed by the states of PbI$_4$. Moreover, a closer inspection to the projected density of states of Fig.~\ref{fig:BandStructure_DOS} reveals the presence of a non-vanishing contribution of states related to organic spacers at energies intermediate between those of PbI$_4$ and of the TMD monolayer close to the valence band edge. This leads to the cascaded band alignment in the valence band, also depicted in the inset of Fig.~\ref{fig:PL_ChargeTransfer_PolarizationIX}(a,c) of the main text, which favours the hole transfer from WSe$_2$ to PbI$_4$ \cite{karpinska2021nonradiative,karpinska2022interlayer}. The spatial separation of a photocreated electron-hole pair that results from this band alignment represents the prerequisite for the formation of the IX.

\section{Excitation power dependence of the PL spectrum}\label{sec:Pdependence}
\begin{figure*}[!ht]
\centering
\includegraphics[width=1.0\linewidth]{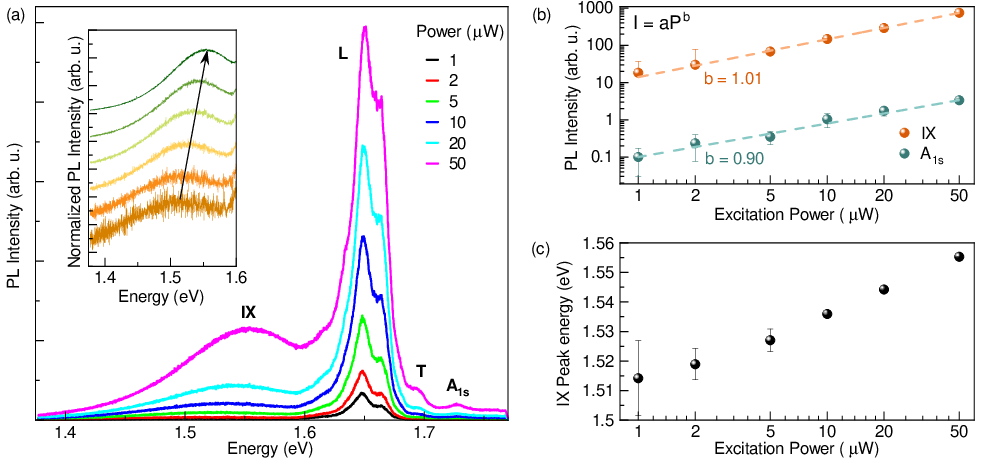}
\caption{(a) Main: Power dependent $\upmu$PL spectrum of the (BA)$_2$PbI$_4$/WSe$_2$ heterostructure. Inset: close-up on the interlayer exciton (IX) transition. Spectra in the inset are normalized and shifted vertically for better visibility. The arrow showcases the blue shift exhibited by this peak with increasing excitation power. (b) PL intensity of the WSe$_2$ A$_{1\text{s}}$ exciton and of the IXs as a function of the excitation power. The dashed lines are fits to a power law. The exponent $b$ of the power law extracted from the fit are indicated. (c) Energy of the IX as a function of the excitation power.}
\label{fig:PowerDependence}
\end{figure*}

An additional argument to support the interlayer origin of the low energy PL peak comes from its excitation power dependence. PL spectra measured at different excitation powers are shown in Fig.\ \ref{fig:PowerDependence}(a). The IX peak displays a notable blue shift with increasing power, which can be better appreciated in the close-up spectra shown in the inset of Fig.\ \ref{fig:PowerDependence}(a). By fitting the PL spectrum with Gaussian peaks, we extracted the intensity of the intralayer WSe$_2$ exciton and of the IX as a function of the excitation power, which we show in Fig.\ \ref{fig:PowerDependence}(b). These trends can be modelled by a power law ($I\propto P^b$), which yields an exponent $b \sim 1$ for both intralayer exciton and IX \cite{li2020dipolar}. This suggests that both these lines originate from the recombination of a single Coulomb bound electron-hole pair. The power used to excite the PL of the heterostructure was limited to \SI{50}{\micro\watt}, which did not allow us to reach the saturation level sometimes reported for IXs \cite{rivera2016valley,montblanch2021confinement}. The power dependence of the IX energy is summarized in Fig.\ \ref{fig:PowerDependence}(c). The approximately linear blue shift with increasing excitation power we observed reflects the expected behaviour of spatially indirect excitons in both coupled epitaxial quantum wells \cite{butov1999magneto} and TMD heterobilayers \cite{jauregui2019electrical}. This blue shift is connected to the out-of-plane dipole moment of IXs, wherein the electron and the hole are spatially separated in two different materials due to the type II band alignment, in combination with an increased density at high excitation power \cite{laikhtman2009exciton}. Thus, the blue shift of the low energy peak with increasing excitation power, characteristic of the dipole-dipole repulsion, confirms its interlayer nature.

\section{Temperature dependence of the PL spectrum}\label{sec:Tdependence}
\begin{figure*}[!ht]
\centering
\includegraphics[width=1.0\linewidth]{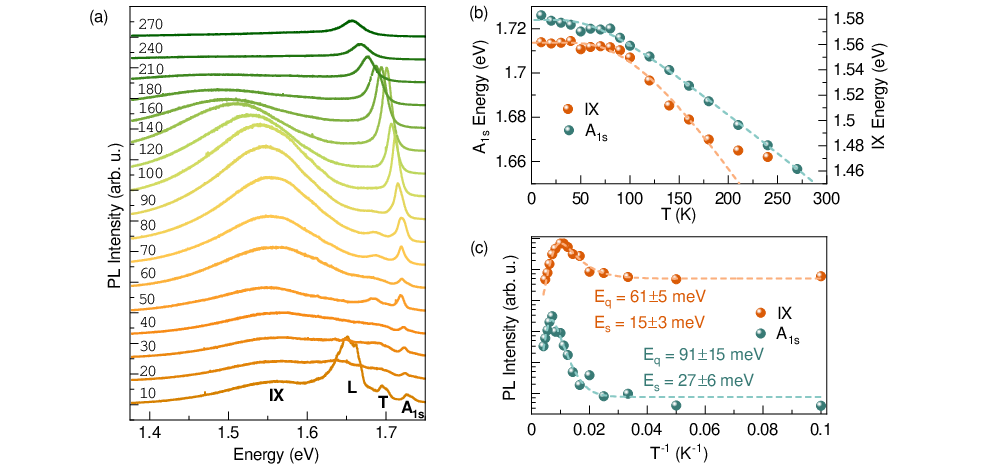}
\caption{(a) PL spectrum of the heterostructure excited below the (BA)$_2$PbI$_4$ bandgap measured at the temperatures indicated on the left of the panel. A$_{1\text{s}}$ identifies the PL peak related to the recombination of the free intralayer exciton, T labels the charged excitons, L the localized excitons and IX the interlayer exciton. (b) Energy of the intralayer exciton (blue points) and of the IX (red points) as a function of the temperature. The lines are fits to the model of Eq.\ \eqref{eq:TdepBandgap}. (c) Arrhenius plot of the intralayer exciton (blue points) and of the IX (red points) as a function of the inverse temperature. The lines are fits to the modified Arrhenius model detailed in Eq.\ \eqref{eq:Arrhenius}.}
\label{fig:TemperatureDependence}
\end{figure*}

The temperature dependence of the PL spectrum can provide complementary information about the origin of the peaks. We compare the salient features of the temperature dependence of the intralayer neutral exciton of WSe$_2$ and of the IX. The PL spectra of the heterostructure measured at temperatures ranging from \SIrange{10}{270}{\kelvin} are shown in Fig.\ \ref{fig:TemperatureDependence}(a). The PL is excited below the bandgap of (BA)$_2$PbI$_4$ to minimize the potentially disrupting influence from the low energy tail of the PL spectrum of the 2D perovskite flake. The temperature dependence of intralayer exciton species displays all the expected features for the PL spectrum of WSe$_2$ monolayers. All peaks red shift with increasing temperature. The temperature dependence of the energy of the intralayer neutral exciton is summarized in Fig.\ \ref{fig:TemperatureDependence}(b). This data is fitted with a function customarily used to analyze the temperature dependent band gap $E(T)$ of layered semiconductors \cite{o1991temperature,tongay2012thermally,ross2013electrical,huang2016probing}:
\begin{equation}
E(T) = E(0) - S\langle \hbar \omega \rangle\left( \coth\frac{\langle \hbar\omega \rangle}{2\kb T} -1 \right),
\label{eq:TdepBandgap}
\end{equation}
where $E(0)$ is the zero-temperature energy of the transition, $S$ is a parameter which describes phenomenologically the strength of the electron-phonon coupling and $\langle \hbar\omega \rangle$ represents an effective acoustic phonon frequency involved in the electron-phonon interaction. Our fit of the energy of the intralayer exciton yields $S \sim 2.15$ and $\langle \hbar\omega \rangle \sim \SI{17.5}{\milli\eV}$, which is comparable to prior reports \cite{huang2016probing}. Interestingly, the IX peak displays a more pronounced decrease of the energy with increasing temperature temperature, which results in  $S \sim 7.7$ and $\langle \hbar\omega \rangle \sim \SI{29.4}{\milli\eV}$. The more pronounced red shift of the IX is due to the combined contribution of the temperature dependence of the WSe$_2$ conduction band and of the (BA)$_2$PbI$_4$ valence band.
%the difference between the shift rates of the band gap of TMDs [$\SI{-3.45e-4}{\eV / \kelvin}$ obtained from the linear part of the measurements shown in Fig.\ \ref{fig:TemperatureDependence}(b)] and of 2D perovskites (\SI{-1.2e-5}{\eV / \kelvin} for encapsulated flake in a configuration similar to our sample, and \SI{-3.4e-4}{\eV / \kelvin} for non encapulated (BA)$_2$PbI$_4$ flakes \cite{gong2024boosting}). 
The larger $S$ and $\langle \hbar\omega \rangle$ found in the case of the IX are indicative of the contribution of the (BA)$_2$PbI$_4$ valence band to the interlayer transition. In 2D perovskites, the electron-phonon coupling is considerably stronger than in other semiconductors \cite{fu2021electronic}, owing to the pronounced ionicity of the lead halide perovskite lattice \cite{wilson2019dielectric}.
%Potential paper to cite: Also PRB 12 3258

An additional confirmation of the strong influence of both constituents on the properties of the IX can be found in the temperature dependence of its PL intensity. By looking at the PL shown in Fig.\ \ref{fig:TemperatureDependence}(a), we notice that the low energy peaks associated to bound excitons are rapidly quenched with increasing temperature \cite{godde2016exciton}. The intensity of the charged exciton peaks initially displays an increase with increasing temperature, and then it is quenched at higher temperatures \cite{godde2016exciton}. We focus here on the temperature dependence of the intralayer neutral exciton and of the IX. In Fig.\ \ref{fig:TemperatureDependence}(a), one can notice that the intensity of the intralayer neutral exciton strongly increases with increasing temperature even when all other lower energy transitions have disappeared. This is observed in systems in which the lowest-lying exciton state is optically dark. Part of the population of this optically inactive state is thermally excited to the higher energy optically bright state, which leads to a stronger PL at higher temperatures \cite{zhang2015experimental}. Importantly, a very similar trend is globally exhibited by the IX. This mirroring behaviour highlights how carriers that relax to the IX state originate from the conduction band of WSe$_2$ and therefore follow by the trends set by this material.

We fitted the temperature dependence of the intensity of the intralayer excitons and of the IX with a modified Arrhenius formula, used when an emissive state is supplied by a finite carrier reservoir \cite{huang2016probing,shibata1998negative}
\begin{equation}
    I(T) = I_0 \frac{1+B_{\text{s}}\text{e}^{-\frac{E_{\text{s}}}{\kb T}}}{1+B_{\text{q}} \text{e}^{-\frac{E_{\text{q}}}{\kb T}}},
    \label{eq:Arrhenius}
\end{equation}
where $B_{\text{s}}$ ($B_{\text{q}}$) is related to the ratio of radiative to non-radiative lifetimes of carriers which supply (are lost from) the emissive state \cite{fang2015investigation}, $E_{\text{s}}$ represents the activation energy of the processes which lead to the increase of the PL intensity, and $E_{\text{q}}$ designates the activation energy of the PL thermal quenching. Using Eq.\ \eqref{eq:Arrhenius}, we extracted the activation energies of the reservoir of the intralayer excitons and IXs, which amount to \SI{27}{\milli\eV} and \SI{15}{\milli\eV}, respectively. These energies do not necessarily correspond to the energy separation between dark and bright intralayer excitons, because this analysis does not account for competing non-radiative recombination channels, for the temperature dependence of the radiative recombination rate and for the presence of a thermal equilibrium between the different populations \cite{zhang2015experimental}. The lower activation energy of the thermal quenching of the IX compared to that of the intralayer exciton (\SI{61}{\milli\eV} versus \SI{91}{\milli\eV}) could be related to the lower binding energy of the IX and the presence of additional non-radiative recombination paths related to the quality of the TMD-2D perovskite interface, which instead likely influences less the PL of intralayer excitons.

%\section{Time-resolved photoluminescence of the interlayer exciton}
%We show in Fig.\ \ref{fig:TRPL_InterlayerExciton} the time-resolved PL signal of the IX measured at \SI{80}{\kelvin}. The decay is characterized by a fast and a slow component, which we evaluated with a biexponential fit to be $\tau_1 = \SI{1.74}{\nano\second}$ and $\tau_2 = \SI{9.70}{\nano\second}$, respectively. These decay times exceed those typical of both TMD monolayers, which are in the range of a few picoseconds \cite{godde2016exciton}, and those of (BA)$_2$PbI$_4$, which are in the range of a nanosecond \cite{guo2016electron}. This long life time is consistent with the large electron-hole spatial separation typical of IX species \cite{rivera2016valley}.
%\begin{figure*}[!ht]
%\centering
%\includegraphics[width=1.0\linewidth]{TRPL_InterlayerExciton}
%\caption{Time-resolved photoluminescence of the IX measured at \SI{80}{\kelvin}. The red curve shows the biexponential fitting of the decay. The two extracted decay times are indicated.}
%\label{fig:TRPL_InterlayerExciton}
%\end{figure*}

\section{Pump-probe reflectivity maps}\label{sec:PumpProbeMaps}
\begin{figure*}[!ht]
\centering
\includegraphics[width=0.5\linewidth]{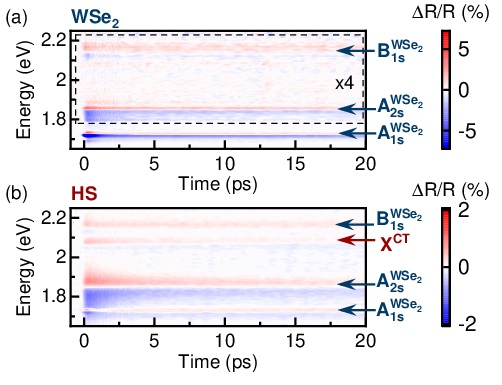}
\caption{Transient differential reflectivity map excited at \SI{2.64}{\eV} (above the perovskite band gap) as a function of the energy and of the pump-probe delay of (a) isolated WSe$_2$ monolayer and of (b) (BA)$_2$PbI$_4$/WSe$_2$ heterostructure. In (a), the high energy part of the transient spectrum has been rescaled to highlight the absence of other resonances than WSe$_2$ intralayer excitons. %(c) Differential reflectivity spectrum of the heterostructure and of the isolated WSe$_2$ monolayer excited at \SI{2.64}{\eV} (above the perovskite band gap) and extracted at a delay $\tau = \SI{500}{\femto\second}$. 
The excitonic resonances are indicated. The resonance corresponding to the WSe$_2$ A exciton (A$_{1\text{s}}^{\text{WSe}_2}$) has been rescaled for increased clarity.}
\label{fig:PumpProbeMapsSpectra}
\end{figure*}

Fig.\ \ref{fig:PumpProbeMapsSpectra}(a) illustrates the transient reflectivity $\Delta R/R$ maps as a function of the pump-probe temporal delay and the probe energy measured on the WSe$_2$ monolayer ($\Delta R$ is the transient variation of reflectivity upon the pump excitation, while $R$ indicates the static reflectivity). For both the measurements the energy of the pump is tuned above the quasi-particle bandgap of both the constituents of the heterostructure. The $\Delta R/R$ map of the WSe$_2$ displays a strong pump-probe signal around the energy of the A exciton. %The transient signal is characterized by a dispersive profile, which originates from the transient reduction of the exciton oscillator strength due to the Pauli blocking process and other processes leading to a transient broadening and energy shift of the excitonic peak \cite{trovatello2022disentangling}. 
At higher probe energies one can also identify the transient signals at the energies of the 2s and B excitons. In Fig.~\ref{fig:PumpProbeMapsSpectra}(b), we show the transient reflectivity map of the heterostructure. Although the WSe$_2$ resonances are still present, they are broadened and have lower intensity compared to those on the isolated WSe$_2$, due to the presence of additional non-radiative recombination channels, such as charge transfer \cite{hill2017exciton}. An additional transient signal appears at the energy of \SI{2.08}{\eV}, assigned to the X$^{\text{CT}}$ resonance. %Pump-probe spectra measured at a fixed delay time of \SI{500}{\femto\second} on the heterostructure and the bare WSe$_2$ flake are shown in Fig.\ \ref{fig:PumpProbeMapsSpectra}(c). Together with the resonance of the interlayer X$^{\text{CT}}$, we notice that the intralayer exciton signals of the heterostructure have lower intensity and increased broadening compared to those of the isolated WSe$_2$. The increased broadening of the excitons has previously been observed in other van der Waals heterostructures. It results from additional non-radiative channels such as the charge transfer \cite{hill2017exciton}.

\section{Dynamics of the charge transfer}\label{sec:ChargeTransferDynamics}
\begin{figure*}[!ht]
\centering
\includegraphics[width=1.0\linewidth]{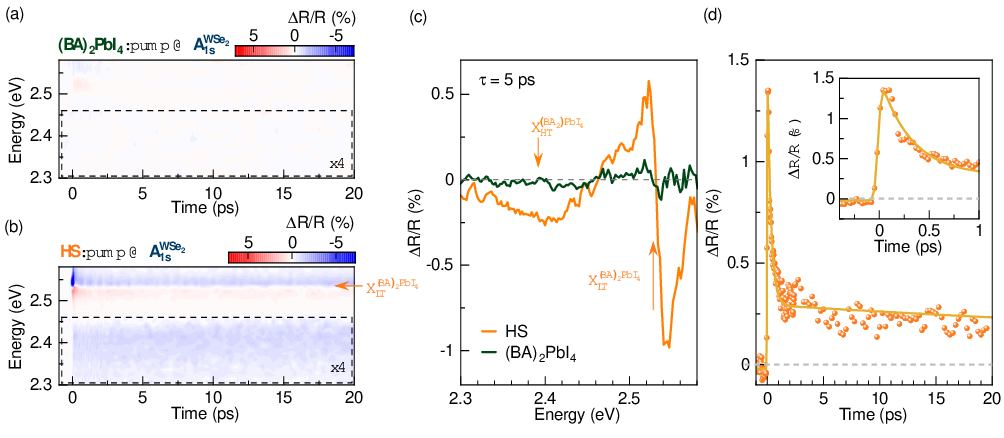}
\caption{Transient differential reflectivity map excited in resonance with the A exciton of WSe$_2$ A$_{1\text{s}}^{\text{WSe}_2}$ as a function of the energy and of the pump-probe delay of (a) bare (BA)$_2$PbI$_4$ flake and of (b) (BA)$_2$PbI$_4$/WSe$_2$ heterostructure. The low energy part of the transient spectrum has been rescaled to highlight the absence of transitions. (c) Differential reflectivity spectrum of the heterostructure and of the bare (BA)$_2$PbI$_4$ flake extracted at a delay $\tau = \SI{5}{\pico\second}$. The excitonic resonances are indicated. (d) Main: transient differential reflectivity of the excitonic transition of the low temperature phase measured on the heterostructure (X$_{\text{LT}}^{\text{(BA)}_2\text{PbI}_4}$) as a function of the pump-probe delay. Inset: Blow-up of the rise of the pump-probe signal of the (BA)$_2$PbI$_4$ exciton to highlight its instantaneous build-up following charge transfer. The line is the fit of an exponential rise and a bi-exponential decay model to the experimental data.}
\label{fig:ChargeTransfer}
\end{figure*}

To get more insight into the mechanisms that lead to the ultrafast formation of the IX, we measure the interlayer charge transfer dynamics using broadband pump-probe optical microscopy measurements. In this experiment, the sample is photoexcited by an ultrashort laser pulse (duration of $\sim\SI{100}{\femto\second}$), while the transient reflectivity spectrum $\Delta R/R$ is measured as a function of the temporal delay and the probe spectral range. In Fig.\ \ref{fig:ChargeTransfer}, we summarize the results obtained when the pump laser pulse is tuned in resonance with the 1s absorption peak of the WSe$_2$ monolayer A$_{1{\text{s}}}^{\text{WSe}_{2}}$, which is below the bandgap of (BA)$_2$PbI$_4$. In Fig.\ \ref{fig:ChargeTransfer}(a), we show the differential reflectivity spectrum measured on the isolated (BA)$_2$PbI$_4$ flake as a function of the energy and of the pump-probe delay. The lack of a transient response is consistent with the fact that excitation is performed in the transparency window. The faint signal detected at energies corresponding to the exciton resonance of the low-temperature phase of (BA)$_2$PbI$_4$ is likely related to the two-photon absorption of the 2D perovskite flake. The situation is radically different when the differential reflectivity spectrum is acquired in the heterostructure area, as shown in Fig.\ \ref{fig:ChargeTransfer}(b). Although the excitation is performed well below the band gap, we notice a distinct resonance at the energy of the (BA)$_2$PbI$_4$ exciton. The substantial difference between the optical responses of the heterostructure and of the isolated (BA)$_4$PbI$_4$ flake can be nicely seen by comparing their transient reflectivity spectra extracted at a delay of \SI{5}{\pico\second}, which are shown in Fig.\ \ref{fig:ChargeTransfer}(c). While the differential reflectivity of the isolated 2D perovskite flake is featureless, the transient spectrum measured on the heterostructure shows a pronounced signal at the energy of the exciton of the low temperature phase X$_{\text{LT}}^{\text{(BA)}_2\text{PbI}_4}$ and an additional broad signal at energies close the resonance associated to the high temperature phase of (BA)$_4$PbI$_4$ X$_{\text{HT}}^{\text{(BA)}_2\text{PbI}_4}$ frozen at low temperature \cite{baranowski2019phase,yaffe2015excitons}. The observation of these additional features is a direct proof of the interlayer hole transfer from the WSe$_2$ monolayer. The build-up of the X$_{\text{LT}}^{\text{(BA)}_2\text{PbI}_4}$ bleaching signal in Fig.\ \ref{fig:ChargeTransfer} is shorter than the instrument response function, which allows us to set an upper limit to the charge transfer time of $\sim\SI{100}{\femto\second}$. %The decay of the X$_{\text{LT}}^{\text{(BA)}_2\text{PbI}_4}$ bleaching is characterized by a fast component, with a time constant of \SI{266}{\femto\second}, possibly related to the relaxation process of the hot IXs after the hole transfer \cite{policht2023time}. The slower decay component present in the time window experimentally accessible is \SI{34}{\pico\second} and it can be interpreted as the fast decay process of the IX population \cite{ceballos2014ultrafast,wang2021phonon,zhou2020controlling,zimmermann2021ultrafast,zheng2021thickness,soni2024long} in the framework of a multiexponential decay \cite{soni2024long}.

\bibliography{Bibliography}